\documentclass[11pt]{article}
\usepackage[british,UKenglish,USenglish,english,american]{babel}
\usepackage[utf8]{inputenc}
\usepackage{latexsym}
\usepackage{graphicx}
\usepackage{amsfonts}
\usepackage{amssymb}
\usepackage{amsmath}
\usepackage{mathrsfs}
\usepackage{dsfont}
\usepackage{amsthm}
\usepackage{mdframed}
\usepackage{lipsum}
\usepackage{bbm}
\usepackage{color}
\usepackage{capt-of}
\usepackage{booktabs}
\usepackage{xcolor}
\usepackage{hyperref}
\usepackage[labelfont=bf]{caption}

\newcommand{\plus}{\raisebox{.4\height}{\scalebox{.75}{$+$}}}
\newcommand{\minus}{\raisebox{.4\height}{\scalebox{.75}{$-$}}}

\usepackage{tabularx}
\usepackage{booktabs}
\newcolumntype{L}[1]{>{\raggedright\arraybackslash}p{#1}} 
\newcolumntype{C}[1]{>{\centering\arraybackslash}p{#1}} 
\newcolumntype{R}[1]{>{\raggedleft\arraybackslash}p{#1}} 
\newcolumntype{Y}{>{\centering\arraybackslash}X}

\definecolor{myGreen}{RGB}{39,120,7}

\usepackage{subcaption}

\newtheorem{Prop}{Proposition}
\newtheorem{Lem}{Lemma}

\newtheorem{Def}{Definition}

\numberwithin{equation}{section}
\makeatletter
\newcommand*{\rom}[1]{\expandafter\@slowromancap\romannumeral #1@}
\makeatother
\title{Intra-Horizon Expected Shortfall and Risk Structure \\ in Models with Jumps}
\author{\sc \Large Walter Farkas\footnote{Email: walter.farkas@bf.uzh.ch} \hspace{1.5em}  Ludovic Mathys\footnote{Email: ludovic.mathys@bf.uzh.ch} \hspace{1.5em} Nikola Vasiljevi\'c\footnote{Email: nikola.vasiljevic@bf.uzh.ch} \vspace{1.0em} \\
       {\it Department of Banking and Finance, University of Zurich, Switzerland.} \vspace{0.2em} \\
			{\it Department of Mathematics, ETH Zurich, Switzerland.} \vspace{0.2em} \\
			{\it Swiss Finance Institute, Switzerland.} }
\date{}
\providecommand{\keywords}[1]{\textbf{Keywords:} #1}
\providecommand{\mscclass}[2]{\textbf{MSC (2010) Classification:} #1}
\providecommand{\jelclass}[3]{\textbf{JEL Classification:} #1}
\providecommand{\acknow}[4]{\textbf{Acknowledgements:} #1}

\begin{document}

\maketitle

\thispagestyle{empty}

\begin{abstract}
\noindent The present article deals with intra-horizon risk in models with jumps. Our general understanding of intra-horizon risk is along the lines of the approach taken in \cite{br04}, \cite{ro08}, \cite{bmk09}, \cite{bp10}, and \cite{lv20}. In particular, we believe that quantifying market risk by strictly relying on point-in-time measures cannot be deemed a satisfactory approach in general. Instead, we argue that complementing this approach by studying measures of risk that capture the magnitude of losses potentially incurred at any time of a trading horizon is necessary when dealing with (m)any financial position(s). To address this issue, we propose an intra-horizon analogue of the expected shortfall for general profit and loss processes and discuss its key properties. Our intra-horizon expected shortfall is well-defined for (m)any popular class(es) of Lévy processes encountered when modeling market dynamics and constitutes a coherent measure of risk, as introduced in \cite{cd04}. On the computational side, we provide a simple method to derive the intra-horizon risk inherent to popular Lévy dynamics. Our general technique relies on results for maturity-randomized first-passage probabilities and allows for a derivation of diffusion and single jump risk contributions. These theoretical results are complemented with an empirical analysis, where popular Lévy dynamics are calibrated to the S\&P~500~index and Brent crude oil data, and an analysis of the resulting intra-horizon risk is presented.
\end{abstract}
$\;$ \vspace{2em} \\
\noindent \keywords{Intra-Horizon Risk, Value at Risk, Expected Shortfall, Lévy Processes, Hyper-Exponential Distribution, Risk Decomposition.} \vspace{0.5em} \\
\noindent \mscclass{91-08, 91B25, 91B30, 91B70, 91B82, 91G60, 91G80.}{} \vspace{0.5em}\\
\noindent \jelclass{C32, C63, G01, G51.} \vspace{-0.5em} \\
\newpage
\section{Introduction}
For the past 20 years, the (point-in-time) value at risk, defined as quantile of the profit and loss distribution at the end of a predefined trading horizon, has been one of the most widely used measure of market risk for regulatory capital allocation (cf.~\cite{bcbs06}, \cite{bcbs19}). Despite its popularity, this risk measure has several major drawbacks that are all known to academics since many years. Firstly, it is mainly concerned with the probability of a loss and not with the actual loss size itself. In particular, when relying on the (point-in-time) value at risk, the distribution of the losses that exceed the quantile of interest is not taken into account. Secondly, it does not satisfy the subadditivity property for monetary risk measures (cf.~\cite{adeh99}, \cite{rt02}, \cite{at02}, \cite{embrechts14}) and therefore does not constitute a coherent measure of risk in the sense of \cite{adeh99}. Lastly, as a measure of market risk, the (point-in-time) value at risk does not capture the full magnitude of losses that may be potentially incurred at any time of a trading horizon (cf.\cite{br04}, \cite{ro08}, \cite{bmk09}). To address some of these issues, two streams have emerged in the academic literature. While certain authors (cf.~\cite{adeh99}, \cite{rt02}, \cite{at02}) introduced the (point-in-time) expected shortfall, a coherent risk measure that additionally depends on the tail of the underlying profit and loss distribution at the end of a predefined trading horizon, other authors (cf.~\cite{br04}, \cite{ro08}, \cite{bmk09}, \cite{bp10}, \cite{lv20}) developed a new path-dependent market risk measure, the intra-horizon value at risk. In light of the recent admission of the (point-in-time) expected shortfall in the new market risk framework of the Basel Accords and of the constant demand therein for sufficient conservatism in the risk estimates (cf.~\cite{bcbs19}), we believe that it is high time to reunify these two branches of the academic literature and to combine their advantages to propose a novel, coherent and path-dependent market risk measure that depends on all extremes in a trading horizon. This is the content of the present article that extensively discusses an intra-horizon version of the (point-in-time) expected shortfall. \vspace{1em} \\
\noindent Our paper's contribution is manifold and has both theoretical and practical relevance. On the theoretical side, we first generalize the current intra-horizon risk quantification approach of the literature (cf.~\cite{br04}, \cite{ro08}, \cite{bmk09}, \cite{bp10}, \cite{lv20}) and propose an intra-horizon analogue of the expected shortfall for general profit and loss processes. The resulting risk measure has several desirable properties and constitutes a coherent measure of risk in the sense of \cite{cd04}. Additionally, we show that our general ansatz is linked to simple (and maturity-randomized) first-passage probabilities of the underlying profit and loss process and subsequently use these relations to prove that the intra-horizon expected shortfall is well-defined for (m)any popular Lévy dynamics encountered in financial modeling. Secondly, we introduce diffusion and jump contributions to first-passage occurrences under Lévy models and present characterizations of diffusion and jump contributions to simple and maturity-randomized first-passage probabilities. These characterizations are then used in the following way: First, diffusion and jump risk contributions to the intra-horizon expected shortfall are inferred. Second, \mbox{(semi-)analytical} results for diffusion and jump contributions to maturity-randomized first-passage probabilities are derived under the class of hyper-exponential jump-diffusion processes by relying on option pricing methods (cf.~among others \cite{ca09}, \cite{cc09}, \cite{ck12}, \cite{cy13}, \cite{hm13}, \cite{ar16}, \cite{lv17}, \cite{lv20}). On the practical side, we introduce a simple and efficient ansatz to compute the intra-horizon risk inherent to popular Lévy dynamics. Our general approach consists in combining hyper-exponential jump-diffusion approximations to (pure jump) Lévy processes having completely monotone jumps with our (semi-)analytical results in this class of processes to recover approximate, though arbitrarily close intra-horizon risk results for the original process. In doing so, we rely on similar ideas to the ones introduced in \cite{amp07}, \cite{jp10}, and \cite{lv20} and subsequently use a mix of our results for maturity-randomized first-passage probabilities and a Laplace inversion algorithm (cf.~\cite{co07}) to arrive at intra-horizon expected shortfall results.\footnote{We choose the Gaver-Stehfest algorithm that has the particularity to allow for an inversion of the transform on the real line and that has been successfully used by several authors in the option pricing literature (cf.~\cite{kw03}, \cite{ki10}, \cite{hm13}, \cite{lv17}).}~Lastly, as an application of the techniques developed in this paper, we calibrate S\&P~500~index and Brent crude oil data to popular Lévy dynamics and investigate the intra-horizon risk inherent to a long position in these underlyings from January 1995 to September 2020. Our empirical findings reveal that even for high loss quantiles (i.e.,~low $\alpha$) the intra-horizon value at risk and the intra-horizon expected shortfall add conservatism to their point-in-time estimates. Additionally, they suggest that these risk measures have a very similar structure across jumps/jump clusters and that already a high contribution of their risk is due to only few, large -- in terms of the absolute jump size -- jump clusters.  \vspace{1em} \\
\noindent The remainder of this paper is structured as follows. In Section~\ref{SecQuant}, we introduce our general intra-horizon risk quantification approach as well as the notation used in the rest of the paper. This section also links intra-horizon risk to first-passage probabilities of the underlying profit and loss process. Section~\ref{IHRMJ} deals with intra-horizon risk in models with jumps and is divided into two parts. Firstly, our intra-horizon risk quantification approach is further developed under the assumption of Lévy dynamics and characterizations of simple (and maturity-randomized) first-passage probabilities are discussed. These characterizations are used in the second step to derive (semi-)analytical results for maturity-randomized first-passage probabilities under the class of hyper-exponential jump-diffusion processes. Section~\ref{SECapproxim} reviews hyper-exponential jump-diffusion approximations to pure jump Lévy processes having a completely monotone jump density as well as few adaptions. All the theoretical results of Sections~\ref{SecQuant}-\ref{SECapproxim} are lastly combined in Section~\ref{NUMres}, where popular Lévy dynamics are calibrated to S\&P~500~index data as well as to Brent crude oil data and the intra-horizon risk inherent to a long position in these underlyings is analyzed from January 1995 to September 2020. The paper concludes with Section~\ref{Paper3CONC}. All proofs and complementary results are presented in the Appendices (Appendix A, B and C).
\section{Fundamental Concepts of Intra-Horizon Risk Quantification}
\label{SecQuant}
\noindent We start by discussing the problem of evaluating the intra-horizon risk inherent to a financial position. To this end, we fix a time horizon $T >0$ and consider a filtered probability space $(\Omega, \mathcal{F}, \mathbf{F}, \mathbb{P})$, whose filtration $\mathbf{F} = \left( \mathcal{F}_{t} \right)_{t \in [0,T]}$ satisfies the usual conditions. We let $({\mathcal{P\&L}}_{t})_{t \in [0,T]}$ be an $\mathbf{F}$-adapted real-valued stochastic process and interpret its realizations as possible discounted profit and loss realizations of a given financial position over the valuation horizon $[0,T]$. Here, we do not necessarily require the process $({\mathcal{P\&L}}_{t})_{t \in [0,T]}$ to start at $\mathcal{P\&L}_{0}=0$ but allow instead for more flexibility in the choice of its initial value, i.e.,~we let $\mathcal{P\&L}_{0} = z$ for general $z \in \mathbb{R}$. This generalization proves useful, when rolling profits/losses of financial positions over multiple valuation periods. In this case, $\mathcal{P\&L}_{0}$ represents the profit/loss accumulated from the establishment of the position until the start of the valuation horizon under consideration.

\subsection{Intra-Horizon Value at Risk}
\noindent Our understanding of intra-horizon risk is in line with the ideas presented in \cite{br04}, \cite{bp10} and \cite{lv20}. As in these papers, our focus is on market risk, i.e.,~we only deal with risk that arises out of movements in the market price of financial assets and fully abstract from other risk types, such as e.g.,~counterparty risk. Additionally, we believe that quantifying market risk by strictly relying on point-in-time measures cannot be deemed a satisfactory approach in general. Instead, complementing this approach by studying measures of risk that capture the magnitude of losses potentially incurred at any time of a trading horizon is necessary for many asset types. This motivates the consideration of the minimum (discounted) profit and loss process, $(I_{t}^\mathcal{P\&L})_{t \in [0,T]}$, that is defined via
\begin{equation}
I_{t}^{\mathcal{P\&L}} := \inf \limits_{0 \leq u \leq t} \mathcal{P\&L}_{u}, \hspace{2em} t \in [0,T].
\end{equation}
\noindent Under this notation, the following definition of the intra-horizon value at risk was presented in \cite{bp10}.
\begin{Def}[Intra-Horizon Value at Risk] \noindent Let $T >0$ and $
\alpha \in (0,1)$ be fixed. The level-$\alpha$ intra-horizon value at risk associated with the (discounted) profit and loss process $(\mathcal{P\&L}_{t})_{t \in [0,T]}$ over the time interval~$[0,T]$,~$\mbox{iV@R}_{\alpha,T}(\mathcal{P\&L})$, is defined as
\begin{equation}
\mbox{iV@R}_{\alpha,T}(\mathcal{P\&L}) := \mbox{V@R}_{\alpha}\big(I_{T}^{\mathcal{P\&L}}
\big),
\end{equation}
\noindent where, for a random variable $Y$, $\mbox{V@R}_{\alpha}(Y)$ denotes the (point-in-time) value at risk to the level $\alpha$ and is defined as 
\begin{equation}
\mbox{V@R}_{\alpha}(Y) := - q_{\alpha}(Y).
\end{equation}
Here, $q_{\alpha}(Y)$ denotes the upper $\alpha$-quantile of $Y$, which is obtained via
\begin{equation}
q_{\alpha}(Y) := \sup \{ y \in \mathbb{R}: \; \mathbb{P} \left(Y \leq y \right
) \leq \alpha \}.
\end{equation}
\end{Def}
\noindent Specifying the intra-horizon value at risk in the above sense is closely linked to the theory of ruin that has received a lot of attention in insurance mathematics.  Indeed, the above definition can be re-stated in terms of first-passage probabilities, as
\begin{equation}
\mbox{\it iV@R}_{\alpha,T}(\mathcal{P\&L}) : = - \sup \left\{ \ell \in \mathbb{R}: \; \mathbb{P}_{z} \big( \tau_{\ell}^{\mathcal{P\&L},-} \leq T \big) \leq \alpha \right\}, 
\label{Repre}
\end{equation}
where we denote by $\mathbb{P}_{z}$ the probability measure under which the process $(\mathcal{P\&L})_{ t \in [0,T]} $ starts at $z \in \mathbb{R}$. For $\ell \in \mathbb{R}$ and a given stochastic process $(Y_{t})_{t \in [0,T]}$, we use the notation
\begin{equation}
\tau_{\ell}^{Y,\pm} := \inf \{ t \geq 0: \, \pm Y_{t} \geq \pm \ell \}, \hspace{1.5em} \mbox{with} \hspace{1.5em} \inf \emptyset = \infty. 
\end{equation}
\noindent This representation will prove useful, as it will allow us to combine properties of first-passage probabilities to subsequently recover intra-horizon value-at-risk results via standard numerical methods. 

\subsection{Intra-Horizon Expected Shortfall}
\noindent Although the intra-horizon value at risk already accounts for intra-horizon features, it suffers from two major drawbacks. Firstly, it is mainly concerned with the probability of a loss and not with the actual loss size itself. In particular, when relying on the intra-horizon value at risk, the distribution of the losses that exceed the quantile of interest is not taken into account. Secondly, it is not subadditive (cf.~\cite{bp10}) and therefore does not constitute a coherent measure of risk in the sense of \cite{cd04}.\footnote{Like its point-in-time counterpart, the intra-horizon value at risk may become superadditive for certain profit and loss processes and therefore defies the notion of diversification.}~This is due to the fact that the intra-horizon value at risk consists of an adaption of the point-in-time value at risk, which is known to have the same deficiencies. To address these major shortcomings, we propose a (market) risk measure that defines an intra-horizon analogue of the expected shortfall. This is the content of the next definition, where we use the notation $\mathbb{E}_{z}^{\mathbb{P}}\left[ \cdot \right]$ to indicate expectation under the measure $\mathbb{P}_{z}$.
\begin{Def}[Intra-Horizon Expected Shortfall] \noindent Let $T>0$ and $\alpha \in (0,1)$ be fixed and assume that $\mathbb{E}^{\mathbb{P}}_{z} \left[ | I_{T}^{\mathcal{P\&L}} | \right] < \infty $. Then, the level-$\alpha$ intra-horizon expected shortfall associated with the (discounted) profit and loss process $(\mathcal{P\&L}_{t})_{t \in [0,T]}$ over the time horizon $[0,T]$, $\mbox{iES}_{\alpha,T}(\mathcal{P\&L})$, is defined as
\begin{equation}
\mbox{iES}_{\alpha,T}(\mathcal{P\&L}) :=  \frac{1}{\alpha} \int \limits_{0}^{
\alpha} \mbox{iV@R}_{\gamma,T}(\mathcal{P\&L}) \, d\gamma .
\label{DefiES}
\end{equation}
\end{Def}
\noindent It is not hard to see that the above specification of the intra-horizon expected shortfall overcomes both major shortcomings of the intra-horizon value at risk. Indeed, a brief look at equation (\ref{DefiES}) reveals that our intra-horizon expected shortfall defines a coherent measure of risk in the sense of \cite{cd04} that additionally depends on the distributional properties in the tail of the underlying profit and loss process. In fact, cash-invariance, monotonicity and positive homogeneity of (\ref{DefiES}) directly follow from the corresponding properties of the point-in-time expected shortfall via the identity
\begin{equation}
\mbox{\it iES}_{\alpha,T}(\mathcal{P\&L}) = \frac{1}{\alpha} \int \limits_{0}^{
\alpha} \mbox{\it V@R}_{\gamma}(I_{T}^{\mathcal{P\&L}}) \, d\gamma =: ES_{\alpha}\big(I_{T}^{\mathcal{P\&L}}\big).
\label{ZZZes}
\end{equation}
\noindent Additionally, subadditivity is obtained by relying on the monotonicity and subadditivity properties of the point-in-time expected shortfall and the fact that for two profit and loss processes $\left( \mathcal{P\&L}_{t}^{1} \right)_{t \in [0,T]}$ and $\left( \mathcal{P\&L}_{t}^{2} \right)_{t \in [0,T]}$ the following identity holds
\begin{equation}
I_{T}^{\mathcal{P\&L}^{1} + \mathcal{P\&L}^{2}} \, \geq I_{T}^{\mathcal{P\&L}^{1}} + I_{T}^{\mathcal{P\&L}^{2}}.
\end{equation}
\noindent Then,
\begin{align}
\mbox{\it iES}_{\alpha,T}\big(\mathcal{P\&L}^{1} + \mathcal{P\&L}^{2}\big) & = \mbox{\it ES}_{\alpha} \big( I_{T}^{\mathcal{P\&L}^{1}+ \mathcal{P\&L}^{2}} \big) \leq \mbox{\it ES}_{\alpha} \big( I_{T}^{\mathcal{P\&L}^{1}} + I_{T}^{\mathcal{P\&L}^{2}}\big) \nonumber \\
& \leq \mbox{\it ES}_{\alpha}\big(I_{T}^{\mathcal{P\&L}^{1}} \big)  +  \mbox{\it ES}_{\alpha}\big( I_{T}^{\mathcal{P\&L}^{2}} \big) = \mbox{\it iES}_{\alpha,T}\big(\mathcal{P\&L}^{1} \big)  +  \mbox{\it iES}_{\alpha,T}\big(\mathcal{P\&L}^{2}\big),
\end{align}
\noindent which is the subadditivity property. \vspace{1em} \\
\noindent The next proposition indicates the link between the intra-horizon expected shortfall and the theory of ruin. In particular, it shows that, whenever well-defined, the difference between intra-horizon expected shortfall and intra-horizon value at risk can be computed for any given profit and loss process based on first-passage probabilities. The proof is provided in Appendix~A.
\begin{Prop}
\label{lem1}
\noindent Let $T>0$ and $\alpha \in (0,1)$ be fixed and assume that $ \mathbb{E}^{\mathbb{P}}_{z} \left[ | I_{T}^{\mathcal{P\&L}} | \right] < \infty $. Then, the level-$\alpha$ intra-horizon expected shortfall can be re-expressed in the form
\begin{equation}
\label{EquLem1}
\mbox{\textit{iES}}_{\alpha,T}(\mathcal{P\&L}) = \frac{1}{\alpha} \int \limits_{-
\infty}^{-iV@R_{\alpha,T}(\mathcal{P\&L})} \mathbb{P}_{z} \big( 
\tau^{\mathcal{P\&L},-}_{\ell} \leq T \big) \, d\ell  + \mbox{\it iV@R}_{\alpha,T}( \mathcal{P\&L}).
\end{equation}
\end{Prop}
\noindent \underline{\bf Remark 1.}
\begin{itemize} \setlength \itemsep{-0.1em}
\item[i)] Combining Proposition \ref{lem1} with Representation (\ref{Repre}) leads to an important implication -- both the intra-horizon value at risk and the intra-horizon expected shortfall can be fully characterized based on first-passage probabilities. Therefore, we will study, for $T >0$, $z \in \mathbb{R}$ and $\ell \leq 0$ the function
\begin{equation}
u(\mathcal{T},z;\ell) := \mathbb{P}_{z} \big( \tau_{\ell}^{\mathcal{P\&L},-} \leq \mathcal{T} \big) = \mathbb{E}_{z}^{\mathbb{P}} \left[ \mathds{1}_{ \{ \tau_{\ell}^{\mathcal{P\&L},-} \leq \mathcal{T} \} }\right] \hspace{1.5em} \mbox{with} \hspace{1.5em} \mathcal{T} \in [0,T],
\label{FPPdef}
\end{equation}
\noindent and subsequently recover intra-horizon measures of risk via numerical techniques. Here, $\mathcal{T} \in [0,T]$ refers to the remaining time to maturity and is linked to any pair of times $(t,T)$ satisfying $0 \leq t \leq T$ via $\mathcal{T}= T-t$.
\item[ii)] Besides providing an important link to first-passage probabilities, Equation (\ref{EquLem1}) formalizes the intuitive property that the intra-horizon expected shortfall always exceeds the intra-horizon value at risk.
\item[iii)] Although the intra-horizon risk approach considered in this article contributes to the theory of risk measures for stochastic processes, we emphasize that, conceptually, it substantially differs from other methods belonging to this theory, especially from the methods of dynamic risk (and performance) measures (cf.~\cite{ap11}, \cite{bcz14}, \cite{kr20}). In fact, while dynamic risk measures aim to quantify, for a fixed maturity $T >0$ and a given terminal position $X_{T}$, at any point in time $t \in [0,T]$ a conditional risk based on the current information $\mathcal{F}_{t}$, our intra-horizon notion of risk aims to statically englobe the full dynamics within the time interval $[0,T]$ under consideration. Nevertheless, since our intra-horizon risk measurement approach is based on the theory initially developed for point-in-time risk measures, we believe that both concepts can be combined to introduce dynamic intra-horizon risk versions. This could be part of future research.
\end{itemize}
\hspace{45em} \scalebox{0.75}{$\blacklozenge$} \\
\section{Intra-Horizon Risk and Models with Jumps}
\label{IHRMJ}
We next turn to a discussion of intra-horizon risk under infinitely divisible distributions, i.e.,~we fix an $\mathbf{F}$-Lévy process $(X_{t})_{t \in [0,T]}$ and consider two different scenarios:
\begin{itemize} \setlength \itemsep{-0.1em}
\item[--] \textit{Scenario 1}, where the dynamics of the (discounted) profit and loss process $(\mathcal{P\&L}_{t})_{t 
\in [0,T]}$ are directly described by $(X_{t})_{t \in [0,T]}$, i.e.,~where
\begin{equation}
 \mathcal{P\&L}_{t} = X_{t}, \hspace{1.5em}  t \in [0,T]. \nonumber
\end{equation}
\item[--] \textit{Scenario 2}, where the (discounted) profit and loss process $(\mathcal{P\&L}_{t})_{t 
\in [0,T]}$ reflects the intrinsic value of a long (\plus) or short~(\minus) position in an asset of ordinary exponential Lévy type, i.e.,~where we have that
\begin{equation}
 \mathcal{P\&L}_{t} = \pm \left( z_{1} e^{X_{t}} - z_{2} \right) , \hspace{1em} t \in [0,T], \hspace{1.5em} \mbox{with} \hspace{0.2em} z_{1},z_{2} \in \mathbb{R}^{+}_{0} . \nonumber
\end{equation}
\end{itemize}
\subsection{Lévy Processes and Notation}
\noindent We recall that a Lévy process $(X_{t})_{t \geq 0}$ on a (filtered) probability space $(\Omega, \mathcal{F}, \mathbf{F}, \mathbb{P}^{X})$ is a càdlàg (right-continuous with left limits) process having independent and stationary increments. The Lévy exponent $\Psi_{X}(\cdot)$ is defined, for $\theta \in \mathbb{R}$, in terms of its characteristic triplet $(b_{X}, \sigma_{X}^{2}, \Pi_{X})$ via
\begin{equation}
\Psi_{X}( \theta ) := - \log \left( \mathbb{E}^{\mathbb{P}^{X}} \left[ e^{i \theta X_{1}} \right] \right)  =  -ib_{X} \theta + \frac{1}{2} \sigma_{X}^{2}\theta^{2} + \int \limits_{ \mathbb{R}} \big(1 - e^{i \theta y} + i \theta y \mathds{1}_{\{ | y | \leq 1\}} \big) \Pi_{X}( dy),
\label{CHARexp}
\end{equation} 
\noindent where $\mathbb{E}^{\mathbb{P}^{X}}\left[ \cdot \right]$ indicates expectation under the measure $\mathbb{P}^{X}$. Well known results in the theory of Lévy processes (cf.~\cite{sa99}, \cite{ap09}) allow to decompose $(X_{t})_{t \geq 0}$ in terms of its jump and diffusion parts as
\begin{equation}
X_{t} = b_{X}t + \sigma_{X} W_{t} + \int \limits_{\mathbb{R}} y \; \bar{N}_{X}(t,dy), \hspace{1.5em} t \geq 0,
\label{DecompX}
\end{equation}
\noindent where $(W_{t})_{t \geq 0}$ denotes an $\mathbf{F}$-Brownian motion and $N_{X}$ refers to an independent Poisson random measure on $[0,\infty) \times \mathbb{R} \setminus \{0\}$ that has intensity measure given by $\Pi_{X}$. Here, we use for any $t \geq 0$ and Borel set $A \in \mathcal{B}(\mathbb{R}\setminus\{0 \})$ the following notation:
\begin{align*}
N_{X}(t,A) & := N_{X}((0,t]\times A), \\
\tilde{N}_{X}(dt,dy) & :=  N_{X}(dt,dy) - \Pi_{X}(dy)dt ,\\
\bar{N}_{X}(dt,dy)&:= \left \{ \begin{array}{cc}
\tilde{N}_{X}(dt,dy), & \mbox{if} \; |y|\leq 1, \\
N_{X}(dt,dy), & \mbox{if} \; |y| > 1. \end{array}
\right. 
\end{align*}
\noindent Additionally, we define the Laplace exponent of $(X_{t})_{t \geq 0}$, for any $ \theta \in \mathbb{R}$ satisfying $ \mathbb{E}^{\mathbb{P}^{X}} \left[ e^{\theta X_{1}} \right] < \infty$, via the following identity 
\begin{align}
\Phi_{X}( \theta )  :=  -\Psi_{X}(-i\theta) & =  b_{X} \theta + \frac{1}{2} \sigma_{X}^{2} \theta^{2} - \int \limits_{ \mathbb{R}} \big(1 - e^{ \theta y} +  \theta y \mathds{1}_{\{ | y | \leq 1\}} \big) \Pi_{X}( dy),
\end{align}
\noindent and recall that $(X_{t})_{t \geq 0}$ has the (strong) Markov property. Therefore, its infinitesimal generator is a partial integro-differential operator given, for sufficiently smooth $V: [0,\infty) \times \mathbb{R} \rightarrow \mathbb{R}$, by
\begin{align}
\mathcal{A}_{X} V(\mathcal{T},x) & :=  \lim \limits_{t \downarrow 0}  \, \frac{\mathbb{E}^{\mathbb{P}^{X}}_{x} \big[ V(\mathcal{T},X_{t})\big] - V(\mathcal{T},x)}{t} \nonumber \\
& =  \frac{1}{2} \sigma^{2}_{X} \partial_{x}^{2} V(\mathcal{T},x) + b_{X} \partial_{x} V(\mathcal{T},x) + \int \limits_{\mathbb{R}} \big[ V(\mathcal{T},x + y) - V(\mathcal{T},x) - y \mathds{1}_{\{ | y | \leq 1\}} \partial_{x} V(\mathcal{T},x)  \big] \Pi_{X}(dy), 
\label{InfGen}
\end{align}
\noindent where the expectation is taken under the measure $\mathbb{P}^{X}_{x}$ having initial distribution $X_{0}=x$. We will extensively make use of these notations in the upcoming sections.
\subsection{First-Passage Decomposition and Intra-Horizon Risk under the Lévy Framework}
\noindent Dealing with first-passage events in any of \textit{Scenario~1} and \textit{Scenario~2} clearly reduces to the consideration of corresponding events for simple Lévy processes. This follows from the properties of the exponential function as well as from the fact that, under both scenarios, the process $(X_{t})_{t \in [0,T]}$ is the only source of uncertainty. Consequently, we only need to study, for $x \in \mathbb{R}$ and $L \in \mathbb{R}$, the first-passage probabilities defined by
\begin{equation}
u_{X}^{\pm}(\mathcal{T},x;L) := \mathbb{P}_{x}^{X} \big( \tau_{L}^{X,\pm} \leq \mathcal{T} \big) = \mathbb{E}_{x}^{\mathbb{P}^{X}} \left[ \mathds{1}_{ \{ \tau_{L}^{X,\pm} \leq \mathcal{T} \} }\right] \hspace{1.5em} \mbox{for} \hspace{1.5em} \mathcal{T} \in [0,T],
\label{FPPdef2}
\end{equation}
\noindent and note that (\ref{FPPdef}) and (\ref{FPPdef2}) are related with each other by means of the following identity
\begin{equation}
u(\mathcal{T},z;\ell) = \left \{ \begin{array}{ll}
u_{X}^{-}(\mathcal{T},z;\ell), & \mbox{for \textit{Scenario 1}}, \\
u_{X}^{-}\big(\mathcal{T},\log(z_{2}+z);\log(z_{2}+\ell) \big), & \mbox{for a long position in \textit{Scenario 2}}, \\
u_{X}^{+}\big(\mathcal{T},\log(z_{2}-z);\log(z_{2}-\ell)\big), & \mbox{for a short position in \textit{Scenario 2}}.
\end{array} \right.
\label{ImpOO7eq}
\end{equation}
\noindent At this point, we should note that not all parameters $z, \ell \in \mathbb{R}$ lead to sensible results when dealing with \textit{Scenario 2}. Therefore, under this scenario, the above formula should be always understood on the respective ranges, i.e.,~we set $\log(0) := -\infty$ and only consider the following values for $z, \ell$:
\begin{itemize} \setlength \itemsep{-0.1em}
\item[i)] $z, \ell \in [-z_{2}, \infty)$ for a long position,
\item[ii)] $z, \ell \in (-\infty, z_{2}]$ for a short position.
\end{itemize}
\noindent The next lemma proves useful when dealing with intra-horizon risk in models with jumps and provides simple conditions on the Laplace exponent of the underlying Lévy process for which intra-horizon expected shortfall measures are well-defined. A proof is provided in Appendix A.
\begin{Lem}
\label{lem2}
Let $(X_{t})_{t \geq 0}$ be a Lévy process and assume that the following condition is satisfied:
\begin{equation}
\exists \theta^{\star} > 1: \hspace{1em}  \mathbb{E}^{\mathbb{P}^{X}}_{0} \left[ e^{\theta^{\star} | X_{1} | } \right] < \infty .
\label{condilem2}
\end{equation}
 Then, there exists a constant $c > 1$ such that for any $x \in \mathbb{R}$ and $\mathcal{T} > 0 $ we have 
\begin{equation}
\lim \limits_{\ell \uparrow \infty} \; e^{c \cdot \ell} \, \mathbb{P}_{x}^{X} \Big( \tau^{X,+}_{\ell} \leq \mathcal{T} \Big) = 0 \hspace{1.5em} \mbox{and} \hspace{1.8em} \lim \limits_{\ell \downarrow -\infty} \; e^{ - c \cdot \ell} \, \mathbb{P}_{x}^{X} \Big( \tau^{X,-}_{\ell} \leq \mathcal{T} \Big) = 0.
\label{lem2equa}
\end{equation}
\noindent In particular, these convergence results ensure that $\mathbb{E}^{\mathbb{P}}_{z} \left[ | I_{\mathcal{T}}^{\mathcal{P\&L}} | \right] < \infty$ holds for any $\mathcal{T} >0$, or, equivalently, that the intra-horizon expected shortfall is well-defined under any of Scenario 1 and Scenario 2.
\end{Lem}
\noindent Although Condition (\ref{condilem2}) slightly restricts the applicability of Lemma \ref{lem2}, many popular classes of Lévy processes encountered in financial applications satisfy this property, at least on a range of parameters that is suitable for market modeling purposes. Important examples include hyper-exponential jump-diffusion models (cf.~\cite{ko02}, \cite{ca09}), Variance Gamma (VG) processes (cf.~\cite{ms90}, \cite{mcc98}), the Carr-Geman-Madan-Yor (CGMY) model (cf.~\cite{cgmy02}) as well as Normal Inverse Gaussian (NIG) processes (cf.~\cite{bn97}). We will deal with the intra-horizon risk inherent to some of these models in our numerical analysis of Section~\ref{NUMres}. \vspace{1em} \\
\noindent \underline{\bf Remark 2.}
\begin{itemize} \setlength \itemsep{-0.1em}
\item[i)] A closer look at the proof of Lemma \ref{lem2} reveals that under each of the scenarios under consideration, only one of the conditions
\begin{equation}
\lim \limits_{\ell \uparrow \infty} \; e^{c \cdot \ell} \, \mathbb{P}_{x}^{X} \Big( \tau^{X,+}_{\ell} \leq \mathcal{T} \Big) = 0, \hspace{1.8em} \mbox{or} \hspace{2.1em} \lim \limits_{\ell \downarrow -\infty} \; e^{ - c \cdot \ell} \, \mathbb{P}_{x}^{X} \Big( \tau^{X,-}_{\ell} \leq \mathcal{T} \Big) = 0,
\nonumber 
\end{equation}
\noindent is sufficient to ensure that $\mathbb{E}^{\mathbb{P}}_{z} \left[ | I_{\mathcal{T}}^{\mathcal{P\&L}} | \right] < \infty$ holds for $\mathcal{T} >0$. Additionally, the proof of Lemma \ref{lem2} indicates that these properties are consequences of the corresponding one-sided conditions 
\begin{equation}
\exists \theta^{\star} > 1: \hspace{1em}  \mathbb{E}^{\mathbb{P}^{X}}_{0} \left[ e^{\theta^{\star} X_{1}  } \right] < \infty  , \hspace{1.8em} \mbox{and} \hspace{2.1em} \exists \theta_{\star} < -1 : \, \mathbb{E}^{\mathbb{P}^{X}}_{0} \left[ e^{\theta_{\star} X_{1}} \right] < \infty ,
\end{equation}
\noindent respectively, which therefore even weaken the requirements on the dynamics of the process $(X_{t})_{t \geq 0}$.
\item[ii)] When considering a long position in \textit{Scenario 2}, well-definedness of the intra-horizon expected shortfall is immediate. In this case, the condition $\mathbb{E}^{\mathbb{P}}_{z} \left[ | I_{\mathcal{T}}^{\mathcal{P\&L}} | \right] < \infty$ directly follows, for any $\mathcal{T} >0$, from the fact that~$I_{\mathcal{T}}^{\mathcal{P\&L}} \geq -z_{2}$, i.e.,~that the maximal possible losses do not exceed the value $z_{2}$. We will dedicate our numerical analysis in Section \ref{NUMres} to exactly this scenario and investigate the intra-horizon risk inherent to a long position in the S\&P~500 index as well as in the Brent crude oil using weekly data ranging from January 1990 to September 2020.
\end{itemize}
$\mbox{}$ \hspace{44.8em} \scalebox{0.75}{$\blacklozenge$} \\
\subsubsection{First-Passage Decomposition and PIDEs}
\noindent For risk management purposes, it may be of great importance to further understand the structure of risk. In particular, one may want to know how often certain shortfall barriers are already exceeded at the time they are breached. When dealing with market risk, this roughly reduces to the question of whether first-passage occurrences are triggered by diffusion or by jumps. This question can be further investigated under the present Lévy framework and a first-passage decomposition can be obtained.\footnote{As we shall see in a moment, the same approach can be adopted for any strong Markov process that is quasi-left-continuous.}~This is discussed next. Here, we start from a similar approach to the one taken in \cite{cc09} and define, for any $\ell \in \mathbb{R}$, the following events 
\begin{align}
\mathcal{E}_{0}^{\pm} := \left \{ X_{\tau_{\ell}^{X,\pm}} = \ell \right \} \hspace{1.5em} \mbox{and} \hspace{1.8em} \mathcal{E}_{\mathcal{J}}^{\pm} := \left \{ X_{\tau_{\ell}^{X,\pm}} \neq \ell \right \}, \label{Decomp}
\end{align}
\noindent i.e.,~we essentially decompose the first-passage times in events that are either triggered by the diffusion part of the process $(X_{t})_{t \in [0,T]}$ or by jumps.\footnote{We emphasize that this interpretation may not be (fully) correct in cases where $\tau_{\ell}^{X,\pm} = \infty $, however, that these cases are subsequently excluded from our analysis, as seen in (\ref{uFuncDEF}). Additionally, we note that for general jump dynamics parts of the event $\mathcal{E}_{0}^{\pm}$ could be due to jumps and that it may be therefore more appropriate to speak of a first-passage decomposition in events with and without overshoot. Nevertheless, since market models usually assume continuous jump distributions, i.e.,~an intensity measure $\Pi_{X}$ of the form $\Pi_{X}(dy) = \pi_{X}(y) dy$, with appropriate jump density $\pi_{X}(\cdot)$, this situation will not occur. This justifies the use of our initial terminology.} Clearly, the first-passage events $\mathcal{E}_{0}^{+}$ and $\mathcal{E}_{\mathcal{J}}^{+}$ are disjoint and the same additionally holds for $\mathcal{E}_{0}^{-}$ and $\mathcal{E}_{\mathcal{J}}^{-}$. Hence, this allows us to obtain, for $\mathcal{T} \in [0,T]$, $x \in \mathbb{R}$ and $\ell \in \mathbb{R}$, a decomposition of the first-passage probabilities as
\begin{align}
u_{X}^{\pm}(\mathcal{T},x;\ell ) = u_{X}^{\mathcal{E}_{0}^{\pm}}(\mathcal{T},x;\ell ) + u_{X}^{\mathcal{E}_{\mathcal{J}}^{\pm}}(\mathcal{T},x;\ell ), \label{DecJump}
\end{align}
\noindent where $u_{X}^{\mathcal{E}_{0}^{\pm}}(\mathcal{T},x;\ell) $ and $u_{X}^{\mathcal{E}_{\mathcal{J}}^{\pm}}(\mathcal{T},x;\ell )$ refer to the functions defined by 
\begin{align}
u_{X}^{\mathcal{E}_{0}^{\pm}}(\mathcal{T},x;\ell) := \mathbb{E}_{x}^{\mathbb{P}^{X}} \left[ \mathds{1}_{ \left\{ \tau_{\ell}^{X,\pm} \leq \mathcal{T} \right\} \, \cap \,\mathcal{E}_{0}^{\pm} }\right] \hspace{1.5em} \mbox{and} \hspace{1.8em} u_{X}^{\mathcal{E}_{\mathcal{J}}^{\pm}}(\mathcal{T},x;\ell ) := \mathbb{E}_{x}^{\mathbb{P}^{X}} \left[ \mathds{1}_{ \left\{ \tau_{\ell}^{X,\pm} \leq \mathcal{T} \right\} \, \cap \,\mathcal{E}_{\mathcal{J}}^{\pm} }\right].
\label{uFuncDEF}
\end{align}
\noindent We now turn to an analysis of these first-passage probabilities and first aim to obtain a PIDE characterization of the functions $u_{X}^{\mathcal{E}_{0}^{\pm}}(\cdot)$ and $u_{X}^{\mathcal{E}_{\mathcal{J}}^{\pm}}(\cdot)$. To this end, we define, for $\ell \in \mathbb{R}$, the following domains
\begin{align}
\mathcal{H}^{+}_{\ell} := ( \ell, \infty) \hspace{2em} \mbox{and} \hspace{2em} \mathcal{H}^{-}_{\ell} := (- \infty, \ell)
\end{align}
\noindent and denote by $\overline{\mathcal{H}^{\pm}}$ the closure of these sets in $\mathbb{R}$. Under this notation, the next proposition is obtained by relying on (strong) Markovian arguments. A proof is provided in Appendix B.
\begin{Prop}
\noindent For any level $\ell \in \mathbb{R}$, the first-passage probability contributed by the diffusion part, $u_{X}^{\mathcal{E}_{0}^{\pm}}(\cdot)$, satisfies the following Cauchy problem:
\begin{align}
-\partial_{\mathcal{T}} u_{X}^{\mathcal{E}_{0}^{\pm}}(\mathcal{T},x;\ell ) + \mathcal{A}_{X} u_{X}^{\mathcal{E}_{0}^{\pm}}(\mathcal{T},x;\ell ) & = 0, \hspace{1.7em} \mbox{on} \hspace{0.5em} (\mathcal{T},x) \in (0,T] \times \left( \mathbb{R}\setminus \overline{\mathcal{H}_{\ell}^{\pm}} \right), \label{Prop2Cauchy11}\\
u_{X}^{\mathcal{E}_{0}^{\pm}}(\mathcal{T},x;\ell ) & = 1, \hspace{1.7em} \mbox{on} \hspace{0.5em} (\mathcal{T},x) \in [0,T] \times \{ \ell \}, \label{Prop2Cauchy12}\\
u_{X}^{\mathcal{E}_{0}^{\pm}}(\mathcal{T},x;\ell ) & = 0, \hspace{1.7em} \mbox{on} \hspace{0.5em} (\mathcal{T},x) \in [0,T] \times \mathcal{H}_{\ell}^{\pm}, \label{Prop2Cauchy13} \\
u_{X}^{\mathcal{E}_{0}^{\pm}}(0,x;\ell ) & = 0, \hspace{1.7em} \mbox{on} \hspace{0.5em} x \in \mathbb{R} \setminus \overline{\mathcal{H}_{\ell}^{\pm}}. \label{Prop2Cauchy14}
\end{align}
\noindent Similarly, the first-passage probability contributed by jumps, $u_{X}^{\mathcal{E}_{\mathcal{J}}^{\pm}}(\cdot)$, solves, for any $\ell \in \mathbb{R}$, the Cauchy problem
\begin{align}
-\partial_{\mathcal{T}} u_{X}^{\mathcal{E}_{\mathcal{J}}^{\pm}}(\mathcal{T},x;\ell ) + \mathcal{A}_{X} u_{X}^{\mathcal{E}_{\mathcal{J}}^{\pm}}(\mathcal{T},x;\ell ) & = 0, \hspace{1.7em} \mbox{on} \hspace{0.5em} (\mathcal{T},x) \in (0,T] \times \left( \mathbb{R}\setminus \overline{\mathcal{H}_{\ell}^{\pm}} \right), \label{Prop2Cauchy21}\\
u_{X}^{\mathcal{E}_{\mathcal{J}}^{\pm}}(\mathcal{T},x;\ell ) & = 0, \hspace{1.7em} \mbox{on} \hspace{0.5em} (\mathcal{T},x) \in [0,T] \times \{ \ell \}, \label{Prop2Cauchy22}\\
u_{X}^{\mathcal{E}_{\mathcal{J}}^{\pm}}(\mathcal{T},x;\ell ) & = 1, \hspace{1.7em} \mbox{on} \hspace{0.5em} (\mathcal{T},x) \in [0,T] \times \mathcal{H}_{\ell}^{\pm}, \label{Prop2Cauchy23}\\
u_{X}^{\mathcal{E}_{\mathcal{J}}^{\pm}}(0,x;\ell ) & = 0, \hspace{1.7em} \mbox{on} \hspace{0.5em} x \in \mathbb{R} \setminus \overline{\mathcal{H}_{\ell}^{\pm}}. \label{Prop2Cauchy24}
\end{align}
\label{prop2}
\end{Prop}
\noindent Combining the characterization in Proposition \ref{prop2} with (standard) numerical techniques already allows for a numerical treatment of the functions $u_{X}^{\mathcal{E}_{0}^{\pm}}(\cdot)$ and $u_{X}^{\mathcal{E}_{\mathcal{J}}^{\pm}}(\cdot)$. Furthermore, the proof of Proposition~\ref{prop2} reveals that our derivations are not restricted to the Lévy framework. Indeed, while the decomposition in (\ref{DecJump}) is a simple consequence of the disjointness of the sets defined in (\ref{Decomp}), the proof of Proposition~\ref{prop2} combines (strong) Markovian arguments with the quasi-left-continuity of the process $(X_{t})_{t \in [0,T]}$. As a consequence, relying on this approach is always possible when dealing with processes that satisfy these two properties and a disentanglement of diffusion and jump contributions can be then obtained via the exact same techniques. \vspace{1em} \\ 
\noindent \underline{\bf Remark 3.} \vspace{0.3em} \\
In general, the above techniques can be applied to subsequently recover intra-horizon risk measures as well as corresponding risk contributions. However, even when dealing with the intra-horizon value at risk to a single level $\alpha \in (0,1)$ numerous iterations of the numerical scheme are needed and the computational costs quickly become high. Therefore, relying on these techniques for expected shortfall measures does not seem to be the best approach. Instead, distributional properties inherent to certain distributions sometimes allow to simplify the problem by switching to maturity-randomization. This holds for instance true when dealing with hyper-exponential jump-diffusion processes that have the particularity to allow for arbitrarily close approximations of Lévy processes with completely monotone jumps (cf.~\cite{jp10}, \cite{ck11}, \cite{hk16}). A discussion of this approach as well as of approximations of Lévy densities via hyper-exponential jump densities is provided in the upcoming sections. \\
$\mbox{}$ \hspace{44.8em} \scalebox{0.75}{$\blacklozenge$} \\

\subsubsection{Maturity-Randomization and OIDEs}
\noindent We next deal with maturity-randomized first-passage probabilities. To this end, we start by defining for any function $g: \mathbb{R}^{+} \rightarrow \mathbb{R}$ satisfying
\begin{equation}
\int \limits_{0}^{\infty} e^{-\vartheta t} |g(t)| \, dt < \infty, \hspace{2em} \forall \vartheta >0,
\label{EquLCTransform0}
\end{equation}
\noindent the Laplace-Carson transform $\mathcal{LC}(g)(\cdot)$ via  
\begin{align}
\mathcal{LC}(g)(\vartheta) & := \int \limits_{0}^{\infty} \vartheta e^{-\vartheta t } \, g(t) \,dt,
\label{EquLCTransform1}
\end{align}
\noindent and note that this transform has several desirable properties. First, the Laplace-Carson transform merely corresponds to a scaled Laplace transform, for which extensive inversion techniques exist (cf.~\cite{co07}). Additionally, as we shall see in a moment, applying the Laplace-Carson transform in the context of mathematical finance allows to randomize the maturity of financial contracts, i.e.,~to switch from objects with deterministic maturity to corresponding objects with stochastic (exponentially distributed) maturity. This last property offers a range of alternative ways to tackle problems related to the valuation of financial positions and has therefore led to a wide adoption of the Laplace-Carson transform in the option pricing literature, with \cite{ca98} being one of the seminal articles in this context. \vspace{1em} \\
\noindent Having computed the transform (either numerically or analytically), the original function $g(\cdot)$ can be recovered from $\mathcal{LC}(g)(\cdot)$ using an inversion algorithm. One possible choice is the Gaver-Stehfest algorithm that has the particularity to allow for an inversion of the transform on the real line and that has been successfully used by several authors for option pricing (cf.~\cite{kw03}, \cite{ki10}, \cite{hm13}, \cite{lv17}, \cite{cv18}). We will also rely on this algorithm, i.e.,~we set
\begin{equation}
g_{N}(t) := \sum \limits_{k=1}^{2N} \zeta_{k,N} \, \mathcal{LC}\big(g\big)\left( \frac{k \log(2)}{t}\right), \hspace{1.5em} N \in \mathbb{N}, \; t > 0,
\end{equation}
\noindent where the coefficients are given by
\begin{equation}
\label{zetaEQUA}
\zeta_{k,N} := \frac{(-1)^{N+k}}{k} \sum \limits_{j = \lfloor (k+1)/2 \rfloor }^{\min \{k,N \}} \frac{j^{N+1}}{N!} \binom{N}{j} \binom{2j}{j} \binom{j}{k-j}, \hspace{1.5em} N \in \mathbb{N}, \; 1 \leq k \leq 2N,
\end{equation}
\noindent with $\lfloor a \rfloor := \sup \{z \in \mathbb{Z}: \, z \leq a \}$, and will recover the original function $g(\cdot)$ by means of the following relation
\begin{equation}
\lim \limits_{N \rightarrow \infty} g_{N}(t) = g(t).
\label{CONVer}
\end{equation}
\noindent More details on the Gaver-Stehfest algorithm as well as formal proofs of the convergence result (\ref{CONVer}) for ``sufficiently well-behaved functions'' are provided in \cite{ku13} and references therein. \vspace{1em} \\
\noindent We now turn to a discussion of Laplace-Carson transformed first-passage probabilities. First, we note that the boundedness of the functions $u_{X}^{\pm}(\cdot)$ and $u_{X}^{\mathcal{E}}(\cdot)$ for $\mathcal{E} \in \{ \mathcal{E}_{0}^{\pm}, \mathcal{E}_{\mathcal{J}}^{\pm} \}$ ensures that these first-passage probabilities satisfy Condition (\ref{EquLCTransform0}) and so that the resulting Laplace-Carson transform is well-defined. Additionally, one easily sees that the first-passage decompositions obtained in (\ref{DecJump}) are preserved under the Laplace-Carson operator, i.e.,~we have for any $\vartheta >0$, $x \in \mathbb{R}$ and $\ell \in \mathbb{R}$ that
\begin{align}
\mathcal{LC}\big(u_{X}^{\pm}\big)(\vartheta,x;\ell ) = \mathcal{LC} \big(u_{X}^{\mathcal{E}_{0}^{\pm}} \big)(\vartheta,x;\ell ) + \mathcal{LC}\big(u_{X}^{\mathcal{E}_{\mathcal{J}}^{\pm}}\big)(\vartheta,x;\ell ). \label{DecJumpLC}
\end{align}
\noindent This property is particularly interesting since it implies that switching back and forth between the original first-passage probabilities and their corresponding Laplace-Carson transforms does not alter the structure of risk across the diffusion and jump parts and therefore allows us to fully concentrate on one or the other. Finally, any of the Laplace-Carson transformed first-passage probabilities can be interpreted as the probability of a respective first-passage occurring before an independent exponentially distributed random time of intensity $\vartheta >0$, $\mathcal{T}_{\vartheta}$, has expired or equivalently before the first jump time of an independent Poisson process $(N_{t})_{t \geq 0}$ having intensity $\vartheta >0$ has happened. This is easily seen from the following identities
\begin{align}
\mathcal{LC} \big(u_{X}^{\pm}\big)(\vartheta,x;\ell) = \int \limits_{0}^{\infty} \vartheta e^{-\vartheta t } \, u_{X}^{\pm}(t,x; \ell) \, dt  = \mathbb{E}_{x}^{\mathbb{P}^{X}} \left[ \, \mathbb{E}_{x}^{\mathbb{P}^{X}} \left[ \mathds{1}_{ \left\{ \tau_{\ell}^{X,\pm} \leq \, \mathcal{T}_{\vartheta} \right\} } \Big | \, \mathcal{T}_{\vartheta} \right] \, \right] = \mathbb{E}_{x}^{\mathbb{P}^{X}} \left[ \mathds{1}_{ \left\{ \tau_{\ell}^{X,\pm} \leq \, \mathcal{T}_{\vartheta} \right\} }\right], 
\label{PoissonAppr1} \\
\mathcal{LC} \big(u_{X}^{\mathcal{E}}\big)(\vartheta,x;\ell) = \int \limits_{0}^{\infty} \vartheta e^{-\vartheta t } \, u_{X}^{\mathcal{E}}(t,x; \ell) \, dt = \mathbb{E}_{x}^{\mathbb{P}^{X}} \left[ \, \mathbb{E}_{x}^{\mathbb{P}^{X}} \left[ \mathds{1}_{ \left\{ \tau_{\ell}^{X,\pm} \leq \, \mathcal{T}_{\vartheta} \right\} \, \cap \,\mathcal{E} } \Big | \, \mathcal{T}_{\vartheta} \right] \, \right] = \mathbb{E}_{x}^{\mathbb{P}^{X}} \left[ \mathds{1}_{ \left\{ \tau_{\ell}^{X,\pm} \leq \, \mathcal{T}_{\vartheta} \right\} \, \cap \,\mathcal{E} } \right], \label{PoissonAppr2}
\end{align}
\noindent where $\mathcal{E} \in \left \{ \mathcal{E}_{0}^{\pm}, \mathcal{E}_{\mathcal{J}}^{\pm} \right \}$. Consequently, any application of the Laplace-Carson operator transforms (in this context) first-passage probabilities into corresponding maturity-randomized quantities and combining these properties with arguments similarly used in the proof of Proposition \ref{prop2} allows us to obtain an OIDE characterization of the maturity-randomized first-passage probabilities contributed by the diffusion part, $\mathcal{LC} \big(u_{X}^{\mathcal{E}_{0}^{\pm}}\big)(\cdot)$, and by jumps, $\mathcal{LC} \big(u_{X}^{\mathcal{E}_{\mathcal{J}}^{\pm}}\big)(\cdot)$. This is the content of the next proposition, whose proof is provided in Appendix~B.
\begin{Prop}
\noindent For any level $\ell \in \mathbb{R}$ and intensity $\vartheta >0$, the maturity-randomized first-passage probability contributed by the diffusion part, $\mathcal{LC} \big(u_{X}^{\mathcal{E}_{0}^{\pm}}\big)(\cdot)$, satisfies the following Cauchy problem:
\begin{align}
\mathcal{A}_{X} \mathcal{LC} \big(u_{X}^{\mathcal{E}_{0}^{\pm}}\big)(\vartheta,x;\ell ) & =  \vartheta \, \mathcal{LC} \big(u_{X}^{\mathcal{E}_{0}^{\pm}}\big)(\vartheta,x;\ell ), \hspace{1.7em} \mbox{on} \hspace{0.5em} x \in \mathbb{R}\setminus \overline{\mathcal{H}_{\ell}^{\pm}}, \label{OIDE1}\\
\mathcal{LC} \big(u_{X}^{\mathcal{E}_{0}^{\pm}}\big)(\vartheta,x;\ell ) & = 1, \hspace{9.0em} \mbox{on} \hspace{0.5em} x = \ell, \label{OIDE2} \\
\mathcal{LC} \big(u_{X}^{\mathcal{E}_{0}^{\pm}}\big)(\vartheta,x;\ell ) & = 0, \hspace{9.0em} \mbox{on} \hspace{0.5em} x \in \mathcal{H}_{\ell}^{\pm}. \label{OIDE3}
\end{align}
\noindent Similarly, the maturity-randomized first-passage probability contributed by jumps, $\mathcal{LC} \big(u_{X}^{\mathcal{E}_{\mathcal{J}}^{\pm}}\big)(\cdot)$, solves, for any $\ell \in \mathbb{R}$ and $\vartheta >0$, the Cauchy problem
\begin{align}
\mathcal{A}_{X} \mathcal{LC} \big(u_{X}^{\mathcal{E}_{\mathcal{J}}^{\pm}}\big)(\vartheta,x;\ell ) & =  \vartheta \, \mathcal{LC} \big(u_{X}^{\mathcal{E}_{\mathcal{J}}^{\pm}}\big)(\vartheta,x;\ell ), \hspace{1.7em} \mbox{on} \hspace{0.5em} x \in \mathbb{R}\setminus \overline{\mathcal{H}_{\ell}^{\pm}}, \label{OIDEj1}\\
\mathcal{LC} \big(u_{X}^{\mathcal{E}_{\mathcal{J}}^{\pm}}\big)(\vartheta,x;\ell ) & = 0, \hspace{9.0em} \mbox{on} \hspace{0.5em} x = \ell, \label{OIDEj2} \\
\mathcal{LC} \big(u_{X}^{\mathcal{E}_{\mathcal{J}}^{\pm}}\big)(\vartheta,x;\ell ) & = 1, \hspace{9.0em} \mbox{on} \hspace{0.5em} x \in \mathcal{H}_{\ell}^{\pm}. \label{OIDEj3}
\end{align}
\label{prop3}
\end{Prop}
\noindent Compared with the results in Proposition \ref{prop2}, Proposition \ref{prop3} offers substantially simpler characterizations. In particular, applying the Laplace-Carson operator to the first-passage probabilities reduces the complexity of the respective problems by transforming the PIDE characterizations of Proposition \ref{prop2} into corresponding OIDE characterizations. Under certain Lévy dynamics $(X_{t})_{t \geq 0}$ the resulting problems (\ref{OIDE1})-(\ref{OIDE3}) and (\ref{OIDEj1})-(\ref{OIDEj3}) even have a simple analytical solution. This is in particular true for the class of hyper-exponential jump-diffusions that is discussed in Section \ref{sechyper}. 

\subsection{Intra-Horizon Risk and Risk Contributions}
\label{IHRRC}
\noindent The analysis developed in the previous sections provided a decomposition of diffusion and jump contributions embodied in first-passage probabilities. Since both the intra-horizon value at risk and the intra-horizon expected shortfall can be fully characterized based on first-passage probabilities (cf.~Section \ref{SecQuant}), these last results can be further extended to infer diffusion and jump risk contributions to the intra-horizon risk measures under consideration. This is discussed next. \vspace{1em} \\
\noindent We start by introducing risk contributions for the intra-horizon value at risk. Here, we follow the ideas in~\cite{lv20} and understand the diffusion and jump risk contributions as the proportions of the $\mbox{\it iV@R}$-first-passage probability contributed by the respective components, i.e.,~we define, for $\alpha \in (0,1)$ and a (discounted) profit and loss process $(\mathcal{P\&L}_{t})_{t \in [0,T]}$ satisfying the dynamics specified in either \textit{Scenario 1} or \textit{Scenario 2}, the diffusion and jump risk contribution inherent to the level-$\alpha$ intra-horizon value at risk over the time interval~$[0,T]$, $\mathcal{R}_{{iV@R}}^{\mathcal{D}}\left(\mathcal{P\&L}; \alpha,T \right)$ and $\mathcal{R}_{{iV@R}}^{\mathcal{J}}\left(\mathcal{P\&L}; \alpha,T \right)$ respectively, via 
\begin{equation}
 \mathcal{R}_{{iV@R}}^{\mathcal{D}}\left(\mathcal{P\&L}; \alpha,T \right) := \left \{ \begin{array}{ll}
\frac{u_{X}^{\mathcal{E}_{0}^{-}}\left( T,\, z; \,-{iV@R}_{\alpha,T}(\mathcal{P\&L})\right)}{u_{X}^{-}\left( T, \, z; \,-{iV@R}_{\alpha,T}(\mathcal{P\&L})\right)}, & \mbox{for \textit{Scenario 1}}, \\
\frac{u_{X}^{\mathcal{E}_{0}^{-}}\left( T, \,\log\left(z_{2}+z\right); \,\log\left(z_{2}- {iV@R}_{\alpha,T}(\mathcal{P\&L})\right)\right)}{u_{X}^{-}\left( T, \, \log\left(z_{2}+z\right); \,\log\left(z_{2}-{iV@R}_{\alpha,T}(\mathcal{P\&L})\right)\right)}, & \mbox{for a long position in \textit{Scenario 2}},\\
\frac{u_{X}^{\mathcal{E}_{0}^{+}}\left( T, \,\log\left(z_{2}-z\right); \,\log\left(z_{2} + {iV@R}_{\alpha,T}(\mathcal{P\&L})\right)\right)}{u_{X}^{+}\left( T, \, \log\left(z_{2}-z\right); \,\log\left(z_{2}+{iV@R}_{\alpha,T}(\mathcal{P\&L})\right)\right)}, & \mbox{for a short position in \textit{Scenario 2}},
\end{array} \right.
\end{equation}
\noindent and
\begin{equation}
 \mathcal{R}_{{iV@R}}^{\mathcal{J}}\left(\mathcal{P\&L}; \alpha,T \right) := \left \{ \begin{array}{ll}
\frac{u_{X}^{\mathcal{E}_{\mathcal{J}}^{-}}\left( T,\, z; \,-{iV@R}_{\alpha,T}(\mathcal{P\&L})\right)}{u_{X}^{-}\left( T, \, z; \,-{iV@R}_{\alpha,T}(\mathcal{P\&L})\right)}, & \mbox{for \textit{Scenario 1}}, \\
\frac{u_{X}^{\mathcal{E}_{\mathcal{J}}^{-}}\left( T, \,\log\left(z_{2}+z\right); \,\log\left(z_{2}- {iV@R}_{\alpha,T}(\mathcal{P\&L})\right)\right)}{u_{X}^{-}\left( T, \, \log\left(z_{2}+z\right); \,\log\left(z_{2}-{iV@R}_{\alpha,T}(\mathcal{P\&L})\right)\right)}, & \mbox{for a long position in \textit{Scenario 2}},\\
\frac{u_{X}^{\mathcal{E}_{\mathcal{J}}^{+}}\left( T, \,\log\left(z_{2}-z\right); \,\log\left(z_{2} + {iV@R}_{\alpha,T}(\mathcal{P\&L})\right)\right)}{u_{X}^{+}\left( T, \, \log\left(z_{2}-z\right); \,\log\left(z_{2}+{iV@R}_{\alpha,T}(\mathcal{P\&L})\right)\right)}, & \mbox{for a short position in \textit{Scenario 2}}.
\end{array} \right. 
\end{equation}
\noindent Defining risk contributions embodied in the intra-horizon expected shortfall can be done via similar techniques and is closely linked to the computation of risk contributions for the intra-horizon value at risk. Indeed, from Proposition \ref{lem1} we already know that (given the intra-horizon value at risk to a certain level) the difference between intra-horizon expected shortfall and intra-horizon value at risk consists in an integral over first-passage probabilities that is given by
\begin{equation}
\mbox{\textit{iES}}_{\alpha,T}(\mathcal{P\&L}) - \mbox{\it iV@R}_{\alpha,T}( \mathcal{P\&L}) = \frac{1}{\alpha} \int \limits_{-\infty}^{-iV@R_{\alpha,T}(\mathcal{P\&L})} u(T,z; \ell)  \, d\ell,
\label{IntPART}
\end{equation}
\noindent where $u(T,z;\ell)$ is specified in each scenario according to Relation (\ref{ImpOO7eq}). Therefore, the diffusion and jump risk contributions inherent to the intra-horizon expected shortfall can be divided into two parts: the respective risk contributions in the intra-horizon value at risk and those of the remaining integral (\ref{IntPART}). For the latter -- which can be interpreted as an average conditional excess intra-horizon tail loss -- diffusion and jump risk contributions can be defined as the proportions of the integral contributed by the respective components, i.e.,~one recovers, for $\alpha \in (0,1)$ and a (discounted) profit and loss process~$(\mathcal{P\&L}_{t})_{t \in [0,T]}$ satisfying the dynamics specified in either \textit{Scenario 1} or \textit{Scenario 2}, the diffusion and jump risk contribution inherent to the integral part (\ref{IntPART}), $\mathcal{R}_{\mathcal{I}}^{\mathcal{D}}\left(\mathcal{P\&L}; \alpha,T \right)$ and $\mathcal{R}_{\mathcal{I}}^{\mathcal{J}}\left(\mathcal{P\&L}; \alpha,T \right)$ respectively, via
\begin{equation}
\mathcal{R}_{\mathcal{I}}^{\mathcal{D}}\left(\mathcal{P\&L}; \alpha,T \right) =  \left \{ \begin{array}{ll} 
 \frac{\int \limits_{-\infty}^{-iV@R_{\alpha,T}(\mathcal{P\&L})} u_{X}^{\mathcal{E}_{0}^{-}}(T,\,z; \,\ell)  \; d\ell}{\int \limits_{-\infty}^{-iV@R_{\alpha,T}(\mathcal{P\&L})} u_{X}^{-}(T,\,z;\, \ell)  \; d\ell}, & \mbox{for \textit{Scenario 1}}, \\
 \frac{\int \limits_{-z_{2}}^{-iV@R_{\alpha,T}(\mathcal{P\&L})} u_{X}^{\mathcal{E}_{0}^{-}}\left(T,\,\log(z_{2}+z); \,\log(z_{2}+\ell) \right)  \; d\ell}{\int \limits_{-z_{2}}^{-iV@R_{\alpha,T}(\mathcal{P\&L})} u_{X}^{-}\left(T,\,\log(z_{2}+z); \,\log(z_{2}+\ell) \right)  \; d\ell}, & \mbox{for a long position in \textit{Scenario 2}},\\
  \frac{\int \limits_{-\infty}^{-iV@R_{\alpha,T}(\mathcal{P\&L})} u_{X}^{\mathcal{E}_{0}^{+}}\left(T,\,\log(z_{2} -z) ; \,\log(z_{2}-\ell) \right)  \; d\ell}{\int \limits_{-\infty}^{-iV@R_{\alpha,T}(\mathcal{P\&L})} u_{X}^{+}\left(T,\,\log(z_{2} -z) ; \,\log(z_{2}-\ell) \right)  \; d\ell}, & \mbox{for a short position in \textit{Scenario 2}},\\
\end{array} \right.
\end{equation}
\noindent and
\begin{equation}
\mathcal{R}_{\mathcal{I}}^{\mathcal{J}}\left(\mathcal{P\&L}; \alpha,T \right) =  \left \{ \begin{array}{ll} 
 \frac{\int \limits_{-\infty}^{-iV@R_{\alpha,T}(\mathcal{P\&L})} u_{X}^{\mathcal{E}_{\mathcal{J}}^{-}}(T,\,z; \,\ell)  \; d\ell}{\int \limits_{-\infty}^{-iV@R_{\alpha,T}(\mathcal{P\&L})} u_{X}^{-}(T,\,z;\, \ell)  \; d\ell}, & \mbox{for \textit{Scenario 1}}, \\
 \frac{\int \limits_{-z_{2}}^{-iV@R_{\alpha,T}(\mathcal{P\&L})} u_{X}^{\mathcal{E}_{\mathcal{J}}^{-}}\left(T,\,\log(z_{2}+z); \,\log(z_{2}+\ell) \right)  \; d\ell}{\int \limits_{-z_{2}}^{-iV@R_{\alpha,T}(\mathcal{P\&L})} u_{X}^{-}\left(T,\,\log(z_{2}+z); \,\log(z_{2}+\ell) \right)  \; d\ell}, & \mbox{for a long position in \textit{Scenario 2}},\\
  \frac{\int \limits_{-\infty}^{-iV@R_{\alpha,T}(\mathcal{P\&L})} u_{X}^{\mathcal{E}_{\mathcal{J}}^{+}}\left(T,\,\log(z_{2} -z) ; \,\log(z_{2}-\ell) \right)  \; d\ell}{\int \limits_{-\infty}^{-iV@R_{\alpha,T}(\mathcal{P\&L})} u_{X}^{+}\left(T,\,\log(z_{2} -z) ; \,\log(z_{2}-\ell) \right)  \; d\ell}, & \mbox{for a short position in \textit{Scenario 2}}.\\
\end{array} \right.
\end{equation}
\noindent Finally, using these definitions, the diffusion and jump risk contributions inherent to the level-$\alpha$ intra-horizon expected shortfall over the time interval $[0,T]$, $\mathcal{R}_{{iES}}^{\mathcal{D}}\left(\mathcal{P\&L}; \alpha,T \right)$ and $\mathcal{R}_{{iES}}^{\mathcal{J}}\left(\mathcal{P\&L}; \alpha,T \right)$ respectively, can be recovered as weighted sums of the corresponding contributions for the intra-horizon value at risk and the integral part (\ref{IntPART}), i.e.,~as
\begin{align}
\mathcal{R}_{{iES}}^{\mathcal{D}}\left(\mathcal{P\&L}; \alpha,T \right) =  \big(1-\omega_{\alpha,T}(\mathcal{P\&L}) \big) \cdot \mathcal{R}_{\mathcal{I}}^{\mathcal{D}}\left(\mathcal{P\&L}; \alpha,T \right) +  \omega_{\alpha,T}(\mathcal{P\&L}) \cdot  \mathcal{R}_{{iV@R}}^{\mathcal{D}}\left(\mathcal{P\&L}; \alpha,T \right) ,\\
\mathcal{R}_{{iES}}^{\mathcal{J}}\left(\mathcal{P\&L}; \alpha,T \right) =  \big(1-\omega_{\alpha,T}(\mathcal{P\&L}) \big) \cdot \mathcal{R}_{\mathcal{I}}^{\mathcal{J}}\left(\mathcal{P\&L}; \alpha,T \right) +  \omega_{\alpha,T}(\mathcal{P\&L}) \cdot  \mathcal{R}_{{iV@R}}^{\mathcal{J}}\left(\mathcal{P\&L}; \alpha,T \right),
\label{RISKc+}
\end{align}
\noindent where $\omega_{\alpha,T}(\mathcal{P\&L}) := \frac{{iV@R}_{\alpha,T}( \mathcal{P\&L})}{{iES}_{\alpha,T}(\mathcal{P\&L})}$ denotes the contribution of the intra-horizon value at risk to the intra-horizon expected shortfall. We will come back to this decomposition when discussing numerical results in Section \ref{NUMres}.

\subsection{First-Passage Decomposition and Intra-Horizon Risk under Hyper-Exponential Jump-Diffusions}
\label{sechyper}
\noindent Having elaborated on our core intra-horizon risk measurement approach under the general Lévy framework, we next discuss (semi-)analytical expressions for hyper-exponential jump-diffusion processes. These results are particularly interesting since they subsequently allow for an approximate, though arbitrarily precise semi-analytical measurement of intra-horizon risk within the important class of Lévy processes having a completely monotone jump density. We will further develop this point in Section \ref{SECapproxim} and lastly provide an application of this approximate approach in the numerical analysis of Section \ref{NUMres}. 
\subsubsection{Generalities on Hyper-Exponential Jump-Diffusions}
\noindent We recall that a hyper-exponential jump-diffusion process $(X_{t})_{t \geq 0}$ is a Lévy process that combines a Brownian diffusion with hyper-exponentially distributed jumps. This process has the usual jump-diffusion structure, i.e.,~it can be characterized on a filtered probability space $(\Omega, \mathcal{F}, \mathbf{F}, \mathbb{P}^{X})$ via
\begin{equation}
X_{t} = \mu t + \sigma_{X} W_{t} + \sum \limits_{i=1}^{N_{t}} J_{i}, \hspace{1.5em} \mbox{for} \hspace{0.8em} t \geq 0,
\label{hypexp}
\end{equation}
\noindent where $(W_{t})_{t \geq 0}$ denotes an $\mathbf{F}$-Brownian motion and $(N_{t})_{t \geq 0}$ is an $\mathbf{F}$-Poisson process that has intensity parameter $\lambda >0$. The constants $\mu \in \mathbb{R}$ and $\sigma_{X} \geq 0$ denote the drift and volatility parameters of the diffusion part respectively. Additionally, the jumps $(J_{i})_{i \in \mathbb{N}}$ are assumed to be independent of $(N_{t})_{t \geq 0}$ and to form a sequence of independent and identically distributed random variables following a hyper-exponential distribution, i.e.,~their (common) density function $f_{J_{1}}(\cdot)$ is given by
\begin{equation}
f_{J_{1}}(y) = \sum \limits_{i=1}^{m} p_{i} \xi_{i}e^{-\xi_{i} y} \mathds{1}_{ \{ y \geq 0 \}} + \sum \limits_{j=1}^{n} q_{j} \eta_{j} e^{\eta_{j} y } \mathds{1}_{ \{ y < 0 \} },
\label{hypexpDens}
\end{equation}
\noindent where $p_{i} >0$ and $\xi_{i} > 1$ for $i \in \{1,\ldots, m\}$ and $q_{j}>0$ and $\eta_{j} >0$ for $j \in \{ 1, \ldots, n \}$. Here, the parameters $(p_{i})_{i \in \{1,\ldots,m \}}$ and $(q_{j})_{j \in \{1,\ldots,n \}}$ represent the proportion of jumps that are attributed to particular jump types and are therefore assumed to satisfy the condition $\sum \limits_{i=1}^{m} p_{i} + \sum \limits_{j=1}^{n} q_{j} = 1$. Finally, we will always assume that the intensity parameters $(\xi_{i})_{i \in \{1, \ldots, m \}}$ and $(\eta_{j})_{j \in \{1, \ldots, n \}}$ are ordered in the sense that
\begin{equation}
\xi_{1} < \xi_{2} < \cdots < \xi_{m} \hspace{2.5em} \mbox{and} \hspace{2.5em} \eta_{1} < \eta_{2} < \cdots < \eta_{n}
\end{equation}
\noindent and note that this does not consist in a loss of generality. \vspace{1em} \\
\noindent As special class of Lévy processes, hyper-exponential jump-diffusions can be equivalently characterized in terms of their Lévy triplet $\left(b_{X},\sigma_{X}^{2},\Pi_{X} \right)$, where $b_{X}$ and $\Pi_{X}$ are then obtained as 
\begin{align}
b_{X}  := \mu + \int \limits_{\{ |y|\leq 1 \}} y \, \Pi_{X}(dy) \hspace{2em} \mbox{and} \hspace{2.3em} \Pi_{X}(dy) := \lambda f_{J_{1}}(y) dy .
\end{align}
\noindent Using these results, their Lévy exponent, $\Psi_{X}(\cdot)$, is easily obtained via (\ref{CHARexp}), as
\begin{align}
\Psi_{X}(\theta) = -i \mu \theta + \frac{1}{2} \sigma_{X}^2 \theta^2  - \lambda \left( \sum \limits_{i=1}^{m} \frac{p_{i} \xi_{i}}{\xi_{i} - i \theta} + \sum \limits_{j=1}^{n} \frac{q_{j} \eta_{j}}{\eta_{j} + i \theta} - 1\right).
\end{align}
\noindent Similarly, the corresponding Laplace exponent, $\Phi_{X}(\cdot)$, is well-defined for $\theta \in (-\eta_{1}, \xi_{1}) $ and equals
\begin{equation}
\Phi_{X}(\theta) = \mu \theta + \frac{1}{2} \sigma_{X}^2 \theta^2  + \lambda \left( \sum \limits_{i=1}^{m} \frac{p_{i} \xi_{i}}{\xi_{i} - \theta} + \sum \limits_{j=1}^{n} \frac{q_{j} \eta_{j}}{\eta_{j} + \theta} - 1\right).
\label{LAP1}
\end{equation}
\noindent In what follows, we will consider the Laplace exponent as standalone function on the extended real domain $\Phi_{X}: \mathbb{R} \setminus \{\xi_{1}, \ldots, \xi_{m}, -\eta_{1}, \ldots, -\eta_{n} \} \rightarrow \mathbb{R}$. This quantity will play a central role in the upcoming derivations. In fact, many distributional properties of hyper-exponential jump-diffusion processes (and of their generalizations) are closely linked to the roots of the equation $\Phi_{X}(\theta) = \alpha$, for $\alpha \geq 0$. This was already used in diverse articles dealing with option pricing and risk management within the class of mixed-exponential jump-diffusion processes (cf.~among others \cite{ca09}, \cite{cc09}, \cite{ck11}, \cite{ck12}). In this context, the following important lemma was partly derived in \cite{ca09} under hyper-exponential jump-diffusion models. Since the proof of all the remaining statements do not substantially differ from the results derived in \cite{ca09}, the reader is referred to the arguments provided in this article.
\begin{Lem}
\label{CAIlem}
For $\Phi_{X}(\cdot)$ defined as in (\ref{LAP1}) and any $\alpha >0$, the following holds:
\begin{itemize} \setlength \itemsep{-0.1em}
\item[1.] If $\sigma_{X} \neq 0$, the equation $\Phi_{X}(\theta) = \alpha$ has $(m+n+2)$ real roots $\beta_{1,\alpha}, \ldots, \beta_{m+1,\alpha}$ and $\gamma_{1,\alpha}, \ldots, \gamma_{n+1,\alpha}$ that satisfy
\begin{align}
-\infty < \gamma_{n+1,\alpha} < -\eta_{n} < \gamma_{n,\alpha} < -\eta_{n-1} < \cdots < \gamma_{2,\alpha} < -\eta_{1} < \gamma_{1,\alpha} < 0, \\
0 < \beta_{1,\alpha} < \xi_{1} < \beta_{2,\alpha} < \cdots < \xi_{m-1} < \beta_{m,\alpha} < \xi_{m} < \beta_{m+1,\alpha} < \infty. \hspace{0.5em} 
\end{align}
\item[2.] If $\sigma_{X} = 0$ and $\mu \neq 0$ the equation $\Phi_{X}(\theta) = \alpha$ has $(m+n+1)$ real roots. Specifically,
\begin{itemize} \setlength \itemsep{-0.1em}
\item[$\bullet$] if $\mu > 0$, there are $m+1$ positive roots $\beta_{1,\alpha}, \ldots, \beta_{m+1,\alpha}$ and $n$ negative roots $\gamma_{1,\alpha}, \ldots, \gamma_{n,\alpha}$ that satisfy
\begin{align}
-\infty <  -\eta_{n} < \gamma_{n,\alpha} < -\eta_{n-1} < \cdots < \gamma_{2,\alpha} < -\eta_{1} < \gamma_{1,\alpha} < 0, \hspace{1em} \\
0 < \beta_{1,\alpha} < \xi_{1} < \beta_{2,\alpha} < \cdots < \xi_{m-1} < \beta_{m,\alpha} < \xi_{m} < \beta_{m+1,\alpha} < \infty.
\end{align}
\item[$\bullet$] if $\mu < 0$, there are $m$ positive roots $\beta_{1,\alpha}, \ldots, \beta_{m,\alpha}$ and $n+1$ negative roots $\gamma_{1,\alpha}, \ldots, \gamma_{n+1,\alpha}$ that satisfy
\begin{align}
-\infty <  \gamma_{n+1,\alpha} < -\eta_{n} < \gamma_{n,\alpha} < -\eta_{n-1} < \cdots < \gamma_{2,\alpha} < -\eta_{1} < \gamma_{1,\alpha} < 0, \\
0 < \beta_{1,\alpha} < \xi_{1} < \beta_{2,\alpha} < \cdots < \xi_{m-1} < \beta_{m,\alpha} < \xi_{m} <  \infty. \hspace{2em} 
\end{align}
\end{itemize}
\item[3.] If $\sigma_{X} = 0$ and $\mu = 0$ the equation $\Phi_{X}(\theta) = \alpha$ has $(m+n)$ real roots $\beta_{1,\alpha}, \ldots, \beta_{m,\alpha}$ and $\gamma_{1,\alpha}, \ldots, \gamma_{n,\alpha}$ that satisfy
\begin{align}
-\infty < -\eta_{n} < \gamma_{n,\alpha} < -\eta_{n-1} < \cdots < \gamma_{2,\alpha} < -\eta_{1} < \gamma_{1,\alpha} < 0, \\
0 < \beta_{1,\alpha} < \xi_{1} < \beta_{2,\alpha} < \cdots < \xi_{m-1} < \beta_{m,\alpha} < \xi_{m}   < \infty. \hspace{0.5em} 
\end{align}
\end{itemize}
\end{Lem}
\noindent At this point, we should mention that the roots in Lemma \ref{CAIlem} are only known in analytical form in very few cases. Nevertheless, this does not impact the importance and practicability of Lemma~\ref{CAIlem} since the roots can be anyway recovered using standard numerical techniques.
\subsubsection{Maturity-Randomization and OIDEs}
\noindent We turn back to the OIDE characterizations of Proposition \ref{prop3} and consider the respective problems (\ref{OIDE1})-(\ref{OIDE3}) and (\ref{OIDEj1})-(\ref{OIDEj3}) under hyper-exponential jump-diffusion processes with non-zero volatility parameter $\sigma_{X} \neq 0$. Switching to the case where $\sigma_{X} = 0$ does not fundamentally change the approach and only few, slight adaptions are needed. We will address some of these adaptions in Section~\ref{SECapproxim}, when discussing hyper-exponential jump-diffusion approximations to infinite-activity pure jump processes. \vspace{1em} \\
\noindent To start, we note that the infinitesimal generator (\ref{InfGen}) simplifies in this case to 
\begin{align}
\mathcal{A}_{X} V(\mathcal{T},x) & = \frac{1}{2} \sigma^{2}_{X} \partial_{x}^{2} V(\mathcal{T},x) + \mu \partial_{x} V(\mathcal{T},x) +  \lambda \int \limits_{\mathbb{R}}  \big( V(\mathcal{T},x+y) - V(\mathcal{T},x) \big) f_{J_{1}}(y)dy,
\end{align}
\noindent which allows us, together with the properties of the hyper-exponential density $f_{J_{1}}(\cdot)$, to uniquely solve Problems (\ref{OIDE1})-(\ref{OIDE3}) and (\ref{OIDEj1})-(\ref{OIDEj3}) and to derive closed-form expressions for the maturity-randomized first-passage probabilities $\mathcal{LC} \big(u_{X}^{\mathcal{E}_{0}^{\pm}}\big)(\cdot)$ and $\mathcal{LC} \big(u_{X}^{\mathcal{E}_{\mathcal{J}}^{\pm}}\big)(\cdot)$. Specifically, we define for any $\vartheta >0$ the $(m+1)\times(m+1)$ and $(n+1)\times(n+1)$ matrices $\underline{\mathbf{A}}_{\vartheta}$ and $\overline{\mathbf{A}}_{\vartheta}$ respectively via
\begin{equation}
\underline{\mathbf{A}}_{\vartheta} := \left( \begin{array}{cccc} 
1 & 1 & \ldots & 1 \\
\frac{\xi_{1}}{\xi_{1}-\beta_{1,\vartheta}} & \frac{\xi_{1}}{\xi_{1}-\beta_{2,\vartheta}} & \cdots & \frac{\xi_{1}}{\xi_{1}-\beta_{m+1,\vartheta}} \\
\frac{\xi_{2}}{\xi_{2}-\beta_{1,\vartheta}} & \frac{\xi_{2}}{\xi_{2}-\beta_{2,\vartheta}} & \cdots & \frac{\xi_{2}}{\xi_{2}-\beta_{m+1,\vartheta}} \\
\vdots & \vdots & \ddots & \vdots \\
\frac{\xi_{m}}{\xi_{m}-\beta_{1,\vartheta}} & \frac{\xi_{m}}{\xi_{m}-\beta_{2,\vartheta}} & \cdots & \frac{\xi_{m}}{\xi_{m}-\beta_{m+1,\vartheta}}
\end{array}\right), 
\hspace{2.5em} 
\overline{\mathbf{A}}_{\vartheta} := \left( \begin{array}{cccc} 
1 & 1 & \ldots & 1 \\
\frac{\eta_{1}}{\eta_{1}+\gamma_{1,\vartheta}} & \frac{\eta_{1}}{\eta_{1}+\gamma_{2,\vartheta}} & \cdots & \frac{\eta_{1}}{\eta_{1}+\gamma_{n+1,\vartheta}} \\
\frac{\eta_{2}}{\eta_{2}+\gamma_{1,\vartheta}} & \frac{\eta_{2}}{\eta_{2}+\gamma_{2,\vartheta}} & \cdots & \frac{\eta_{2}}{\eta_{2}+\gamma_{n+1,\vartheta}} \\
\vdots & \vdots & \ddots & \vdots \\
\frac{\eta_{n}}{\eta_{n}+\gamma_{1,\vartheta}} & \frac{\eta_{n}}{\eta_{n}+\gamma_{2,\vartheta}} & \cdots & \frac{\eta_{n}}{\eta_{n}+\gamma_{n+1,\vartheta}}
\end{array}\right), 
\label{MATRIX2}
\end{equation}
\noindent and observe that these matrices are invertible.\footnote{The invertibility of $\underline{\mathbf{A}}_{\vartheta}$ and $\overline{\mathbf{A}}_{\vartheta}$ can be proved as in \cite{ck11} and the reader is referred to this article.}~Additionally, we denote for any $k \in \mathbb{N}$ the $1 \times k$ vectors of zeros and ones as
\begin{equation}
\mathbf{0}_{k} := (\underbrace{0, \ldots, 0}_{k}) \hspace{2em} \mbox{and} \hspace{2.3em} \mathbf{1}_{k} := (\underbrace{1, \ldots, 1}_{k} ) .
\end{equation}
\noindent Using the above notation, the following proposition can be derived. The proof is presented in Appendix C.
\begin{Prop}
\label{prop4}
\noindent Assume that $(X_{t})_{t \geq 0}$ follows a hyper-exponential jump-diffusion process with non-zero diffusion component, as described in (\ref{hypexp}), (\ref{hypexpDens}) with $\sigma_{X} \neq 0$. Then, for any level $\ell \in \mathbb{R}$ and intensity $\vartheta > 0$, the maturity-randomized upside first-passage probabilities $\mathcal{LC} \big(u_{X}^{\mathcal{E}_{0}^{+}}\big)(\cdot)$ and $\mathcal{LC} \big(u_{X}^{\mathcal{E}_{\mathcal{J}}^{+}}\big)(\cdot)$ take the form
\begin{equation}
\mathcal{LC} \big(u_{X}^{\mathcal{E}_{0}^{+}}\big)(\vartheta,x;\ell) = \left \{ \begin{array}{lc}
0, &  x > \ell, \\
1, &   x = \ell, \\
\sum \limits_{k=1}^{m+1} \underline{v}_{0,k} \, e^{\beta_{k,\vartheta} \cdot (x-\ell)}, &   x < \ell,
\end{array} \right. \hspace{0.5em} \mbox{and} \hspace{0.9em} \mathcal{LC} \big(u_{X}^{\mathcal{E}_{\mathcal{J}}^{+}}\big)(\vartheta,x;\ell) = \left \{ \begin{array}{lc}
1, &  x > \ell, \\
0, &  x = \ell, \\
\sum \limits_{k=1}^{m+1} \underline{v}_{\mathcal{J},k} \, e^{\beta_{k,\vartheta} \cdot (x-\ell)}, &  x < \ell. \\
\end{array} \right. \label{STru1}
\end{equation}
\noindent Here, $\underline{\mathbf{v}}_{0} := \left(\underline{v}_{0,1}, \ldots, \underline{v}_{0,m+1} \right)^{\intercal}$ and $\underline{\mathbf{v}}_{\mathcal{J}} := \left(\underline{v}_{\mathcal{J},1}, \ldots, \underline{v}_{\mathcal{J},m+1} \right)^{\intercal}$ are weight vectors uniquely determined by the system of linear equations 
\begin{equation}
\underline{\mathbf{A}}_{\vartheta} \underline{\mathbf{v}}_{0} = \left(1,\mathbf{0}_{m} \right)^{\intercal} \hspace{2em} \mbox{and} \hspace{2.3em} \underline{\mathbf{A}}_{\vartheta} \underline{\mathbf{v}}_{\mathcal{J}} = \left(0,\mathbf{1}_{m} \right)^{\intercal} .
\label{CoefEq1}
\end{equation}
\noindent Similarly, for $\ell \in \mathbb{R}$ and $\vartheta > 0$, the maturity-randomized downside first-passage probabilities $\mathcal{LC} \big(u_{X}^{\mathcal{E}_{0}^{-}}\big)(\cdot)$ and $\mathcal{LC} \big(u_{X}^{\mathcal{E}_{\mathcal{J}}^{-}}\big)(\cdot)$ are given by
\begin{equation}
\mathcal{LC} \big(u_{X}^{\mathcal{E}_{0}^{-}}\big)(\vartheta,x;\ell) = \left \{ \begin{array}{lc}
\sum \limits_{k=1}^{n+1} \overline{v}_{0,k} \, e^{\gamma_{k,\vartheta} \cdot (x-\ell)}, &  x > \ell, \\
1, &  x = \ell, \\
0, &  x < \ell,
\end{array} \right. \hspace{0.5em} \mbox{and} \hspace{0.9em} \mathcal{LC} \big(u_{X}^{\mathcal{E}_{\mathcal{J}}^{-}}\big)(\vartheta,x;\ell) = \left \{ \begin{array}{lc}
\sum \limits_{k=1}^{n+1} \overline{v}_{\mathcal{J},k} \, e^{\gamma_{k,\vartheta} \cdot (x-\ell)}, &  x > \ell, \\
0, &  x = \ell, \\
1, &  x < \ell, \\
\end{array} \right. \label{STru2}
\end{equation}
\noindent where, $\overline{\mathbf{v}}_{0} := \left(\overline{v}_{0,1}, \ldots, \overline{v}_{0,n+1} \right)^{\intercal}$ and $\overline{\mathbf{v}}_{\mathcal{J}} := \left(\overline{v}_{\mathcal{J},1}, \ldots, \overline{v}_{\mathcal{J},n+1} \right)^{\intercal}$ are uniquely determined by the system of linear equations
\begin{equation}
\overline{\mathbf{A}}_{\vartheta} \overline{\mathbf{v}}_{0} = \left( 1, \mathbf{0}_{n} \right)^{\intercal} \hspace{2em} \mbox{and} \hspace{2.3em} \overline{\mathbf{A}}_{\vartheta} \overline{\mathbf{v}}_{\mathcal{J}} = \left( 0, \mathbf{1}_{n} \right)^{\intercal}.
\label{CoefEq2}
\end{equation}
\end{Prop}
\noindent Proposition \ref{prop4} already provides an important analytical disentanglement of the diffusion and jump contributions underlying first-passage probabilities. Nevertheless, it may be additionally insightful to understand how the jump risk is further distributed across jump types. Under hyper-exponential jump-diffusion processes such a decomposition can be derived and this is discussed next. To start, we define for $i \in \{1, \ldots, m \}$ and $j \in \{1, \ldots, n \}$ the following jump-events
\begin{align}
\mathcal{E}_{i}^{+} := \left \{ X_{\tau_{\ell}^{X,+}} > \ell, \, J_{N_{\tau_{\ell}^{X,+}}} \sim \mbox{Exp}(\xi_{i}) \right \} \hspace{1em} \mbox{and} \hspace{1.3em} \mathcal{E}_{j}^{-} := \left \{ X_{\tau_{\ell}^{X,-}} < \ell, \, J_{N_{\tau_{\ell}^{X,-}}} \sim \mbox{Exp}(\eta_{j}) \right \}, 
\label{DecompEnhancedJump} 
\end{align}
\noindent and see that
\begin{equation}
\mathcal{E}_{\mathcal{J}}^{+} \setminus \left( \{ X_{0} > \ell \} \cup \{ \tau_{\ell}^{X,+} = \infty \} \right) = \bigcup \limits_{i=1}^{m} \mathcal{E}_{i}^{+}  \hspace{1.5em} \mbox{and} \hspace{1.7em} \mathcal{E}_{\mathcal{J}}^{-} \setminus \left( \{ X_{0} < \ell \} \cup \{ \tau_{\ell}^{X,-} = \infty \} \right) =  \bigcup \limits_{j = 1}^{n}  \mathcal{E}_{j}^{-} ,
\end{equation}
\noindent i.e.,~we essentially decompose the first-passage events contributed by jumps in events that are triggered by jumps of certain types. Additionally, we note that the first-passage events $\mathcal{E}_{i}^{+}$ and $\mathcal{E}_{j}^{-}$ for $i \in \{1,\ldots, m \}$ and $j \in \{1, \ldots, n \}$ are disjoint among each others. In particular, this gives that the first-passage probabilities contributed by jumps, $u_{X}^{\mathcal{E}_{\mathcal{J}}^{\pm}}(\cdot)$, and the respective maturity-randomized quantities, $\mathcal{LC} \big(u_{X}^{\mathcal{E}_{\mathcal{J}}^{\pm}}\big)(\cdot)$, have, for $\ell \in \mathbb{R}$, $x \in \mathbb{R}\setminus \mathcal{H}_{\ell}^{\pm}$, and $\mathcal{T} \in [0,T]$ and $\vartheta > 0$ respectively the following decomposition
\begin{align}
u_{X}^{\mathcal{E}_{\mathcal{J}}^{+}}(\mathcal{T},x;\ell ) = \sum \limits_{i=1}^{m} u_{X}^{\mathcal{E}_{i}^{+}}(\mathcal{T},x;\ell ),  & \hspace{2em} u_{X}^{\mathcal{E}_{\mathcal{J}}^{-}}(\mathcal{T},x;\ell ) = \sum \limits_{j=1}^{n} u_{X}^{\mathcal{E}_{j}^{-}}(\mathcal{T},x;\ell ), \label{DecJump+}\\
\mathcal{LC} \big(u_{X}^{\mathcal{E}_{\mathcal{J}}^{+}}\big)(\vartheta,x;\ell ) =  \sum \limits_{i=1}^{m} \mathcal{LC} \big(u_{X}^{\mathcal{E}_{i}^{+}}\big)(\vartheta,x;\ell ), & \hspace{2em}  \mathcal{LC} \big(u_{X}^{\mathcal{E}_{\mathcal{J}}^{-}}\big)(\vartheta,x;\ell ) = \sum \limits_{j=1}^{n} \mathcal{LC} \big(u_{X}^{\mathcal{E}_{j}^{-}}\big)(\vartheta,x;\ell ), \label{DecJump-}
\end{align}
\noindent where $u_{X}^{\mathcal{E}_{i}^{+}}(\mathcal{T},x;\ell)$ for $i \in \{1, \ldots, m \}$ and $u_{X}^{\mathcal{E}_{j}^{-}}(\mathcal{T},x;\ell)$ for $j \in \{1, \ldots, n \}$ refer to the functions defined by 
\begin{align}
u_{X}^{\mathcal{E}_{i}^{+}}(\mathcal{T},x;\ell) := \mathbb{E}_{x}^{\mathbb{P}^{X}} \left[ \mathds{1}_{ \left\{ \tau_{\ell}^{X,+} \leq \mathcal{T} \right\} \, \cap \,\mathcal{E}_{i}^{+} }\right] & \hspace{1.5em} \mbox{and} \hspace{1.7em} u_{X}^{\mathcal{E}_{j}^{-}}(\mathcal{T},x;\ell) := \mathbb{E}_{x}^{\mathbb{P}^{X}} \left[ \mathds{1}_{ \left\{ \tau_{\ell}^{X,-} \leq \mathcal{T} \right\} \, \cap \,\mathcal{E}_{j}^{-} }\right].
\end{align}
\noindent Next, we note as in \cite{kw03} that the monotonicity of the cumulative distribution functions $t \mapsto u_{X}^{\mathcal{E}}(t,x;\ell)$ for $\mathcal{E} \in \{\mathcal{E}_{0}^{\pm}, \mathcal{E}_{\mathcal{J}}^{\pm}, \mathcal{E}_{1}^{+}, \ldots, \mathcal{E}_{m}^{+}, \mathcal{E}_{1}^{-}, \ldots, \mathcal{E}_{n}^{-} \} $ implies that we can rewrite each of the maturity-randomized versions $\mathcal{LC}(u_{X}^{\mathcal{E}})(\cdot)$, for $\ell \in \mathbb{R}$, $x \in \mathbb{R} \setminus \overline{\mathcal{H}_{\ell}^{\pm}}$ and $\vartheta >0$, in the form
\begin{align}
\mathcal{LC}(u_{X}^{\mathcal{E}})(\vartheta, x; \ell) & := \int \limits_{0}^{\infty} \vartheta e^{-\vartheta t } \, u_{X}^{\mathcal{E}}(t,x; \ell) dt \nonumber \\
& = \int \limits_{0}^{\infty} e^{-\vartheta t} \, \partial_{t} u_{X}^{\mathcal{E}}(t,x; \ell) dt  \label{IntEqBup}\\
& = \int \limits_{0}^{\infty} e^{-\vartheta t} \, u_{X}^{\mathcal{E}}(dt,x; \ell) =  \mathbb{E}_{x}^{\mathbb{P}^{X}} \left[ e^{-\vartheta \, \tau_{\ell}^{X,\pm}} \, \mathds{1}_{\mathcal{E}} \right], \nonumber 
\end{align}
\noindent where $\tau_{\ell}^{X,\pm} = \tau_{\ell}^{X,+}$ for $\mathcal{E} \in \{\mathcal{E}_{1}^{+}, \ldots, \mathcal{E}_{m}^{+} \}$ and $\tau_{\ell}^{X,\pm} = \tau_{\ell}^{X,-}$ for $\mathcal{E} \in \{\mathcal{E}_{1}^{-}, \ldots, \mathcal{E}_{n}^{-} \}$. Combining this representation with the fact that the overshoot distribution is conditionally memoryless and independent of the first-passage time, given that the overshoot is greater than zero and that the exponential type of the jump distribution is specified (cf.~\cite{ca09}), finally allows us to arrive at the next proposition. The proof is provided in Appendix C.
\begin{Prop}
\label{prop5}
\noindent Assume that $(X_{t})_{t \geq 0}$ follows a hyper-exponential jump-diffusion process with non-zero diffusion component, as described in (\ref{hypexp}), (\ref{hypexpDens}) with $\sigma_{X} \neq 0$ and define for any level $\ell \in \mathbb{R}$, $x \in \mathbb{R}$ and intensity $\vartheta > 0$ the following vectors
\begin{align}
\underline{\mathbf{LC}}_{\vartheta, \ell}(x) := \Big(\mathcal{LC}(u_{X}^{\mathcal{E}_{0}^{+}})(\vartheta, x; \ell), \ldots, \mathcal{LC}(u_{X}^{\mathcal{E}_{m}^{+}})(\vartheta, x; \ell) \Big)^{\intercal}, & \hspace{1.5em} \underline{\mathbf{e}}_{\vartheta, \ell}(x) := \big( e^{\beta_{1,\vartheta} \cdot (x-\ell)}, \ldots, e^{\beta_{m+1,\vartheta} \cdot (x-\ell)}\big)^{\intercal} ,\\
\overline{\mathbf{LC}}_{\vartheta, \ell}(x) := \Big(\mathcal{LC}(u_{X}^{\mathcal{E}_{0}^{-}})(\vartheta, x; \ell), \ldots, \mathcal{LC}(u_{X}^{\mathcal{E}_{n}^{-}})(\vartheta, x; \ell) \Big)^{\intercal}, & \hspace{1.5em} \overline{\mathbf{e}}_{\vartheta, \ell}(x) := \big( e^{\gamma_{1,\vartheta} \cdot (x-\ell)}, \ldots, e^{\gamma_{n+1,\vartheta} \cdot (x-\ell)}\big)^{\intercal}.
\end{align}
\noindent Then, for $x \in \mathbb{R}\setminus \overline{\mathcal{H}_{\ell}^{+}}$, the vector of maturity-randomized upside jump contributions, $\underline{\mathbf{LC}}_{\vartheta, \ell}(x)$, is uniquely determined by the system of linear equations
\begin{equation}
\underline{\mathbf{A}}_{\vartheta}^{\intercal} \, \underline{\mathbf{LC}}_{\vartheta, \ell}(x) = \underline{\mathbf{e}}_{\vartheta, \ell}(x).
\label{SYSbap1}
\end{equation}
\noindent Similarly, for $x \in \mathbb{R}\setminus \overline{\mathcal{H}_{\ell}^{-}}$ the vector of maturity-randomized downside jump contributions, $\overline{\mathbf{LC}}_{\vartheta, \ell}(x)$, is uniquely determined by the system of linear equations
\begin{equation}
\overline{\mathbf{A}}_{\vartheta}^{\intercal} \, \overline{\mathbf{LC}}_{\vartheta, \ell}(x) = \overline{\mathbf{e}}_{\vartheta, \ell}(x).
\label{SYSbap2}
\end{equation}
\end{Prop}
\noindent Proposition \ref{prop5} provides an implicit characterization of diffusion and jump contributions underlying (maturity-randomized) first-passage probabilities and already allows for a derivation of the full vectors $\underline{\mathbf{LC}}_{\vartheta, \ell}(x)$ or $\overline{\mathbf{LC}}_{\vartheta, \ell}(x)$ using standard numerical methods. Nevertheless, the systems (\ref{SYSbap1}) and (\ref{SYSbap2}) can be explicitly solved to derive analytical expressions for each of the functions $\mathcal{LC}(u_{X}^{\mathcal{E}})(\cdot)$ with $\mathcal{E} \in \{\mathcal{E}_{0}^{+}, \mathcal{E}_{1}^{+}, \ldots, \mathcal{E}_{m}^{+} \} $ or $\mathcal{E} \in \{\mathcal{E}_{0}^{-}, \mathcal{E}_{1}^{-}, \ldots, \mathcal{E}_{n}^{-} \} $. This was already derived in a different context in \cite{cy13}. In particular, their results can be refined to arrive at the following useful proposition.
\begin{Prop}
\label{prop6}
\noindent Assume that $(X_{t})_{t \geq 0}$ follows a hyper-exponential jump-diffusion process with non-zero diffusion component, as described in (\ref{hypexp}), (\ref{hypexpDens}) with $\sigma_{X} \neq 0$. Then, for any level $\ell \in \mathbb{R}$, $x \in \mathbb{R}\setminus \overline{\mathcal{H}_{\ell}^{+}}$ and intensity $ \vartheta >0$ we have that
\begin{equation}
\label{JuMpEq1}
\mathcal{LC}\big(u_{X}^{\mathcal{E}_{i}^{+}}\big)(\vartheta, x; \ell) = \sum \limits_{k=1}^{m+1} v_{\mathcal{E}_{i}^{+},k} \cdot e^{\beta_{k,\vartheta} \cdot (x- \ell)} , \hspace{1.5em} i \in \{1, \ldots, m \},
\end{equation}
\noindent where, for $k \in \{1, \ldots, m+1 \}$, the coefficients $v_{\mathcal{E}_{i}^{+},k}$ can be expressed in terms of the coefficients $\underline{v}_{0,k}$ by
\begin{equation}
\label{JumpCO1}
v_{\mathcal{E}_{i}^{+},1} = - \frac{1}{\xi_{i}} \frac{\mathbf{C}_{\vartheta}^{+}(\xi_{i})}{ \big(\mathbf{B}^{+} \big)'(\xi_{i})} \, \underline{v}_{0,1}, \hspace{1.5em} \mbox{and} \hspace{1.8em} v_{\mathcal{E}_{i}^{+},k} = -\frac{1}{\xi_{i}} \frac{\mathbf{C}_{\vartheta}^{+}(\xi_{i})}{ \big(\mathbf{B}^{+} \big)'(\xi_{i})} \, \underline{v}_{0,k} + \frac{1}{\xi_{i}} d_{k,i}^{+}, \hspace{1em} k \in \{2, \ldots, m+1 \}, 
\end{equation}
\noindent with 
\begin{equation}
\mathbf{B}^{+}(x) := \prod \limits_{s=1}^{m} \big(\xi_{s} - x \big), \hspace{2em} \mathbf{C}_{\vartheta}^{+}(x) := \prod \limits_{s=2}^{m+1} \big(\beta_{s,\vartheta} - x \big),
\label{EqB&C}
\end{equation}
\noindent and
\begin{equation}
d_{i,j}^{+} := - \frac{\mathbf{B}^{+}(\beta_{i,\vartheta} ) \, \mathbf{C}^{+}_{\vartheta}(\xi_{j})}{(\xi_{j} - \beta_{i,\vartheta}) \, \big(\mathbf{B}^{+} \big)'(\xi_{j}) \, \big(\mathbf{C}_{\vartheta}^{+} \big)'(\beta_{i,\vartheta})}, \hspace{1.5em} i \in \{2, \ldots, m+1 \}, \,\;  j \in \{1, \ldots, m \}.
\label{Eqdplus}
\end{equation}
\noindent Similarly, for $\ell \in \mathbb{R}$, $x \in \mathbb{R}\setminus \overline{\mathcal{H}_{\ell}^{-}}$ and $ \vartheta >0$ we have that
\begin{equation}
\label{JuMpEq2}
\mathcal{LC}\big(u_{X}^{\mathcal{E}_{j}^{-}}\big)(\vartheta, x; \ell) = \sum \limits_{k=1}^{n+1} v_{\mathcal{E}_{j}^{-},k} \cdot e^{\gamma_{k,\vartheta} \cdot (x- \ell)} , \hspace{1.5em} j \in \{1, \ldots, n \},
\end{equation}
\noindent where the coefficients $v_{\mathcal{E}_{j}^{-},k}$ are given, for $k \in \{1, \ldots, n+1 \}$, by
\begin{equation}
\label{JumpCO2}
v_{\mathcal{E}_{j}^{-},1} = (-1)^{n} \, \frac{1}{\eta_{j}} \frac{\mathbf{C}_{\vartheta}^{-}(\eta_{j})}{ \big(\mathbf{B}^{-} \big)'(-\eta_{j})} \, \overline{v}_{0,1}, \hspace{1.5em} \mbox{and} \hspace{1.8em} v_{\mathcal{E}_{j}^{-},k} = (-1)^{n} \, \frac{1}{\eta_{j}} \frac{\mathbf{C}_{\vartheta}^{-}(\eta_{j})}{ \big(\mathbf{B}^{-} \big)'(-\eta_{j})} \, \overline{v}_{0,k} + \frac{1}{\eta_{j}} d_{k,j}^{-}, \hspace{1em} k \in \{2, \ldots, n+1 \}, 
\end{equation}
\noindent with 
\begin{equation}
\mathbf{B}^{-}(x) := \prod \limits_{s=1}^{n} \big(\eta_{s} + x \big), \hspace{2em} \mathbf{C}_{\vartheta}^{-}(x) := \prod \limits_{s=2}^{n+1} \big(\gamma_{s,\vartheta} + x \big),
\label{EqB&Cdown}
\end{equation}
\noindent and
\begin{equation}
d_{i,j}^{-} := \frac{\mathbf{B}^{-}(\gamma_{i,\vartheta} ) \, \mathbf{C}^{-}_{\vartheta}(\eta_{j})}{(\eta_{j} + \gamma_{i,\vartheta}) \, \big(\mathbf{B}^{-} \big)'(-\eta_{j}) \, \big(\mathbf{C}_{\vartheta}^{-} \big)'(-\gamma_{i,\vartheta})}, \hspace{1.5em} i \in \{2, \ldots, n+1 \}, \; \, j \in \{1, \ldots, n \} .
\label{Eqdplusdown}
\end{equation}
\end{Prop}
$\mbox{ }$ \vspace{1em} \\
\noindent \underline{\bf Remark 4.}
\begin{itemize} \setlength \itemsep{-0.1em}
\item[i)] We re-emphasize that the (full) results in Proposition \ref{prop4}, Proposition \ref{prop5} and Proposition \ref{prop6} only hold under non-zero diffusion component. In fact, when $\sigma_{X} = 0$ the maturity-randomized first-passage probabilities $\mathcal{LC} \big(u_{X}^{\pm}\big)(\cdot)$ reduce (for $x \neq \ell$) to the jump contributions $\mathcal{LC} \big(u_{X}^{\mathcal{E}_{\mathcal{J}}^{\pm}}\big)(\cdot)$ and the finite activity of the underlying jump process implies that the continuous-fit conditions
$$ \mathcal{LC} \big(u_{X}^{\mathcal{E}_{\mathcal{J}}^{+}}\big)(\vartheta,\ell-;\ell) = 0 \hspace{2.5em} \mbox{and} \hspace{2.7em} \mathcal{LC} \big(u_{X}^{\mathcal{E}_{\mathcal{J}}^{-}}\big)(\vartheta,\ell+;\ell) = 0 $$
\noindent do not anymore hold. In addition, in view of Lemma \ref{CAIlem}, the matrices defined in (\ref{MATRIX2}) may have to be replaced by corresponding $(m+1) \times m$, $(n+1) \times n$, $m\times m$ or $n \times n$ matrices. Therefore, the resulting systems of equations (\ref{CoefEq1}), (\ref{CoefEq2}) and (\ref{SYSbap1}), (\ref{SYSbap2}) need to be adjusted accordingly and this may finally impact our derivations in Proposition \ref{prop6}. We will deal with these adaptions in more details in Section~\ref{SECapproxim}.
\item[ii)] In addition to obtaining (semi-)analytical expressions for $\mathcal{LC}(u_{X}^{\mathcal{E}})(\cdot)$ with $\mathcal{E} \in \{\mathcal{E}_{0}^{+}, \mathcal{E}_{1}^{+}, \ldots, \mathcal{E}_{m}^{+} \} $ or $\mathcal{E} \in \{\mathcal{E}_{0}^{-}, \mathcal{E}_{1}^{-}, \ldots, \mathcal{E}_{n}^{-} \} $, Proposition~\ref{prop6} reveals, together with (\ref{STru1}) and (\ref{STru2}), that for any values $x \in \mathbb{R}\setminus \overline{\mathcal{H}_{\ell}^{\pm}}$ and intensity $ \vartheta >0$ the functions $\ell \mapsto \mathcal{LC}(u_{X}^{\mathcal{E}})(\vartheta, x ; \ell)$ with $\mathcal{E} \in \{\mathcal{E}_{0}^{+}, \mathcal{E}_{1}^{+}, \ldots, \mathcal{E}_{m}^{+} \} $ or $\mathcal{E} \in \{\mathcal{E}_{0}^{-}, \mathcal{E}_{1}^{-}, \ldots, \mathcal{E}_{n}^{-} \} $ consist in linear combinations of exponentials. We will combine this particularly simple form with the structure of the Gaver-Stehfest algorithm in Section \ref{NUMres} to derive a simple inversion algorithm for the integral part of Proposition \ref{lem1}. 
\end{itemize}
$\mbox{}$ \hspace{44.8em} \scalebox{0.75}{$\blacklozenge$} \\
\section{Approximating Models with Jumps via Hyper-Exponential Jump-Diffusions}
\label{SECapproxim}
\noindent In this section, we complement the theory developed in the previous parts by discussing hyper-exponential approximations to (infinite-activity) pure jump processes having a completely monotone jump density. We slightly adapt the approach followed in \cite{amp07}, \cite{jp10}, briefly discuss the resulting approximations and comment on how they relate to our final aim of intra-horizon risk quantification. The results will play a central role in the upcoming numerical analysis of Section \ref{NUMres}. 
\subsection{General Approximation Scheme}
\noindent To start, we recall that a one-sided density $f: [0,\infty) \rightarrow \mathbb{R}$ is said to be completely monotone if for any $k \in \mathbb{N}$ its $k$-th derivative $f^{(k)}(\cdot)$ exists and $(-1)^k \, f^{(k)}(x) \geq 0$ holds on $x \in (0,\infty)$. Additionally, we recall that when dealing instead with a two-sided density $g: \mathbb{R} \rightarrow \mathbb{R}$ this definition naturally extends by requiring that both of the functions $x \mapsto g(x) \, \mathds{1}_{[0,\infty)}(x) $ and $x \mapsto g(-x) \, \mathds{1}_{[0,\infty)}(x)$ are completely monotone. It can be shown that many Lévy processes employed in financial modeling have a completely monotone jump density $\pi_{X}(\cdot)$, i.e.,~that their intensity measure $\Pi_{X}$ takes the particular form
\begin{equation}
\Pi_{X}(dy) = \pi_{X}(y)\,  dy
\end{equation}
\noindent with $\pi_{X}(\cdot)$ being a (two-sided) completely monotone density. This includes among others hyper-exponential jump-diffusion models (cf.~\cite{ko02}, \cite{ca09}), Normal Inverse Gaussian (NIG) processes (cf.~\cite{bn97}) as well as the whole class of stable and tempered stable processes (cf.~\cite{kt13}, \cite{hk16}), containing the very popular Variance-Gamma (VG) (cf.~\cite{ms90}, \cite{mcc98}) and Carr-Geman-Madan-Yor (CGMY) models (cf.~\cite{cgmy02}). We are going to deal with some of these dynamics in Section~\ref{NUMres}. \vspace{1em} \\
\noindent In view of Bernstein's theorem, a Lévy process $(X_{t})_{t \geq 0}$ has a completely monotone jump density if and only if its density can be decomposed as 
\begin{equation}
\pi_{X}(y) = \mathds{1}_{(0,\infty)}(y) \int \limits_{0}^{\infty} e^{-u y} \, \mu^{+}(du) + \mathds{1}_{(-\infty,0)}(y) \int \limits_{-\infty}^{0} e^{-|v y|} \, \mu^{-}(dv),
\end{equation}
\noindent where $\mu^{+}(\cdot)$ and $\mu^{-}(\cdot)$ are (non-negative and finite) measures defined on $(0,\infty)$ and $(-\infty,0)$, respectively. In particular, discretizing these integrals allows to approximate the completely monotone density $\pi_{X}(\cdot)$ by a sequence $\big(\pi^{(n)}_{X}(\cdot)\big)_{n \in \mathbb{N}}$ of densities having the form
\begin{equation}
\label{HYPERapproxDENS1}
\pi_{X}^{(n)}(y) := \Lambda_{n} f_{n}(y), \hspace{1.5em} n \in \mathbb{N},
\end{equation}
\noindent where $f_{n}(\cdot), n \in \mathbb{N}$, are hyper-exponential densities defined, for partitions $(u_{i}^{(n)})_{i \in \{0, \ldots,N_{n}\}}$ and $(v_{j}^{(n)})_{j \in \{0, \ldots,M_{n}\}}$ of the sets $(0,\infty)$ and $(-\infty,0)$, respectively\footnote{Our convention is that the elements of the partitions are increasing in their index, i.e.,~we assume that for any $n \in \mathbb{N}$ the relations $0<|v_{M_{n}}^{(n)}| < \ldots < |v_{0}^{(n)}|< \infty$ and $0<u_{0}^{(n)} < \ldots < u_{N_{n}}^{(n)} < \infty$ hold.}, having vanishing mesh\footnote{Recall that the mesh of a partition $(u_{i}^{(n)})_{i \in \{0, \ldots,N_{n}\}}$ is defined by $\max \limits_{1\leq i \leq N_{n}} |u_{i}^{(n)} - u_{i-1}^{(n)}|$.}, by
\begin{equation}
\label{HYPERapproxDENS2}
f_{n}(y) := \sum \limits_{i =1}^{N_{n}} p_{i}^{(n)} \tilde{u}_{i}^{(n)} e^{-\tilde{u}_{i}^{(n)} y} \, \mathds{1}_{\{ y \geq 0 \}} + \sum \limits_{j=1}^{M_{n}} q_{j}^{(n)} |\tilde{v}_{j}^{(n)}| \, e^{|\tilde{v}_{j}^{(n)}| \,y} \, \mathds{1}_{\{ y < 0 \}},
\end{equation}                             
 and, for $n \in \mathbb{N}$, 
\begin{align}
\tilde{u}_{i}^{(n)} := \frac{1}{2} \left( u_{i-1}^{(n)} + u_{i}^{(n)} \right), \hspace{0.5em} i=1, \ldots, N_{n}, \hspace{2em} & \tilde{v}_{j}^{(n)} := \frac{1}{2} \left( v_{j-1}^{(n)} + v_{j}^{(n)} \right), \hspace{0.5em} j=1, \ldots, M_{n}, \\
p_{i}^{(n)} := \frac{\mu^{+}\big(\big[u_{i-1}^{(n)},u_{i}^{(n)} \big) \big)}{\Lambda_{n} \cdot \tilde{u}_{i}^{(n)}}, \hspace{0.5em} i=1, \ldots, N_{n}, \hspace{2em} & q_{j}^{(n)} := \frac{\mu^{-}\big(\big(v_{j-1}^{(n)},v_{j}^{(n)} \big] \big)}{\Lambda_{n} \cdot |\tilde{v}_{j}^{(n)}|}, \hspace{0.5em} j=1, \ldots, M_{n}, \\
\Lambda_{n} := \sum \limits_{i=1}^{N_{n}} \frac{\mu^{+}\big(\big[u_{i-1}^{(n)},u_{i}^{(n)} \big) \big)}{\tilde{u}_{i}^{(n)}} & + \sum \limits_{j=1}^{M_{n}} \frac{\mu^{-}\big(\big(v_{j-1}^{(n)},v_{j}^{(n)} \big] \big)}{|\tilde{v}_{j}^{(n)}|}.
\end{align}
\noindent This intuitive idea can be further developed and hyper-exponential jump-diffusion approximations to Lévy processes having completely monotone jumps can be derived. For this purpose, let $(X_{t})_{t \geq 0}$ be such a process whose Lévy triplet is denoted by $\left(b_{X}, \sigma_{X}^{2}, \Pi_{X} \right)$. Then, for any sequence $(\epsilon_{n})_{n \in \mathbb{N}}$ of positive numbers converging to zero corresponding sequences of partitions $\big((u_{i}^{(n)})_{i \in \{0, \ldots,N_{n}\}}\big)_{n \in \mathbb{N}}$ and $\big((v_{j}^{(n)})_{j \in \{0, \ldots,M_{n}\}}\big)_{n \in \mathbb{N}}$ can be constructed such that the following conditions hold for each $n \in \mathbb{N}$:
\begin{align}
\int \limits_{(-\infty,v_{0}^{(n)} ] \, \cup \, [u_{N_{n}}^{(n)},\infty )} \pi_{X}(y) \, dy & < \epsilon_{n},  \hspace{2em} \int \limits_{(v_{0}^{(n)}, u_{N_{n}}^{(n)}) \, \setminus \, (v_{M_{n}}^{(n)}, u_{0}^{(n)})} \big(\pi_{X}(y) - \pi_{X}^{(n)}(y) \big)^2 \, dy < \epsilon_{n}, \hspace{3em} \\
& \int \limits_{v_{M_{n}}^{(n)}}^{u_{0}^{(n)}} y^2 \big( \pi_{X}(y) - \pi_{X}^{(n)}(y) \big) \, dy < \epsilon_{n} .
\label{REQuirlast}
\end{align}
\noindent In particular, Requirement (\ref{REQuirlast}) can be fulfilled since any Lévy process' intensity measure satisfies
\begin{equation}
\int \limits_{\mathbb{R}} \left( 1 \wedge y^2 \right) \Pi_{X}(dy) < \infty.
\end{equation}
\noindent We denote by $(X^{n}_{t})_{t \geq 0}$ the resulting, approximating Lévy process having jump density $\pi_{X}^{(n)}(\cdot)$, Gaussian parameter $\sigma_{n}^{2} := \sigma_{X}^{2}$ and drift $b_{n} \in \mathbb{R}$ defined by the identity $\Phi_{{X}^{n}}(1) = \Phi_{X}(1)$. Then, following the lines of argument in \cite{jp10} allows to see that the sequence of approximating processes constructed this way, $\big( (X^{n}_{t})_{t \geq 0} \big)_{n \in \mathbb{N}}$, converges weakly, in the Skorokhod topology, to the true process $(X_{t})_{t \geq 0}$ and additionally that for any $ \mathcal{T} \geq 0$, $x \in \mathbb{R}$ and $\ell \in \mathbb{R}$
\begin{equation}
\mathbb{P}_{x}^{X^{n}} \left( \tau_{\ell}^{X^{n},\pm} \leq \mathcal{T} \right) \, \rightarrow \, \mathbb{P}_{x}^{X} \left( \tau_{\ell}^{X,\pm} \leq \mathcal{T} \right), \hspace{1.5em} n \rightarrow \infty.
\end{equation}
\noindent In view of our general intra-horizon risk measurement approach, the latter convergence has an important implication. Not only can Lévy processes with completely monotone jumps be approximated by hyper-exponential jump-diffusion models, but the same approximating sequence can be also used for intra-horizon risk quantification. We will rely on this approach in Section~\ref{NUMres}, when discussing the intra-horizon risk inherent to certain infinite-activity pure jump Lévy dynamics, i.e.,~we will derive intra-horizon risk results to these Lévy models by relying on their hyper-exponential approximations $(X_t^{n})_{t \geq 0}$ with $\sigma_{n}^{2} = \sigma_{X}^{2} = 0$. However, since our main results in Section \ref{sechyper} made explicitly use of the assumption $\sigma_{X} \neq 0$, a few adaptions need to be discussed. This is the content of the next section. \vspace{1em} \\
\noindent \underline{\bf Remark 5.} \vspace{0.2em} \\
\noindent Although our general approximation scheme relies on ideas similarly employed in~\cite{amp07} and \cite{jp10}, the resulting approximating processes $\big((X^{n}_{t})_{t \geq 0} \big)_{n \in \mathbb{N}}$ substantially deviate in their structure from the ones presented in these papers. This comes from the fact that the authors in~\cite{amp07} and \cite{jp10} choose to aggregate small jumps into an additional diffusion factor by taking
\begin{equation}
\sigma_n^{2} := \sigma_{X}^{2} + \int \limits_{v_{M_{n}}^{(n)}}^{u_{0}^{(n)}} y^2 \big( \pi_{X}(y) - \pi_{X}^{(n)}(y) \big)^{+} \, dy,
\end{equation}
\noindent while we prefer to stick with the diffusion coefficient of the original process $(X_{t})_{t \geq 0}$ and rely instead on (\ref{HYPERapproxDENS1}), (\ref{HYPERapproxDENS2}), and $\sigma_{n}^{2} := \sigma_{X}^{2}$. Since the difference between the two diffusion coefficients does not exceed~$\epsilon_{n}$, choosing one or the other approximation scheme may seem equivalent. However, aggregating small jumps into an additional diffusion factor transforms in particular infinite-activity pure jump processes into approximating hyper-exponential jump-diffusion processes with non-zero diffusion. When dealing with first-passage problems, this additional diffusion factor is known to artificially imply a smooth-pasting condition at the barrier level, which subsequently leads, near the barrier, to qualitative differences in the solutions to the first-passage problem under the original process and under the approximating processes (cf.~\cite{bl09}, \cite{bl12}). As we are particularly interested in quantifying intra-horizon risk for small~$\alpha$, we only need to compute first-passage probabilities for starting values far from the barrier. Consequently, relying on the same approach used in~\cite{amp07} and \cite{jp10} may still provide reasonable results (cf.~\cite{lv20}). Nevertheless, we prefer to follow a more natural approach and keep the pure jump structure of the original process by relying on (\ref{HYPERapproxDENS1}), (\ref{HYPERapproxDENS2}), and $\sigma_{n}^{2} := \sigma_{X}^{2}$. \\
$\mbox{}$ \hspace{44.8em} \scalebox{0.75}{$\blacklozenge$} \\
\subsection{Adaptions for Pure Jump Lévy Models}
\noindent At this point, we have already emphasized that the structure of pure jump processes implies for the maturity-randomized first-passage probabilities $\mathcal{LC} \big(u_{X}^{\mathcal{E}_{0}^{\pm}}\big)(\cdot)$ and $\mathcal{LC} \big(u_{X}^{\mathcal{E}_{\mathcal{J}}^{\pm}}\big)(\cdot)$ that the continuous-pasting conditions
\begin{equation}
\mathcal{LC} \big(u_{X}^{\mathcal{E}_{0}^{\pm}}\big)(\vartheta, \ell \mp; \ell) = \mathcal{LC} \big(u_{X}^{\mathcal{E}_{0}^{\pm}}\big)(\vartheta, \ell ; \ell) \hspace{1.5em} \mbox{and} \hspace{1.5em} \mathcal{LC} \big(u_{X}^{\mathcal{E}_{\mathcal{J}}^{\pm}}\big)(\vartheta, \ell \mp; \ell) = \mathcal{LC} \big(u_{X}^{\mathcal{E}_{\mathcal{J}}^{\pm}}\big)(\vartheta, \ell ; \ell)
\end{equation}
\noindent do not anymore hold. Instead, when dealing with pure jump processes of infinite variation, these conditions need to be replaced by
\begin{equation}
\mathcal{LC} \big(u_{X}^{\mathcal{E}_{0}^{\pm}}\big)(\vartheta, \ell \mp; \ell) = \mathcal{LC} \big(u_{X}^{\mathcal{E}_{0}^{\pm}}\big)(\vartheta, \ell \pm ; \ell) \hspace{1.5em} \mbox{and} \hspace{1.5em} \mathcal{LC} \big(u_{X}^{\mathcal{E}_{\mathcal{J}}^{\pm}}\big)(\vartheta, \ell \mp; \ell) = \mathcal{LC} \big(u_{X}^{\mathcal{E}_{\mathcal{J}}^{\pm}}\big)(\vartheta, \ell \pm ; \ell).
\label{NEWcontFit}
\end{equation} 
\noindent Although (\ref{NEWcontFit}) is not anymore satisfied under hyper-exponential jump-diffusion approximations to pure jump processes of infinite variation, one may want to impose them anyway and analogous results to the ones in~Proposition~\ref{prop4} can be derived under (\ref{NEWcontFit}). Alternatively, jump contributions to (maturity-randomized) first-passage probabilities can be obtained by following the approach taken in Proposition \ref{prop5} and subsequently solving the resulting systems of equations. This leads to the following analogue of Proposition~\ref{prop6}. The reader is referred for a proof in a slightly different context to \cite{cy13}.
\begin{Prop}
\label{prop7}
\noindent Assume that $(X_{t})_{t \geq 0}$ follows a hyper-exponential jump-diffusion process, as described in (\ref{hypexp}), (\ref{hypexpDens}) and with $\sigma_{X} = 0$ and $\mu \leq 0$. Then, for any level $\ell \in \mathbb{R}$, $x \in \mathbb{R}\setminus \overline{\mathcal{H}_{\ell}^{+}}$ and intensity $ \vartheta >0$ we have that
\begin{equation}
\label{JuMpEq1Vol0}
\mathcal{LC}\big(u_{X}^{\mathcal{E}_{i}^{+}}\big)(\vartheta, x; \ell) = \sum \limits_{k=1}^{m} \tilde{v}_{\mathcal{E}_{i}^{+},k} \cdot e^{\beta_{k,\vartheta} \cdot (x- \ell)} , \hspace{1.5em} i \in \{1, \ldots, m \},
\end{equation}
\noindent where the coefficients $\tilde{v}_{\mathcal{E}_{i}^{+},k}$ are given by
\begin{equation}
\label{JumpCO1Vol0}
\tilde{v}_{\mathcal{E}_{i}^{+},k} = - \frac{\mathbf{\tilde{B}}^{+}(\beta_{k,\vartheta} ) \, \mathbf{\tilde{C}}^{+}_{\vartheta}(\xi_{i})}{\xi_{i} \, (\xi_{i} - \beta_{k,\vartheta}) \, \big(\mathbf{\tilde{B}}^{+} \big)'(\xi_{i}) \, \big(\mathbf{\tilde{C}}_{\vartheta}^{+} \big)'(\beta_{k,\vartheta})},  \hspace{1.5em} k \in \{1, \ldots, m \}, 
\end{equation}
\noindent with 
\begin{equation}
\mathbf{\tilde{B}}^{+}(x) := \prod \limits_{s=1}^{m} \big(\xi_{s} - x \big), \hspace{2em} \mathbf{\tilde{C}}_{\vartheta}^{+}(x) := \prod \limits_{s=1}^{m} \big(\beta_{s,\vartheta} - x \big) .
\label{EqB&CVol0}
\end{equation}
\noindent Similarly, if $\sigma_{X} = 0$ and $\mu \geq 0$, we obtain, for $\ell \in \mathbb{R}$, $x \in \mathbb{R}\setminus \overline{\mathcal{H}_{\ell}^{-}}$ and $ \vartheta >0$, that
\begin{equation}
\label{JuMpEq2Vol0}
\mathcal{LC}\big(u_{X}^{\mathcal{E}_{j}^{-}}\big)(\vartheta, x; \ell) = \sum \limits_{k=1}^{n} \tilde{v}_{\mathcal{E}_{j}^{-},k} \cdot e^{\gamma_{k,\vartheta} \cdot (x- \ell)} , \hspace{1.5em} j \in \{1, \ldots, n \}
\end{equation}
\noindent where the coefficients $\tilde{v}_{\mathcal{E}_{j}^{-},k}$ are given by
\begin{equation}
\label{JumpCO2Vol0}
\tilde{v}_{\mathcal{E}_{j}^{-},k} = \frac{\mathbf{\tilde{B}}^{-}(\gamma_{k,\vartheta} ) \, \mathbf{\tilde{C}}^{-}_{\vartheta}(\eta_{j})}{\eta_{j} \, (\eta_{j} + \gamma_{k,\vartheta}) \, \big(\mathbf{\tilde{B}}^{-} \big)'(-\eta_{j}) \, \big(\mathbf{\tilde{C}}_{\vartheta}^{-} \big)'(-\gamma_{k,\vartheta})}, \hspace{1.5em} k \in \{1, \ldots, n \}, 
\end{equation}
\noindent with 
\begin{equation}
\mathbf{\tilde{B}}^{-}(x) := \prod \limits_{s=1}^{n} \big(\eta_{s} + x \big) , \hspace{2em} \mathbf{\tilde{C}}_{\vartheta}^{-}(x) := \prod \limits_{s=1}^{n} \big(\gamma_{s,\vartheta} + x \big).
\label{EqB&C2Vol0}
\end{equation}
\end{Prop}
$ \mbox{ }$ \vspace{0.3em} \\
\noindent \underline{\bf Remark 6.}
\begin{itemize} \setlength \itemsep{-0.1em}
\item[i)] We note that the coefficients $v_{\mathcal{E}_{i}^{+},k}$, $\tilde{v}_{\mathcal{E}_{i}^{+},k}$ and $v_{\mathcal{E}_{j}^{-},k}$, $\tilde{v}_{\mathcal{E}_{j}^{-},k}$ in Proposition~\ref{prop6} and Proposition~\ref{prop7} depend on the intensity $\vartheta > 0$ chosen in the Laplace-Carson transform. In order to bear this dependency in mind when dealing with applications of the Gaver-Stehfest inversion algorithm in the next section, we will sometimes express these coefficients as functions of $\vartheta >0$ and write $v_{\mathcal{E}_{i}^{+},k}(\vartheta)$, $\tilde{v}_{\mathcal{E}_{i}^{+},k}(\vartheta)$, and $v_{\mathcal{E}_{j}^{-},k}(\vartheta)$, $\tilde{v}_{\mathcal{E}_{j}^{-},k}(\vartheta)$ instead.
\item[ii)] To conclude our analysis, we observe that, for $\sigma_{X} =0$ and $\mu >0$ (or $\sigma_{X}=0$ and $\mu < 0$), the same results as in Proposition \ref{prop6} hold, i.e.,~we have that, for $\ell \in \mathbb{R}$, $x \in \mathbb{R}\setminus \overline{\mathcal{H}_{\ell}^{+}}$ (or $x \in \mathbb{R}\setminus \overline{\mathcal{H}_{\ell}^{-}}$) and $ \vartheta >0$, the functions $\mathcal{LC}\big(u_{X}^{\mathcal{E}_{0}^{+}}\big)(\cdot)$ and $\mathcal{LC}\big(u_{X}^{\mathcal{E}_{i}^{+}}\big)(\cdot)$, for $i \in \{1, \ldots, m \}$, (or $\mathcal{LC}\big(u_{X}^{\mathcal{E}_{0}^{-}}\big)(\cdot)$ and $\mathcal{LC}\big(u_{X}^{\mathcal{E}_{j}^{-}}\big)(\cdot)$, for $j \in \{1, \ldots, n \}$,) are given via (\ref{RESULT1}) and  (\ref{JuMpEq1}), (\ref{JumpCO1}) (or (\ref{RESULT1down}) and (\ref{JuMpEq2}), (\ref{JumpCO2}) respectively). This follows by combining the results in Lemma \ref{CAIlem} with Proposition \ref{prop5}, since considering these two cases do not alter the number of positive (or negative) roots to the equation $\Phi_{X}(\theta) = \vartheta$ and, therefore, the system of equations (\ref{SYSbap1}) (or (\ref{SYSbap2})). Further details can be also found in \cite{cy13}.
\end{itemize}
$\mbox{}$ \hspace{44.8em} \scalebox{0.75}{$\blacklozenge$} \\
\section{Calibration and Numerical Results}
\label{NUMres}
\noindent To illustrate the practicability of the intra-horizon risk measurement approach developed in the previous sections, we lastly analyze the 10-days intra-horizon risk inherent to a long position in the S\&P~500 index as well as in the Brent crude oil over (slightly more than)~25~years. More specifically, we focus on the case where the (discounted) profit and loss process reflects the intrinsic value of a long position in either of these underlyings (cf.~\textit{Scenario~2} in Section~\ref{IHRMJ}) and derive historical intra-horizon risk results by calibrating double-exponential~(Kou), Variance-Gamma (VG), and Carr-Geman-Madan-Yor (CGMY) dynamics to S\&P~500 index and Brent crude oil data and subsequently approximating the intra-horizon risk in these models by combining the general approach of Section~\ref{IHRMJ} with the methods presented in Section~\ref{SECapproxim}.
\subsection{Data}
\noindent Our data set comprises historical returns of the S\&P~500 index and the Brent crude oil from January~1990 until September~2020, therefore spanning more than three decades. During this period, a wide variety of macroeconomic, financial, and political risk factors have influenced the performance of the US equity and commodity market. Following \cite{bp10}, we consider weekly frequency in our empirical study, which gives us in total 1,344 calibration dates starting from January~1995 (i.e., the initial five years of weekly returns are used to construct the first rolling-window sample). As discussed in \cite{lv20}, weekly returns are suboptimal in the sense that there is a mismatch between the sampling frequency and the 10-days horizon that is typically considered in risk management applications. However, using biweekly returns would halve the number of observations, which would further exacerbate estimation problems regarding the risk measures considered in this paper. Therefore, similarly to the previous studies, our decision to rely on weekly returns represents a trade-off between the quality of our estimation results and the accuracy of the sampling frequency.
\subsection{Calibration Method and Results}
\label{SECcaliRES}
Our approach closely follows \cite{bp10}. We estimate parameters of certain L\'{e}vy models on a rolling-window basis using a maximum likelihood estimation (MLE) procedure. The only difference is that we rely on the Fourier cosine method of \cite{fo08} to estimate the probability density function of the weekly index returns. This method is very fast and has been already recommended for a similar application in \cite{lv20}.\footnote{Details around the calibration are thoroughly discussed in the two cited papers, hence we do not elaborate here further on the estimation procedure.} \vspace{1em} \\
\noindent We consider the Kou (double-exponential jump-diffusion) model as well as two L\'{e}vy models that are well established in the literature and widely applied in practice: the Variance-Gamma (VG) model and the Carr-Geman-Madan-Yor (CGMY) model. In the case of the CGMY model, we set the fine structure parameter, $Y$, to $Y=0.5$. This particular choice was proposed in \cite{bp10}, mainly for two reasons. First, the resulting model has an infinite-activity-and-finite-variation property which is known to describe time series of equity returns very well. Second, having this parameter fixed allows for a better identification of the remaining model parameters, i.e., the jump arrival rate $C$, and the exponential decay parameters $G$ and $M$. For the VG model, we note that the resulting dynamics corresponds to a special case of the CGMY dynamics -- the fine structure parameter is given by $Y = 0$. \vspace{1em} \\
\noindent The next table summarizes our calibration results for the two models under consideration. Besides reporting average values and standard deviations for all model parameters, we have chosen to include the median as well as the median absolute deviation. This is important as the mean may be influenced by outliers and the non-normality of the data. \vspace{1em}\\
\noindent Several patterns can be observed over time and across the estimates for the S\&P~500 index (Bloomberg ticker: SPX). First, the jump intensity parameter -- denoted $C$ for the VG and CGMY models, and $\lambda$ for the Kou model -- spiked during the 1997 Asian Financial Crisis and the 2008-2009 Global Financial Crisis. Interestingly, the jump intensity was rather high during the 2002-2007 Bull Market, however it was not accompanied by elevated positive and/or negative jumps. Second, the expected size of positive jumps -- i.e.,~the inverse of the parameter $M$ for the VG and CGMY model, and the inverse of the parameter $\eta$ for the Kou model -- is rather stable at the level of about 1-2\% for the whole sample and tends to become smaller during the crisis periods. Third, the expected size of negative jumps -- i.e.,~the inverse of the parameter $G$ for the VG and CGMY model, and the inverse of the parameter $\theta$ for the Kou model -- was particularly high (around 3.5-4.0\%) during the 2000 Dot-Com Bubble Burst and following the 2008-2009 Global Financial~Crisis. This parameter exhibits high persistence and clustering behavior as it remains elevated long after a bear market is over. In recent years, this parameter has been increasing against the backdrop of slower global economic growth and increased political risks. Except for the 2002-2007 Bull Market, the left tail of the returns distribution was always heavier than the right tail, i.e.,~$M>G$ and $\eta>\theta$ for the respective models. Last but not least, we stress that the key driver of the differences between the reported parameter estimates is the fine structure parameter $Y$ which is fixed at $Y=0$ and $Y=0.5$ for VG and CGMY, respectively. Overall, we observe that the VG model exhibits a higher activity of small jumps and a more symmetric distribution of returns.
\begin{center}
  \renewcommand{\arraystretch}{1.2}
  \captionof{table}{\textbf{Summary statistics for the calibrated parameters.} We estimate parameters of the Kou, Variance-Gamma (VG) and, Carr-Geman-Madan-Yor (CGMY) models using weekly historical returns of the S\&P~500 index and the Brent crude oil from January~1995 until September~2020 (the total number of calibration dates is 1,344) on a five-years rolling-window basis. The table reports the average and median values, the standard deviations and the mean absolute deviations (MAD) for the estimated parameters and the negative log-likelihood (MLE) over time. The parameters $C$, $G$ and $M$ are based on unconstrained calibrations. We set the values of the parameter $Y$ to $Y=0$ and $Y=0.5$ for VG and CGMY, respectively.}
  \label{tbl:calibration_parameters}
	\scalebox{0.9}{
  \begin{tabularx}{\linewidth}{c*{13}{Y}}
    \toprule 
			\multicolumn{13}{l}{\bf Panel A: Summary Statistics for the Kou model} \\
			\midrule
     \bf{} & \multicolumn{2}{c}{$\sigma (\%)$} & \multicolumn{2}{c}{$\lambda$} & \multicolumn{2}{c}{$p$} & \multicolumn{2}{c}{$\eta$}  & \multicolumn{2}{c}{$\theta$}  & \multicolumn{2}{c}{MLE} \\[0pt]
     \cmidrule(lr){2-3} \cmidrule(lr){4-5} \cmidrule(lr){6-7} \cmidrule(lr){8-9} \cmidrule(lr){10-11} \cmidrule(lr){12-13}
      \bf{} & \tt SPX & \tt CO1 & \tt SPX & \tt CO1 & \tt SPX & \tt CO1 & \tt SPX & \tt CO1 & \tt SPX & \tt CO1 & \tt SPX & \tt CO1
\\
    \midrule
																	
    Mean    &  $6.94$ &  $16.41$ &  $127.89$ &  $264.78$ &  $0.40$ &  $0.25$ & $163.43$ &  $151.23$ &  $90.94$ &  $79.41$ &  $4.41$ &  $5.16$ \\ 
    Median  &  $6.23$ &  $19.83$ &  $103.72$ &  $169.26$ &  $0.32$ &  $0.12$ & $100.08$ &   $61.52$ &  $77.00$ &  $64.10$ &  $4.39$ &  $5.18$ \\        
    STD     &  $2.83$ &  $11.60$ &  $116.68$ &  $319.41$ &  $0.25$ &  $0.28$ & $249.24$ &  $195.83$ &  $44.13$ &  $86.44$ &  $0.23$ &  $0.15$ \\           
    MAD     &  $2.36$ &  $10.73$ &   $72.19$ &  $250.93$ &  $0.22$ &  $0.24$ & $111.29$ &  $145.74$ &  $34.37$ &  $54.47$ &  $0.20$ &  $0.12$ \\

  \end{tabularx}
                }
  \scalebox{0.9}{                
  \begin{tabularx}{\linewidth}{c*{11}{Y}}
    \toprule 
			\multicolumn{11}{l}{\bf Panel B: Summary Statistics for the VG model} \\
			\midrule
     \bf{} & \multicolumn{2}{c}{$C$} & \multicolumn{2}{c}{$G$} & \multicolumn{2}{c}{$M$} & \multicolumn{2}{c}{$Y$}  & \multicolumn{2}{c}{MLE} \\[0pt]
     \cmidrule(lr){2-3} \cmidrule(lr){4-5} \cmidrule(lr){6-7} \cmidrule(lr){8-9} \cmidrule(lr){10-11} 
      \bf{} & \tt SPX & \tt CO1 & \tt SPX & \tt CO1 & \tt SPX & \tt CO1 & \tt SPX & \tt CO1 & \tt SPX & \tt CO1
\\
    \midrule
																	
    Mean    &  $92.56$ &  $243.60$ &  $74.80$ &  $55.23$ & $112.62$ &  $96.94$ &  $0$     &  $0$     &  $4.41$ &  $5.17$ \\ 
    Median  &  $71.21$ &  $136.42$ &  $72.85$ &  $57.81$ & $105.41$ &  $68.88$ &  $0$     &  $0$     &  $4.39$ &  $5.19$ \\        
    STD     &  $47.09$ &  $190.60$ &  $23.85$ &  $18.95$ &  $46.05$ &  $72.92$ &  \bf{--} &  \bf{--} &  $0.23$ &  $0.15$ \\           
    MAD     &  $37.33$ &  $169.78$ &  $19.81$ &  $27.04$ &  $31.37$ &  $58.64$ &  \bf{--} &  \bf{--} &  $0.20$ &  $0.18$ \\
  \end{tabularx}
                }
  \scalebox{0.9}{                
  \begin{tabularx}{\linewidth}{c*{11}{Y}}
    \toprule 
			\multicolumn{11}{l}{\bf Panel C: Summary Statistics for the CGMY model} \\
			\midrule
     \bf{} & \multicolumn{2}{c}{$C$} & \multicolumn{2}{c}{$G$} & \multicolumn{2}{c}{$M$} & \multicolumn{2}{c}{$Y$}  & \multicolumn{2}{c}{MLE} \\[0pt]
     \cmidrule(lr){2-3} \cmidrule(lr){4-5} \cmidrule(lr){6-7} \cmidrule(lr){8-9} \cmidrule(lr){10-11} 
      \bf{} & \tt SPX & \tt CO1 & \tt SPX & \tt CO1 & \tt SPX & \tt CO1 & \tt SPX & \tt CO1 & \tt SPX & \tt CO1
\\
    \midrule
				    
	Mean    &   $6.59$ &  $39.69$ &  $48.56$ &  $52.88$ &  $85.00$ &  $196.87$ &  $0.5$   &  $0.5$   &  $4.41$ &  $5.16$ \\  
    Median  &   $5.23$ &  $14.77$ &  $44.84$ &  $39.95$ &  $77.05$ &   $55.36$ &  $0.5$   &  $0.5$   &  $4.39$ &  $5.18$ \\     
    STD     &   $3.32$ &  $66.25$ &  $19.03$ &  $51.26$ &  $38.32$ &  $745.51$ &  \bf{--} &  \bf{--} &  $0.23$ &  $0.15$ \\     
    MAD     &   $2.53$ &  $38.67$ &  $16.09$ &  $29.28$ &  $26.89$ &  $216.65$ &  \bf{--} &  \bf{--} &  $0.20$ &  $0.12$ \\
    \bottomrule

  \end{tabularx}
		}
\end{center}
$\mbox{}$ \vspace{1.2em} \\
\noindent In addition to the US large-cap equity index, we calibrate the models to Brent crude oil data (Bloomberg ticker: CO1). First, we note that the Brent crude oil historical returns exhibit annualized volatility of about 36 percent, which is approximately twice the volatility of the S\&P~500 index. Due to more frequent and violent negative jumps, the distribution of the Brent crude oil returns features more pronounced negative skewness and fatter tails. These stylized facts make the Brent crude oil an interesting candidate for our empirical exercise. Depending on the model, we obtain a jump intensity parameter that is two to six times higher than in the case of the equity index. The average magnitude of the positive (negative) jumps is somewhat smaller (larger) for the Brent crude oil. Moreover, these parameters were subject to pronounced swings over time. The most dramatic episodes were observed during the 2008-2009 Global Financial Crisis, then between 2014 and 2016 and in November 2018 due to the excess crude oil supply, and finally in 2020 as a result of the demand shock driven by the pandemic-induced economic slowdown. Last but not least, we remark that the MLE score is higher for the Brent crude oil than for the S\&P~500 index, which is indicative of somewhat elevated estimation and model risk. This result is not surprising given the extreme market moves in the oil market over the last three decades.

\subsection{Empirical Intra-Horizon Risk Results}

\begin{figure}
\begin{subfigure}{.5\linewidth}
\centering
\includegraphics[scale=.145]{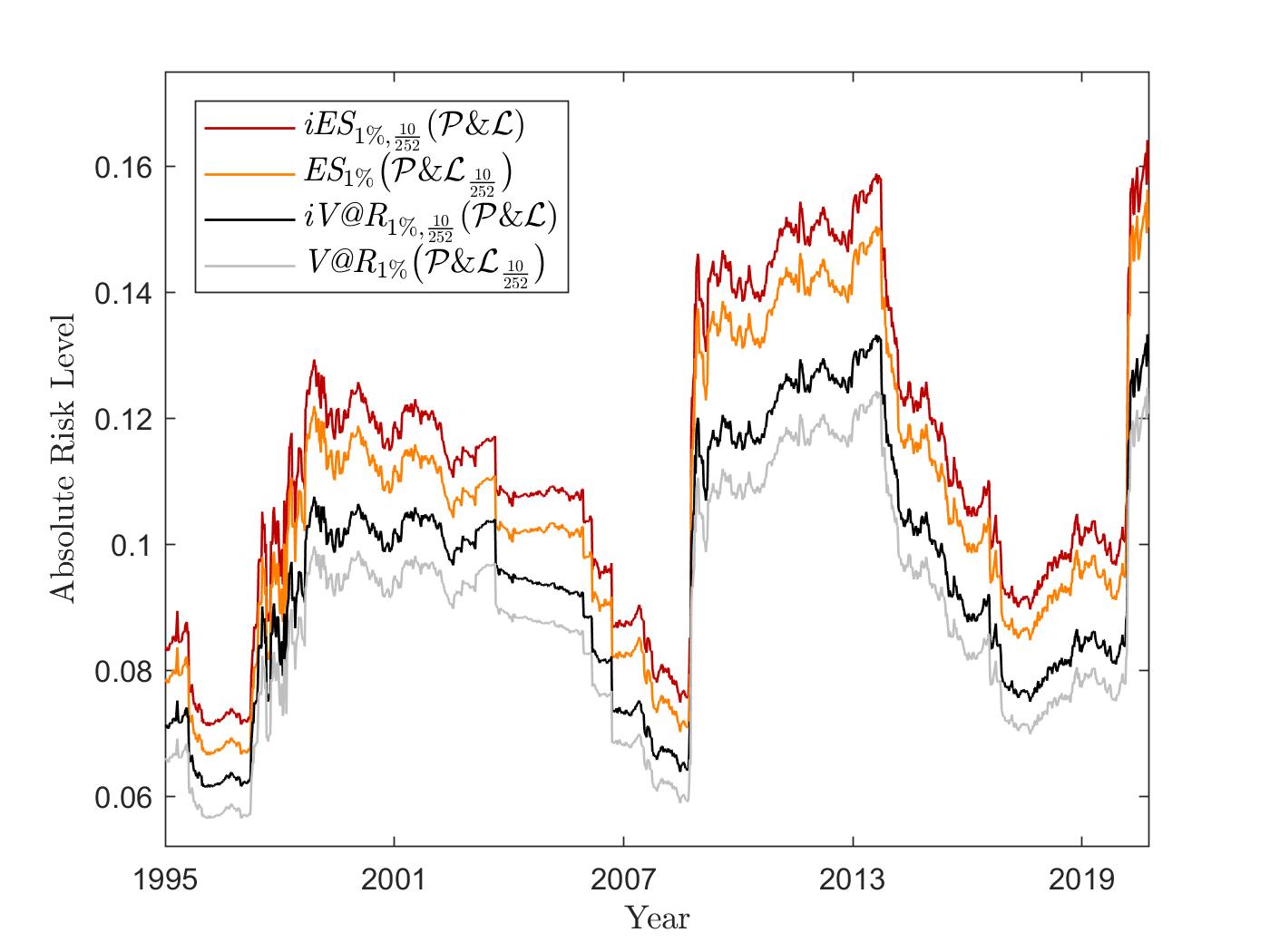}
\caption{Evolution of Absolute Risk (Kou)}
\label{fig:iVaRiESEvol:sub1}
\end{subfigure}%
\begin{subfigure}{.5\linewidth}
\centering
\includegraphics[scale=.145]{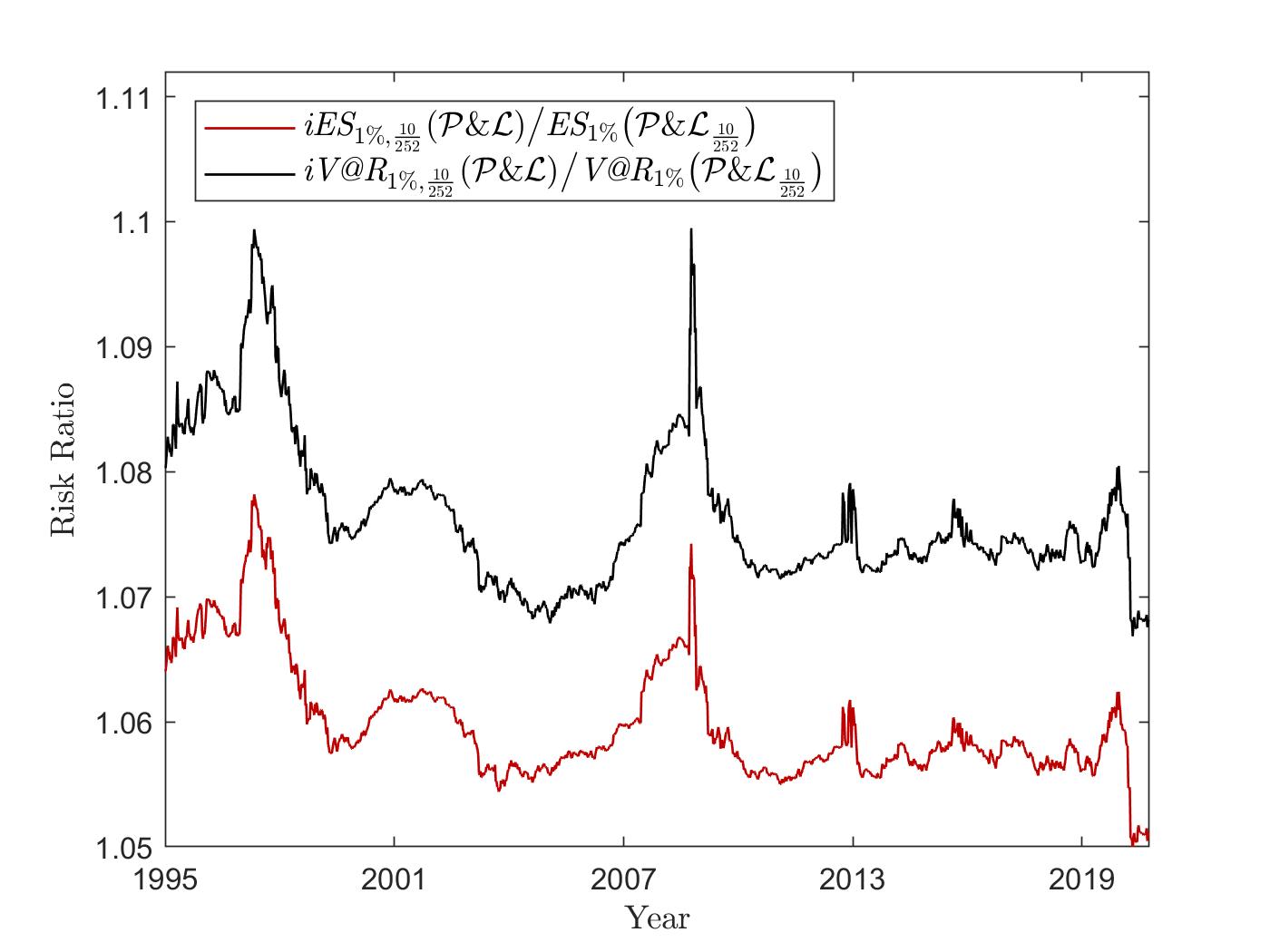}
\caption{Intra-Horizon to Point-in-Time Risk Ratio (Kou)}
\label{fig:iVaRiESEvol:sub2}
\end{subfigure}\\[1ex]
\begin{subfigure}{.5\linewidth}
\centering
\includegraphics[scale=.145]{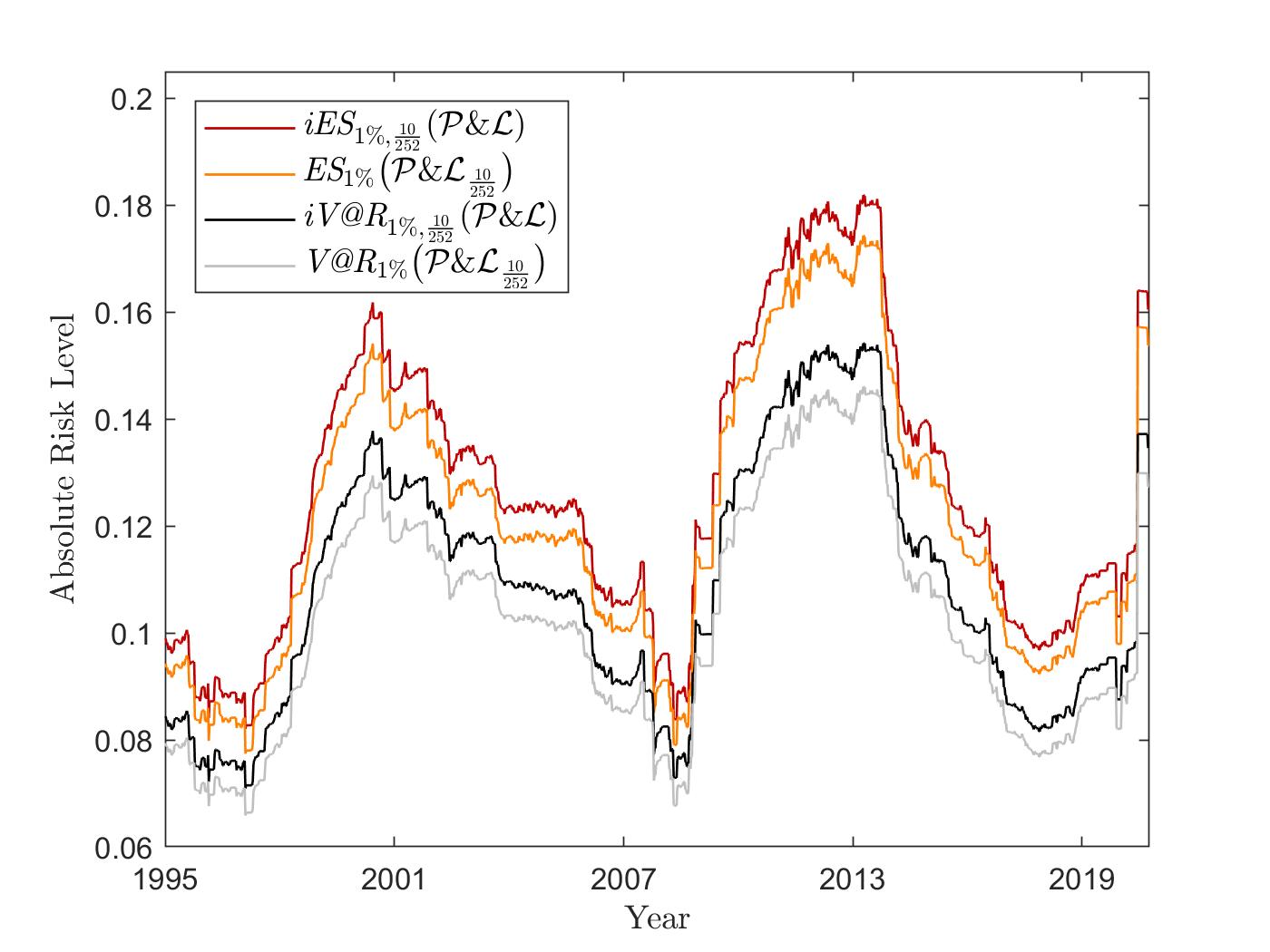}
\caption{Evolution of Absolute Risk (VG)}
\label{fig:iVaRiESEvol:sub3}
\end{subfigure}%
\begin{subfigure}{.5\linewidth}
\centering
\includegraphics[scale=.145]{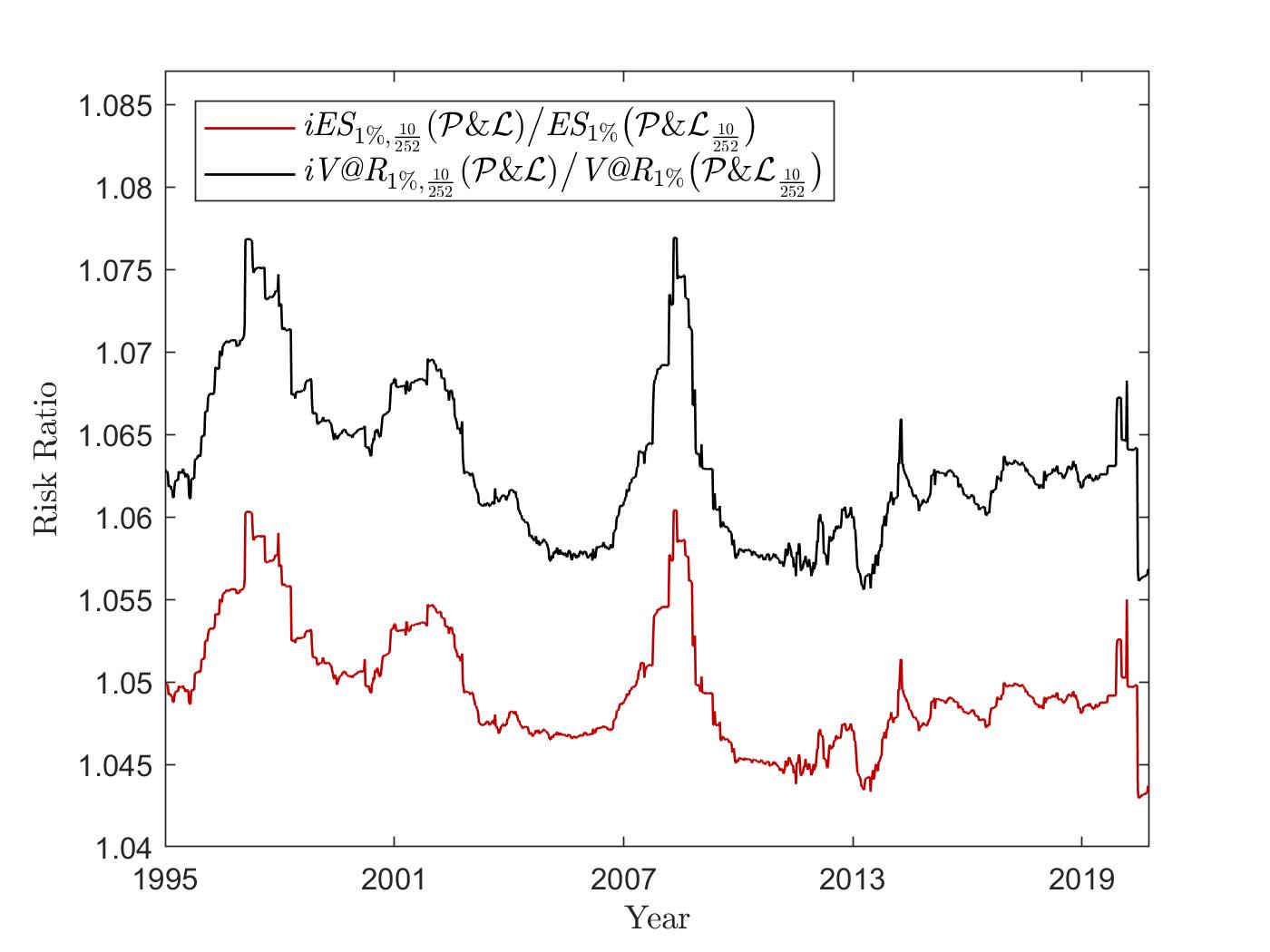}
\caption{Intra-Horizon to Point-in-Time Risk Ratio (VG)}
\label{fig:iVaRiESEvol:sub4}
\end{subfigure}\\[1ex]
\begin{subfigure}{.5\linewidth}
\centering
\includegraphics[scale=.145]{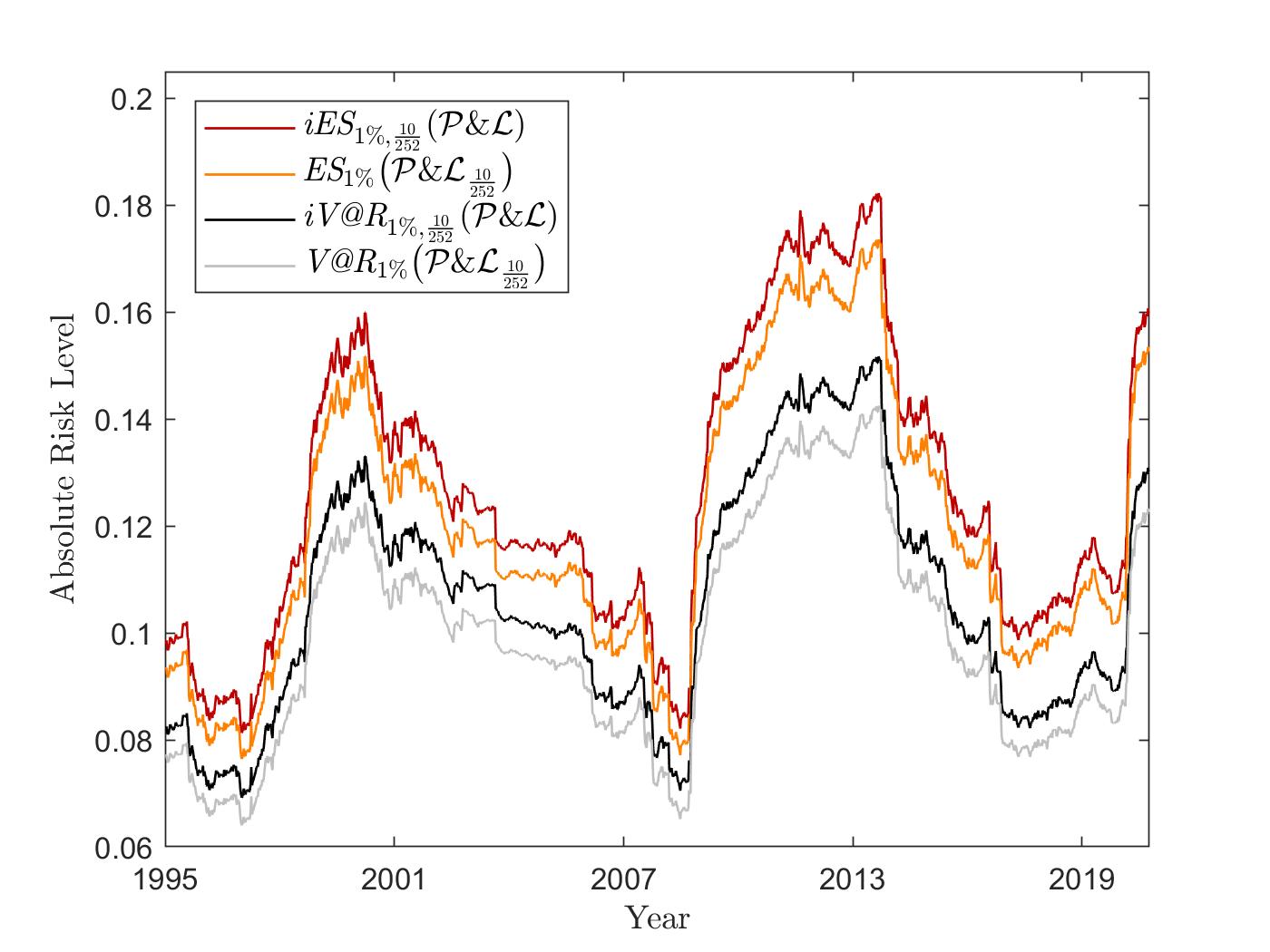}
\caption{Evolution of Absolute Risk (CGMY)}
\label{fig:iVaRiESEvol:sub5}
\end{subfigure}
\begin{subfigure}{.5\linewidth}
\centering
\includegraphics[scale=.145]{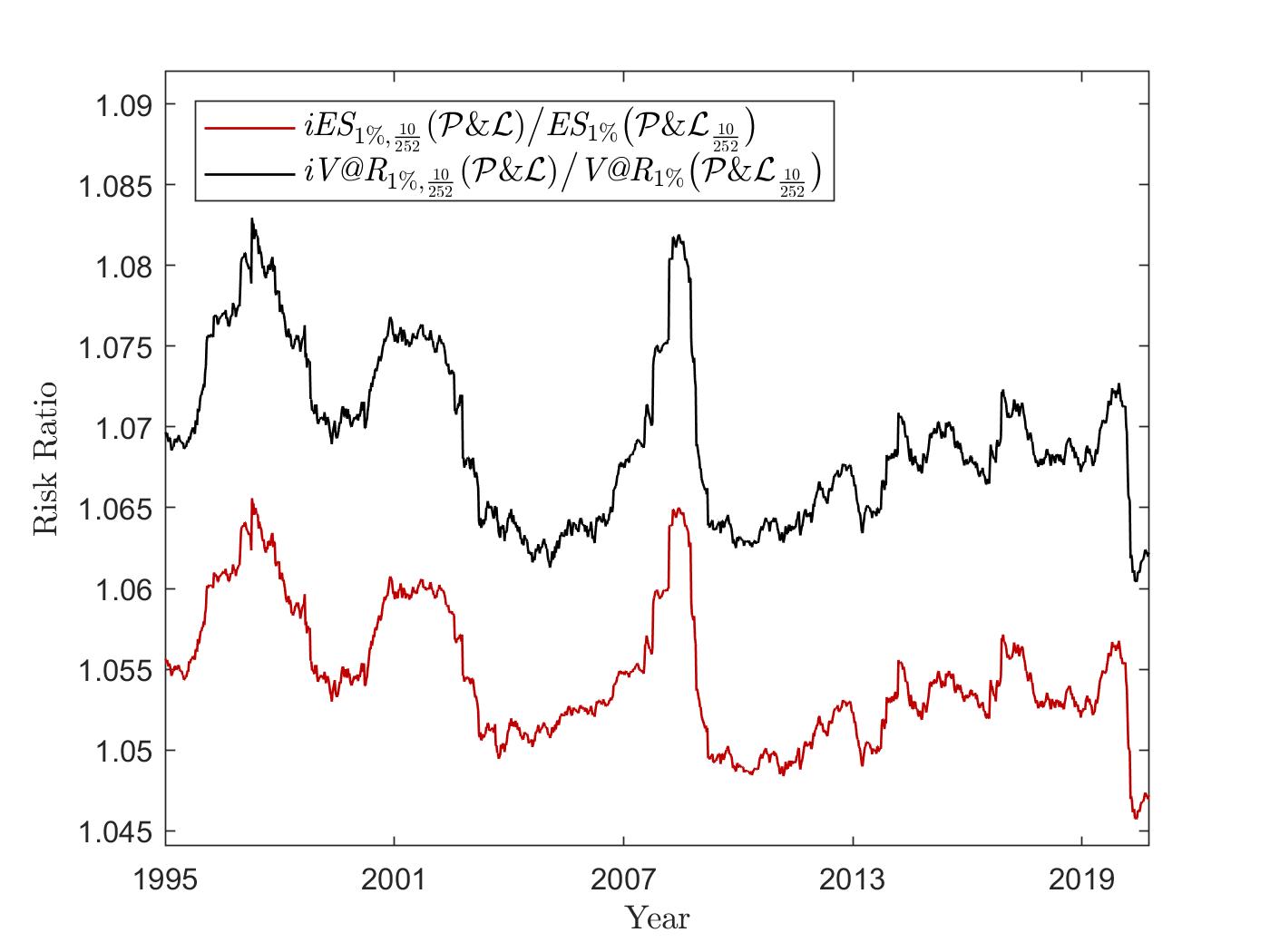}
\caption{Intra-Horizon to Point-in-Time Risk Ratio (CGMY)}
\label{fig:iVaRiESEvol:sub6}
\end{subfigure}
\caption{\textbf{Comparison of the intra-horizon and point-in-time risk for the S\&P~500 index.} Figure~\ref{fig:iVaRiESEvol:sub1}, Figure~\ref{fig:iVaRiESEvol:sub3}, and Figure~\ref{fig:iVaRiESEvol:sub5} show the time evolution of the 10-days intra-horizon risk $\big($intra-horizon value at risk, $\mbox{\it iV@R}_{1\%,\frac{10}{252}}(\mathcal{P\&L})$, and intra-horizon expected shortfall, $\mbox{\it iES}_{1\%,\frac{10}{252}}(\mathcal{P\&L})$$\big)$ and 10-days point-in-time risk $\big($(standard) value at risk, $\mbox{\it V@R}_{1\%}\big(\mathcal{P\&L}_{\frac{10}{252}}\big)$, and (standard) expected shortfall, $\mbox{\it ES}_{1\%}\big(\mathcal{P\&L}_{\frac{10}{252}}\big)$$\big)$ to the 99\% loss quantile from January 1995 until September 2020. The resulting absolute risk levels correspond to negative return levels under the respective dynamics. Additionally, Figure~\ref{fig:iVaRiESEvol:sub2}, Figure~\ref{fig:iVaRiESEvol:sub4}, and Figure~\ref{fig:iVaRiESEvol:sub6} provide the risk ratio of intra-horizon risk to point-in-time risk under the respective Lévy models.}
\label{fig:iVaRiESEvol}
\end{figure}

\begin{figure}
\begin{subfigure}{.5\linewidth}
\centering
\includegraphics[scale=.145]{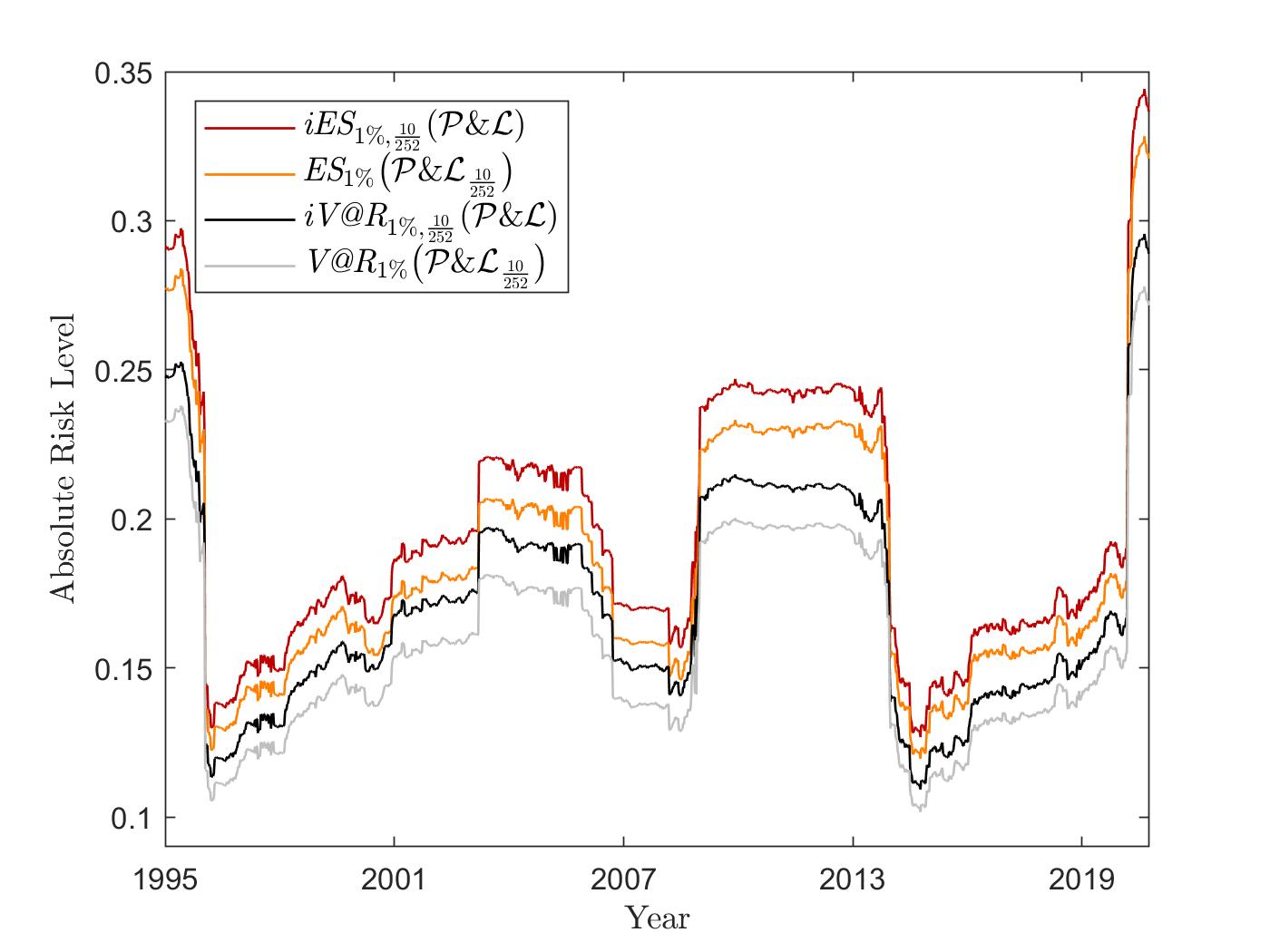}
\caption{Evolution of Absolute Risk (Kou)}
\label{fig:CO1iVaRiESEvol:sub1}
\end{subfigure}%
\begin{subfigure}{.5\linewidth}
\centering
\includegraphics[scale=.145]{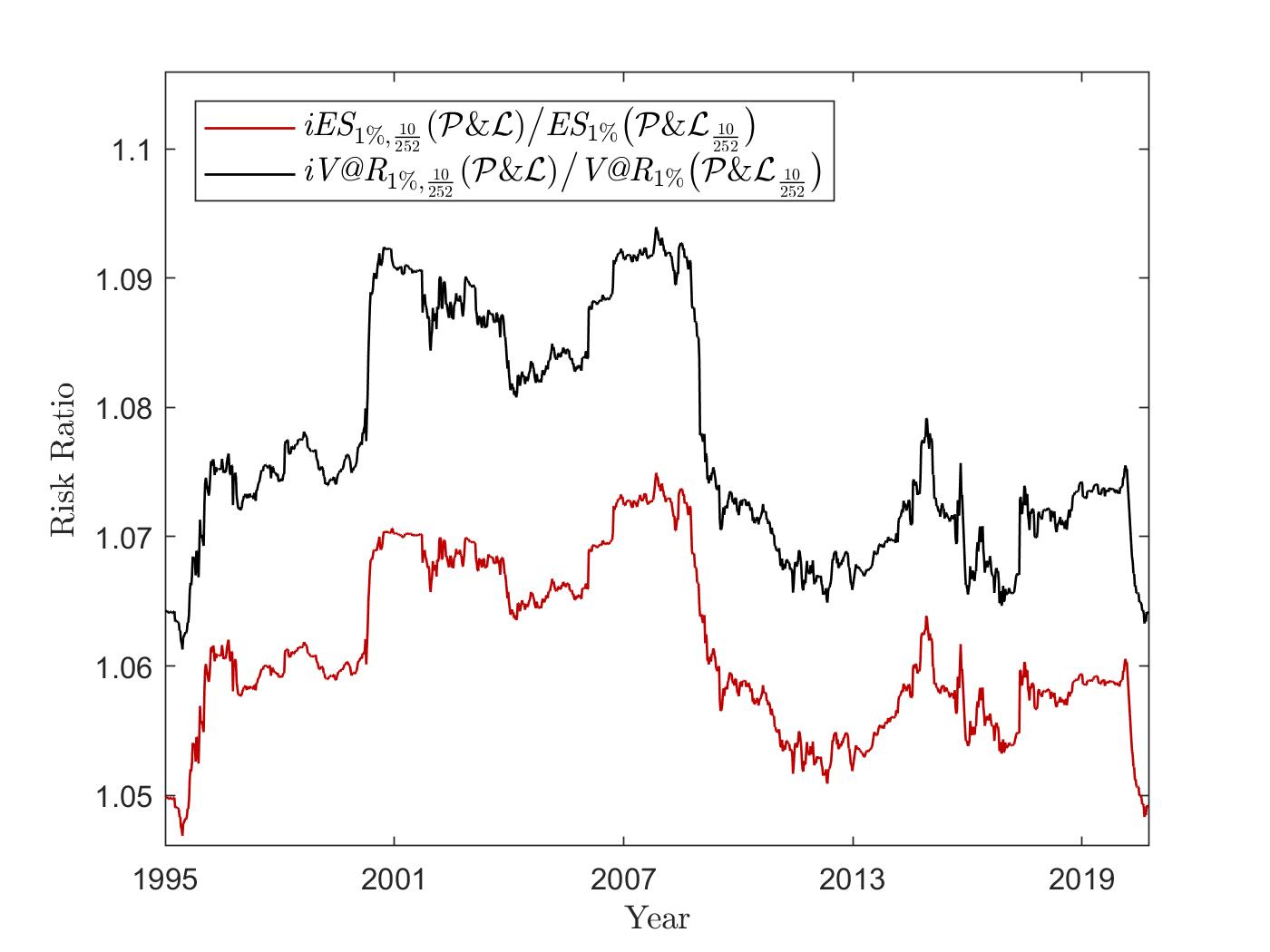}
\caption{Intra-Horizon to Point-in-Time Risk Ratio (Kou)}
\label{fig:CO1iVaRiESEvol:sub2}
\end{subfigure}\\[1ex]
\begin{subfigure}{.5\linewidth}
\centering
\includegraphics[scale=.145]{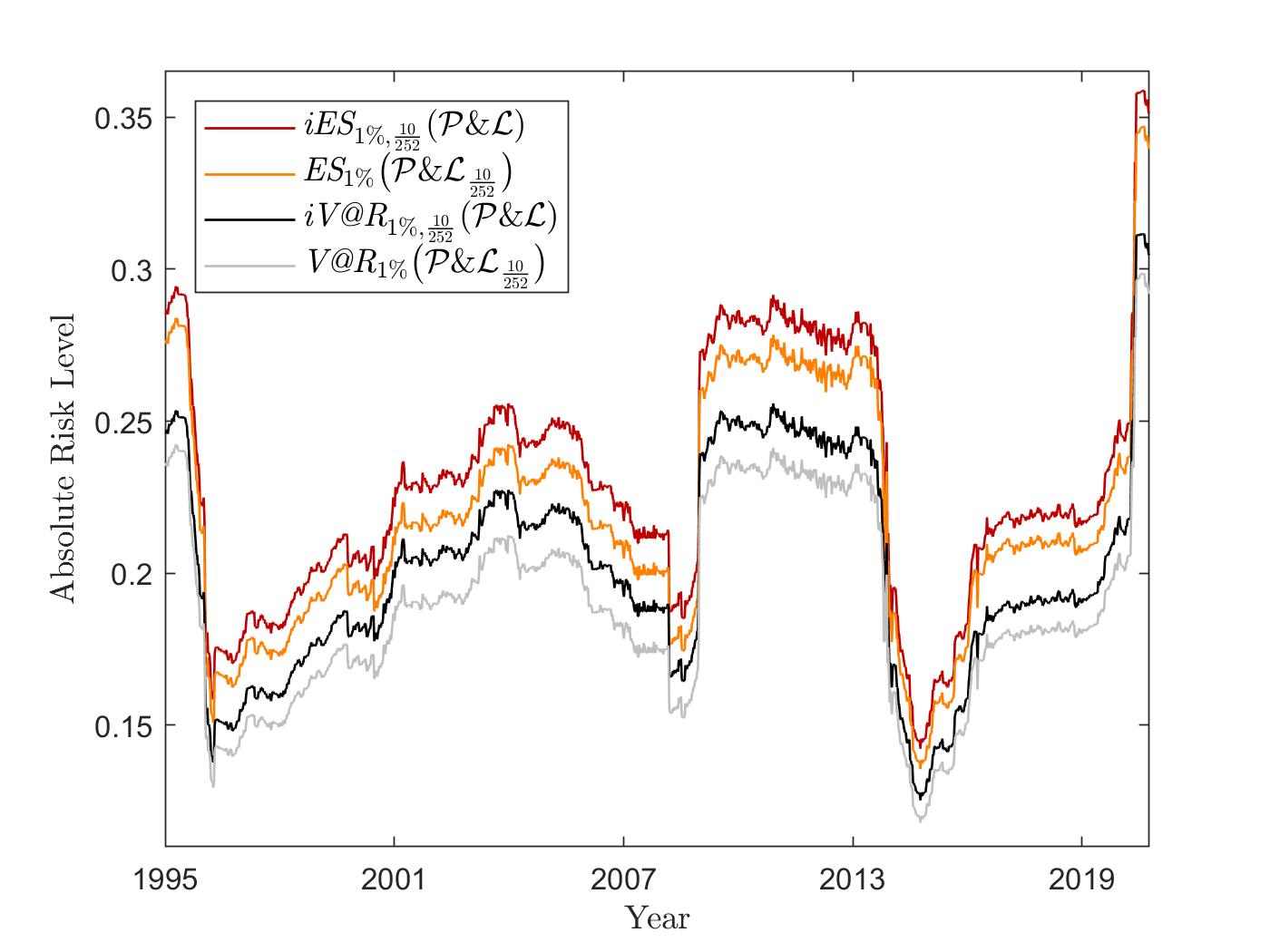}
\caption{Evolution of Absolute Risk (VG)}
\label{fig:CO1iVaRiESEvol:sub3}
\end{subfigure}%
\begin{subfigure}{.5\linewidth}
\centering
\includegraphics[scale=.145]{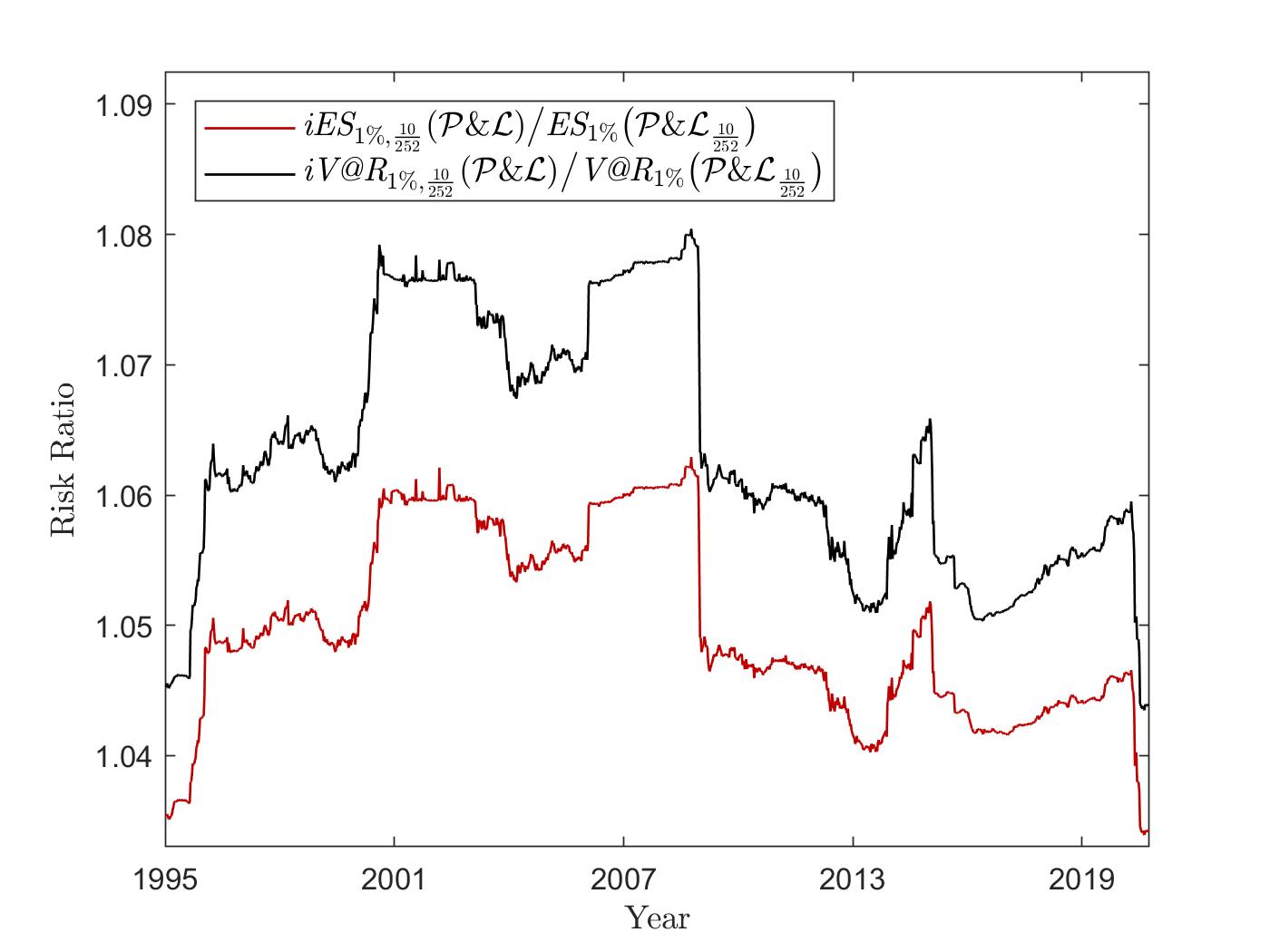}
\caption{Intra-Horizon to Point-in-Time Risk Ratio (VG)}
\label{fig:CO1iVaRiESEvol:sub4}
\end{subfigure}\\[1ex]
\begin{subfigure}{.5\linewidth}
\centering
\includegraphics[scale=.145]{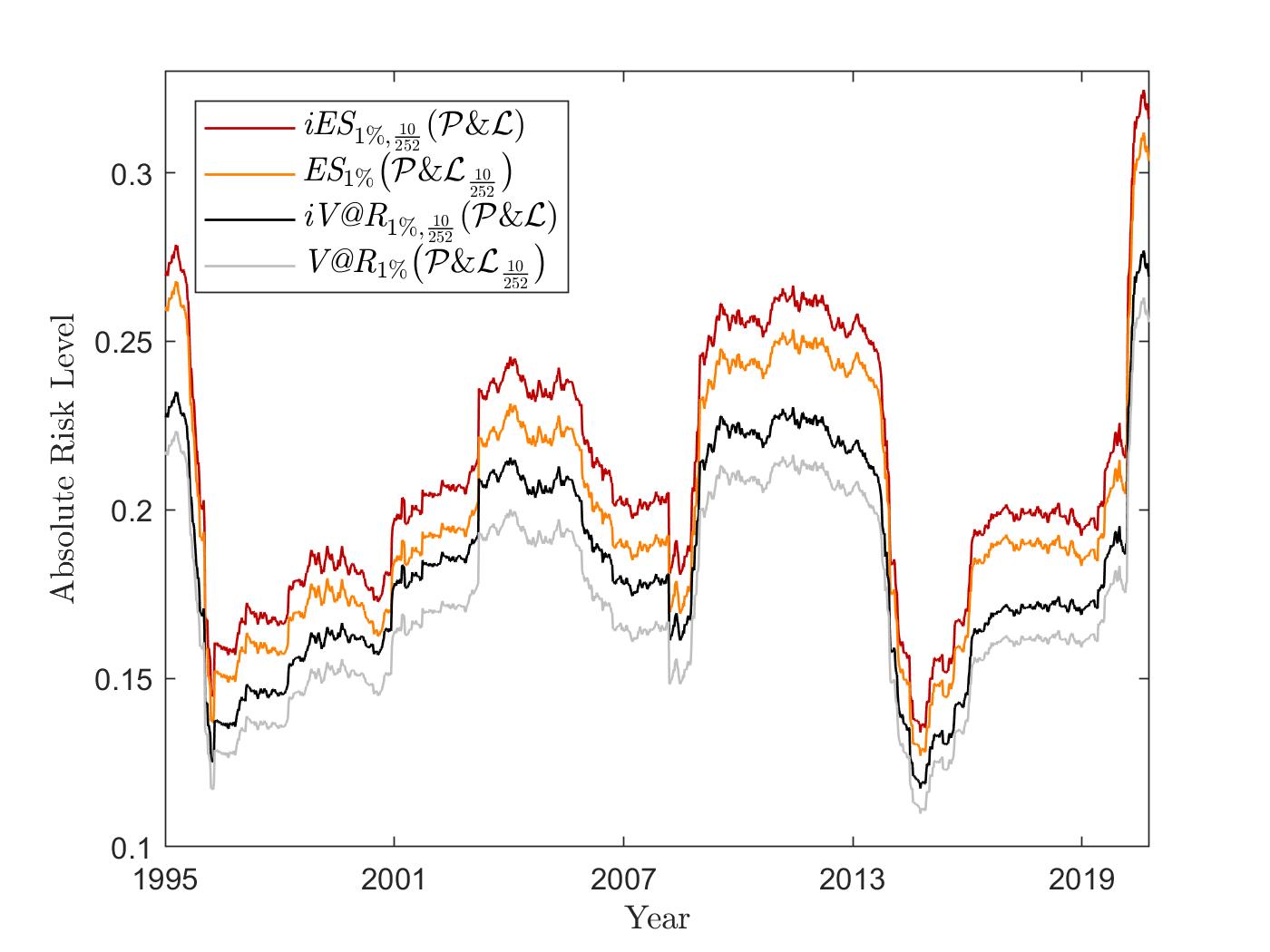}
\caption{Evolution of Absolute Risk (CGMY)}
\label{fig:CO1iVaRiESEvol:sub5}
\end{subfigure}
\begin{subfigure}{.5\linewidth}
\centering
\includegraphics[scale=.145]{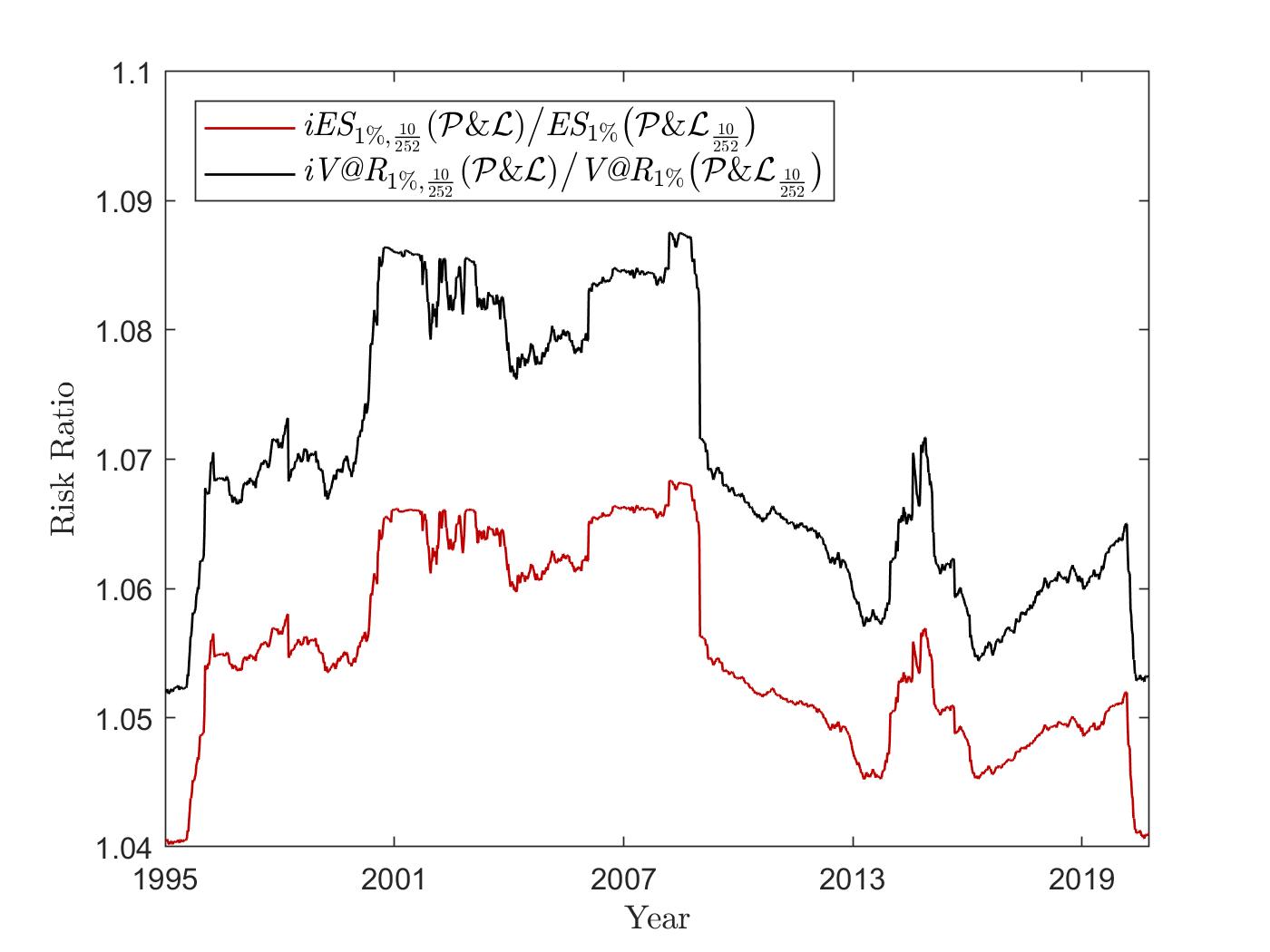}
\caption{Intra-Horizon to Point-in-Time Risk Ratio (CGMY)}
\label{fig:CO1iVaRiESEvol:sub6}
\end{subfigure}
\caption{ \textbf{Comparison of the intra-horizon and point-in-time risk for the Brent crude oil.} Figure~\ref{fig:CO1iVaRiESEvol:sub1}, Figure~\ref{fig:CO1iVaRiESEvol:sub3}, and Figure~\ref{fig:CO1iVaRiESEvol:sub5} show the time evolution of the 10-days intra-horizon risk $\big($intra-horizon value at risk, $\mbox{\it iV@R}_{1\%,\frac{10}{252}}(\mathcal{P\&L})$, and intra-horizon expected shortfall, $\mbox{\it iES}_{1\%,\frac{10}{252}}(\mathcal{P\&L})$$\big)$ and 10-days point-in-time risk $\big($(standard) value at risk, $\mbox{\it V@R}_{1\%}\big(\mathcal{P\&L}_{\frac{10}{252}}\big)$, and (standard) expected shortfall, $\mbox{\it ES}_{1\%}\big(\mathcal{P\&L}_{\frac{10}{252}}\big)$$\big)$ to the 99\% loss quantile from January 1995 until September 2020. The resulting absolute risk levels correspond to negative return levels under the respective dynamics. Additionally, Figure~\ref{fig:CO1iVaRiESEvol:sub2}, Figure~\ref{fig:CO1iVaRiESEvol:sub4}, and Figure~\ref{fig:CO1iVaRiESEvol:sub6} provide the risk ratio of intra-horizon risk to point-in-time risk under the respective Lévy models.}
\label{fig:CO1iVaRiESEvol}
\end{figure}

\noindent Having calibrated the Kou, VG, and CGMY dynamics to S\&P~500 index and to Brent crude oil data, both ranging from January 1990 to September 2020, we next turn to a derivation of the 10-days intra-horizon risk in these models. While the Kou model allows for a direct application of the results derived in Section~\ref{IHRMJ}, we compute intra-horizon risk results for VG and CGMY dynamics based on hyper-exponential jump-diffusion approximations and the combination of the theory developed in Section~\ref{IHRMJ}, Proposition~\ref{prop7}, and the Gaver-Stehfest inversion algorithm. More specifically, we follow the ideas of \cite{amp07} (cf.~also \cite{lv20}), i.e.,~we fix in advance the number of exponentials $N_{n}$ and $M_{n}$ in the approximating density (\ref{HYPERapproxDENS1}), (\ref{HYPERapproxDENS2}) and minimize the distance between the approximating and the true Lévy densities by optimally choosing the partition of the integration intervals. This slightly differs from the approach used in \cite{jp10}, where the authors additionally fix all mean jump sizes $\left(\xi_{i}\right)^{-1}$ and $\left(\eta_{j} \right)^{-1}$ and subsequently use least-squares optimization to determine the values of the remaining (mixing) parameters. However, while these authors only work with few exponentials,\footnote{The authors in \cite{jp10} only use a total of 14 exponentials, i.e.,~7 exponentials for both the positive and the negative parts of the distribution.}~we choose $N_{n} = 100$ and $M_{n} = 100$ and incorporate this way $200$ exponentials. This additionally ensures that we approximate small jumps sufficiently well, as we have decided to keep the pure jump structure of the approximated processes by refraining from converting small jumps into an extra diffusion factor (cf.~Remark 5 in Section \ref{SECapproxim}).\footnote{Our numerical tests show that the suggested procedure is fast and stable. However, we emphasize that other approaches exist in the literature (cf.~for instance~\cite{cl10}) and that investigating the performance of all these algorithms is not the sake of this article, but constitutes a separate research topic.} \vspace{1em} \\
\noindent As soon as hyper-exponential jump-diffusion processes are fixed, we make use of the derivations in the previous sections to derive intra-horizon risk results in the following way. First, 10-days intra-horizon value-at-risk measures as well as corresponding risk contributions per jump type are obtained by combining Proposition~\ref{prop7} with Relations (\ref{Repre}), (\ref{ImpOO7eq}) and (\ref{CONVer}), i.e.,~we invert the functions $\mathcal{LC}\big(u_{X}^{\mathcal{E}_{j}^{-}}\big)(\cdot)$, for any $j \in \{1, \ldots, n \}$, via
\begin{align}
u_{X}^{\mathcal{E}_{j}^{-}} ( T,x; \ell) & = \lim \limits_{N \rightarrow \infty} \Big( u_{X}^{\mathcal{E}_{j}^{-}} \Big)_{N} ( T,x; \ell), \label{AsSuMpTiOn}\\
\Big( u_{X}^{\mathcal{E}_{j}^{-}} \Big)_{N} ( T,x; \ell) & := \sum \limits_{k = 1}^{2N} \, \zeta_{k,N} \, \mathcal{LC}\big(u_{X}^{\mathcal{E}_{j}^{-}}\big)\Big(\frac{k \log(2)}{T}, x ; \ell \Big)  \nonumber \\
& = \sum \limits_{k=1}^{2N} \sum \limits_{s=1}^{n} \, \zeta_{k,N} \, \tilde{v}_{\mathcal{E}_{j}^{-},s} \Big( \frac{k \log(2)}{T} \Big) \, \exp \left\{ \gamma_{s,\frac{k \log(2)}{T}} \cdot (x - \ell) \right \} ,
\label{RARA}
\end{align}
\noindent with $\zeta_{k,N}$ defined as in (\ref{zetaEQUA}), and derive the respective results based on Relations (\ref{Repre}), (\ref{ImpOO7eq}) and the ideas introduced in Section \ref{IHRRC}. Once these quantities are obtained, recovering 10-days intra-horizon expected shortfall results reduces to the evaluation of integrals and of fractions of integrals of the form of (\ref{IntPART}) and (\ref{RISKc+}). Here, combining (\ref{AsSuMpTiOn}), (\ref{RARA}) with the monotonicity of the function $\ell \mapsto u_{X}^{\mathcal{E}_{j}^{-}} ( T,x; \ell)$ allows us to derive, for each $j \in \{1, \ldots, n \}$, that
\begin{align}
\int \limits_{-z_{2}}^{-iV@R_{\alpha,T}(\mathcal{P\&L})} u_{X}^{\mathcal{E}_{j}^{-}} \big(T, \log(z_{2}+z); \log(z_{2}+\ell) \big) \, d\ell & = \lim \limits_{N \rightarrow \infty} \Big( \mathcal{I}_{\alpha,T}^{\mathcal{E}_{j}^{-}} \Big)_{N} (z ,z_{2}),
\end{align} 
\noindent where $\Big( \mathcal{I}_{\alpha,T}^{\mathcal{E}_{j}^{-}} \Big)_{N} (z ,z_{2})$ is given, for any $N \in \mathbb{N}$, via
\begin{align}
\Big(\mathcal{I}_{\alpha,T}^{\mathcal{E}_{j}^{-}} \Big)_{N} (z,z_{2}) & := \int \limits_{-z_{2}}^{-iV@R_{\alpha,T}(\mathcal{P\&L})} \Big(u_{X}^{\mathcal{E}_{j}^{-}}\Big)_{N} \big(T, \log(z_{2}+z); \log(z_{2}+\ell) \big) \, d \ell \nonumber \\
& = \sum \limits_{k=1}^{2N} \sum \limits_{s=1}^{n} \, \zeta_{k,N} \, \tilde{v}_{\mathcal{E}_{j}^{-},s} \Big( \frac{k \log(2)}{T} \Big) \, \int \limits_{-z_{2}}^{-iV@R_{\alpha,T}(\mathcal{P\&L})} \left( \frac{z_{2} + z}{z_{2} + \ell} \right)^{\gamma_{s,\frac{k \log(2)}{T}}} \, d\ell \nonumber \\
& = \sum \limits_{k=1}^{2N} \sum \limits_{s=1}^{n} \, \zeta_{k,N} \, \tilde{v}_{\mathcal{E}_{j}^{-},s} \Big( \frac{k \log(2)}{T} \Big) \, \frac{ z_{2} + z }{1-\gamma_{s,\frac{k \log(2)}{T}}} \,  \left(\frac{z_{2}-{iV@R}_{\alpha,T}(\mathcal{P\&L})}{z_{2} + z} \right)^{1-\gamma_{s,\frac{k \log(2)}{T}}}.  
\end{align}
\noindent This finally provides us with a simple numerical scheme to compute 10-days intra-horizon expected shortfall measures as well as, based on the ideas outlined in Section \ref{IHRRC}, corresponding risk contributions per jump type inherent to any long position in either the S\&P~500 index or the Brent crude oil.

\begin{figure}[h!]
\begin{subfigure}{.5\linewidth}
\centering
\includegraphics[scale=.16]{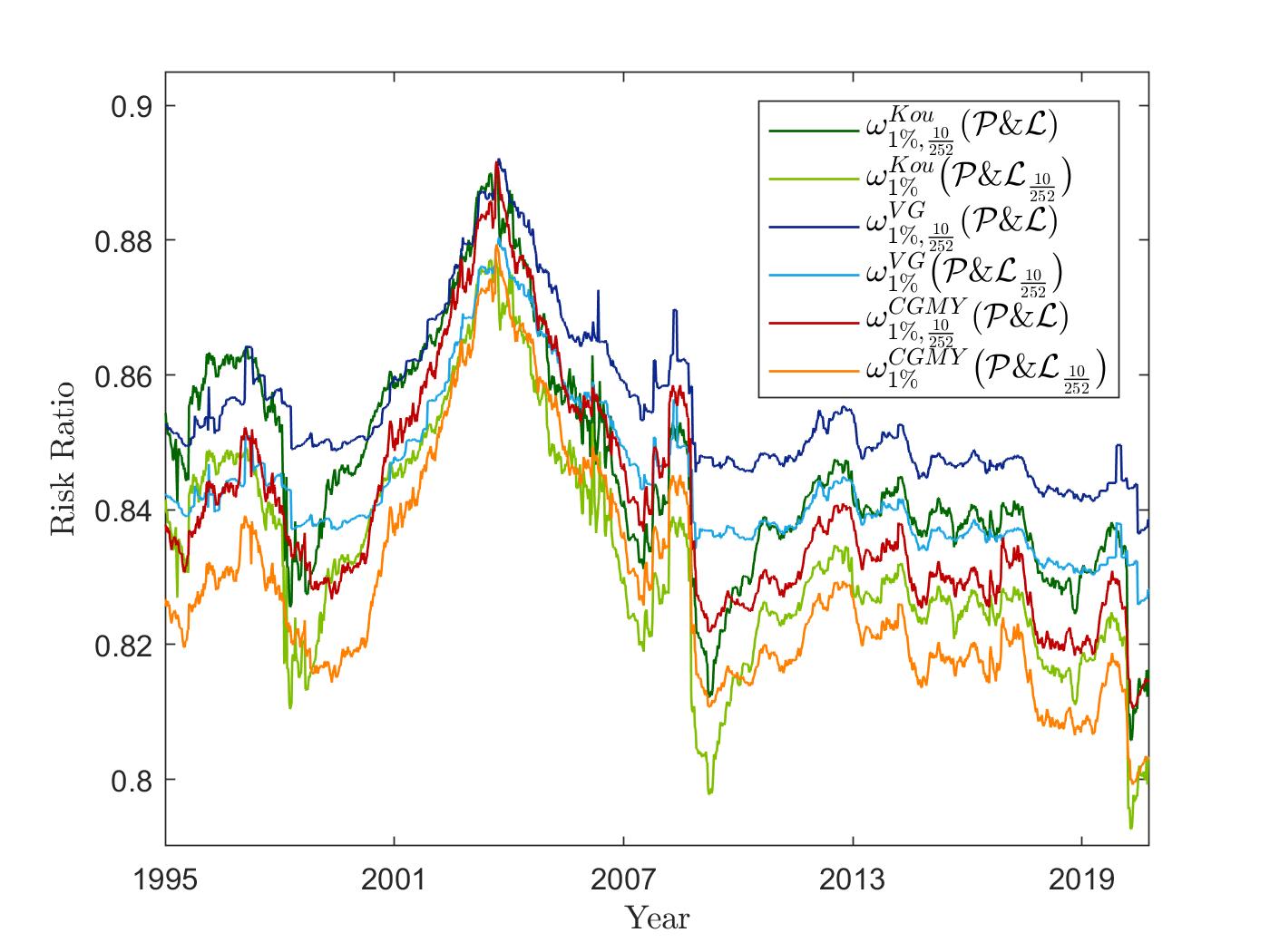}
\caption{Ratios for the S\&P~500 index}
\label{fig:JumpPlot:sub3}
\end{subfigure}
\begin{subfigure}{.5\linewidth}
\centering
\includegraphics[scale=.16]{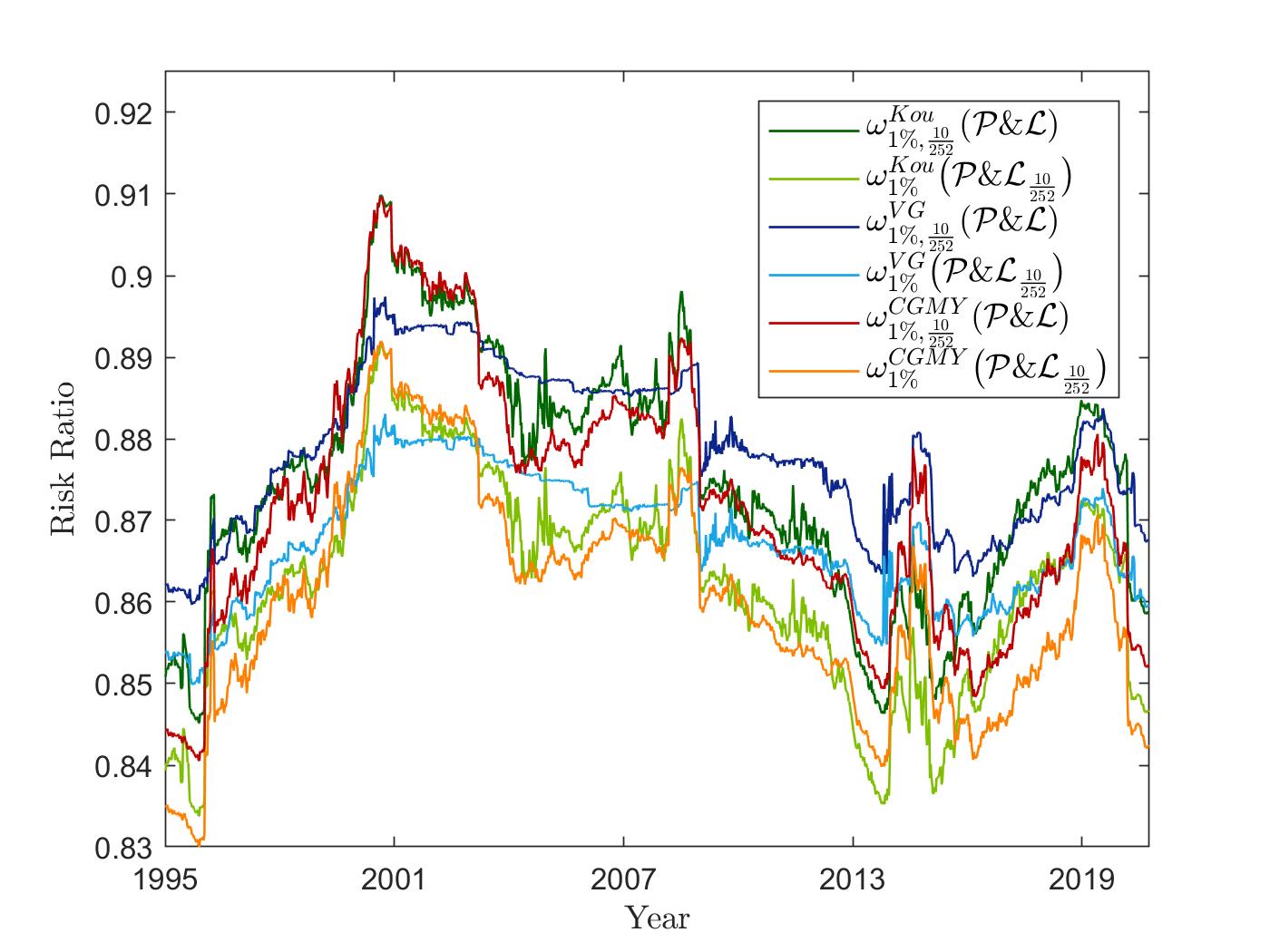}
\caption{Ratios for the Brent crude oil}
\label{fig:JumpPlot:sub4}
\end{subfigure}
\caption{\textbf{Time evolution of the intra-horizon/point-in-time value at risk to expected shortfall ratios.} We have plotted for Kou, VG, and CGMY the time evolution of the intra-horizon and point-in-time risk ratios $\omega_{1\%,\frac{10}{252}}(\mathcal{P\&L}) := \mbox{\it iV@R}_{1\%,\frac{10}{252}}(\mathcal{P\&L}) \big/ \mbox{\it iES}_{1\%,\frac{10}{252}}(\mathcal{P\&L})$ and $\omega_{1\%}\big(\mathcal{P\&L}_{\frac{10}{252}}\big) := \mbox{\it V@R}_{1\%}\big(\mathcal{P\&L}_{\frac{10}{252}}\big) \big/ \mbox{\it ES}_{1\%}\big(\mathcal{P\&L}_{\frac{10}{252}}\big)$, respectively.} These ratios give the relative contribution of the 10-days intra-horizon/point-in-time value at risk to the 10-days intra-horizon/point-in-time expected shortfall to the $99\%$ quantile of the loss distribution under the respective Lévy dynamics.
\label{fig:iVaRRATIO}
\end{figure}

\begin{figure}
\begin{subfigure}{.5\linewidth}
\centering
\includegraphics[scale=.16]{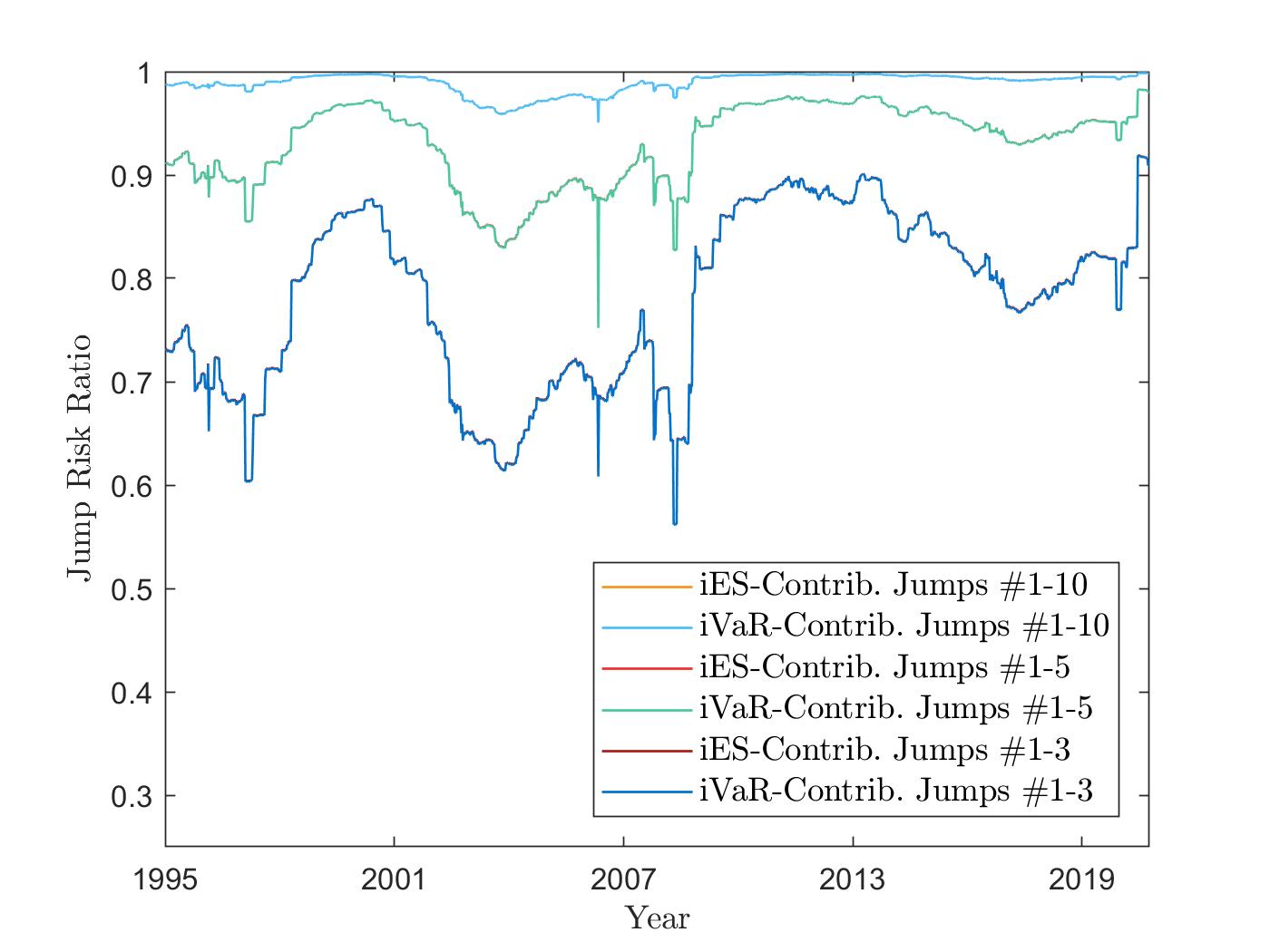}
\caption{Evolution of Jump Contributions (VG)}
\label{fig:JumpPlot:sub1}
\end{subfigure}%
\begin{subfigure}{.5\linewidth}
\centering
\includegraphics[scale=.16]{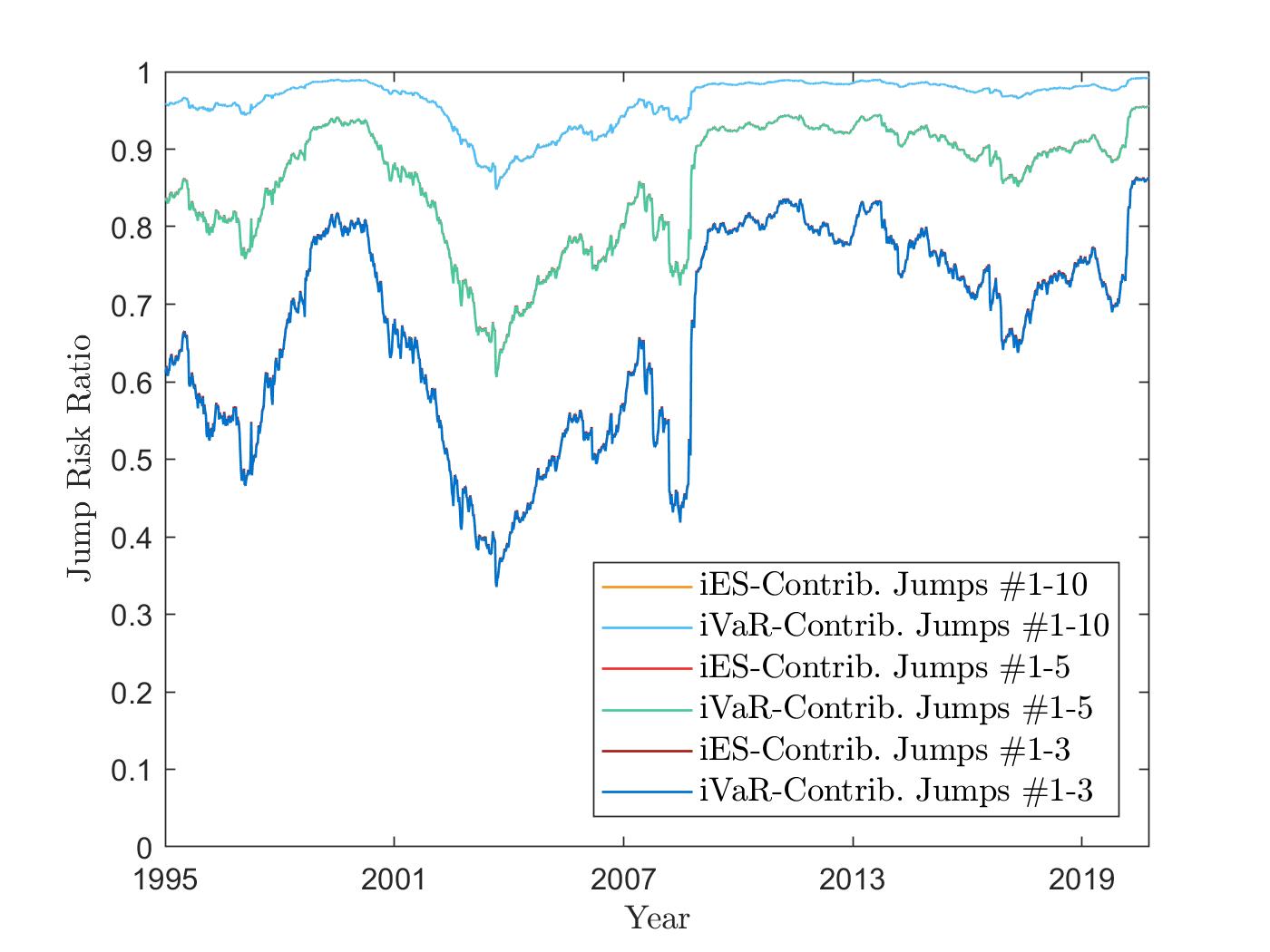}
\caption{Evolution of Jump Contributions (CGMY)}
\label{fig:JumpPlot:sub2}
\end{subfigure}\\[1ex]
\begin{subfigure}{.5\linewidth}
\centering
\includegraphics[scale=.16]{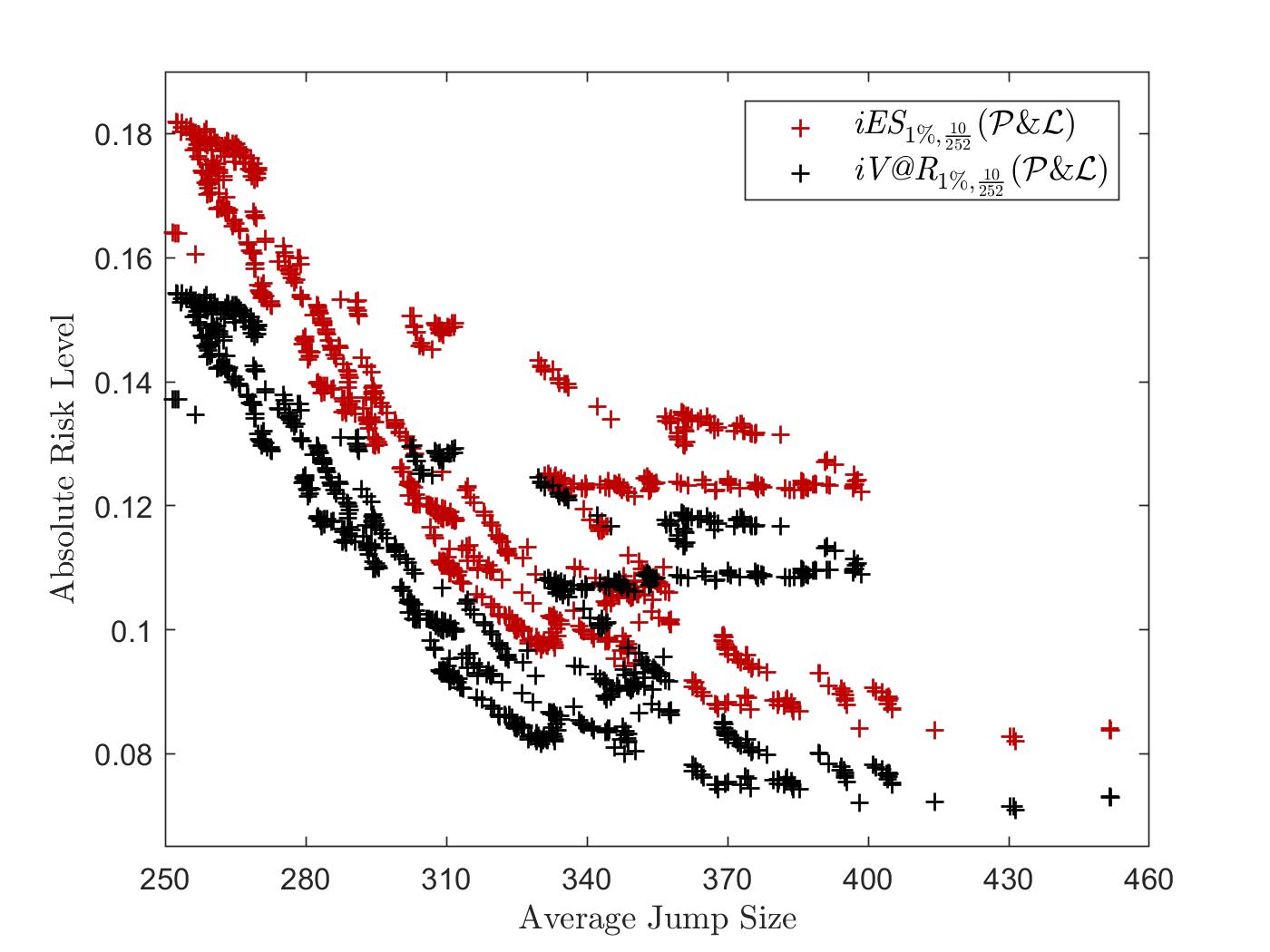}
\caption{Average Jump Size and Absolute Risk (VG)}
\label{fig:JumpPlot:sub3}
\end{subfigure}
\begin{subfigure}{.5\linewidth}
\centering
\includegraphics[scale=.16]{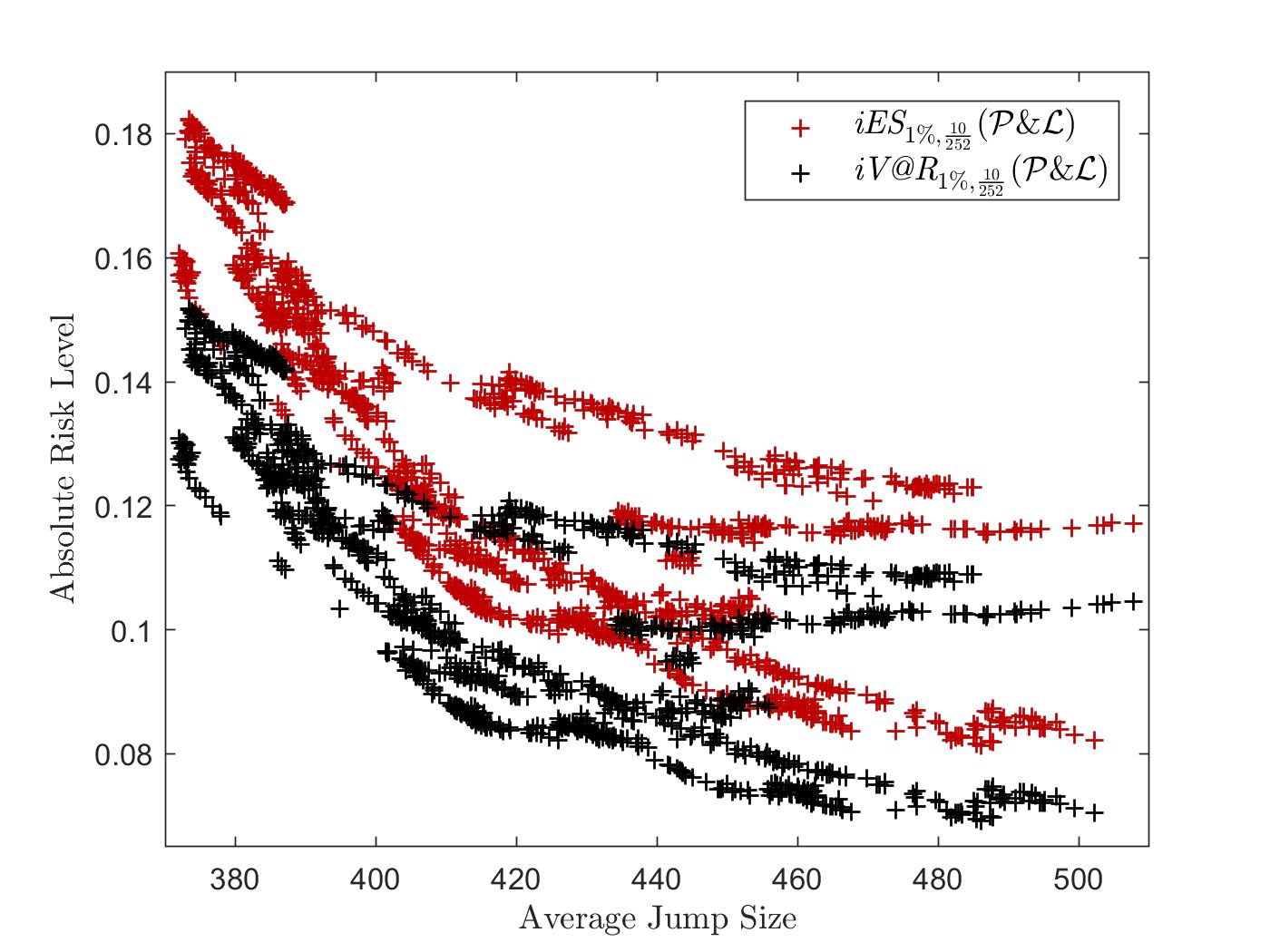}
\caption{Average Jump Size and Absolute Risk (CGMY)}
\label{fig:JumpPlot:sub4}
\end{subfigure}
\caption{\textbf{Comparison of the intra-horizon risk and the contribution of certain jumps for the S\&P~500 index.} Figure~\ref{fig:JumpPlot:sub1} and Figure~\ref{fig:JumpPlot:sub2} show the time evolution of the 10-days intra-horizon risk contributions to the 99\% loss quantile for the greatest -- in absolute size -- 3~down jumps, the greatest 5~down jumps, and the greatest 10~down jumps in the hyper-exponential jump-diffusion approximations. Additionally, Figure~\ref{fig:JumpPlot:sub3} and Figure~\ref{fig:JumpPlot:sub4} present the relation of the (absolute) average (down) jump size -- weighted by the probability of occurrence of each jump in the hyper-exponential jump-diffusion approximations -- to the absolute intra-horizon risk level. As earlier, the absolute risk levels correspond to negative return levels under the respective dynamics.}
\label{fig:JumpPlot}
\end{figure}

\begin{figure}
\begin{subfigure}{.5\linewidth}
\centering
\includegraphics[scale=.16]{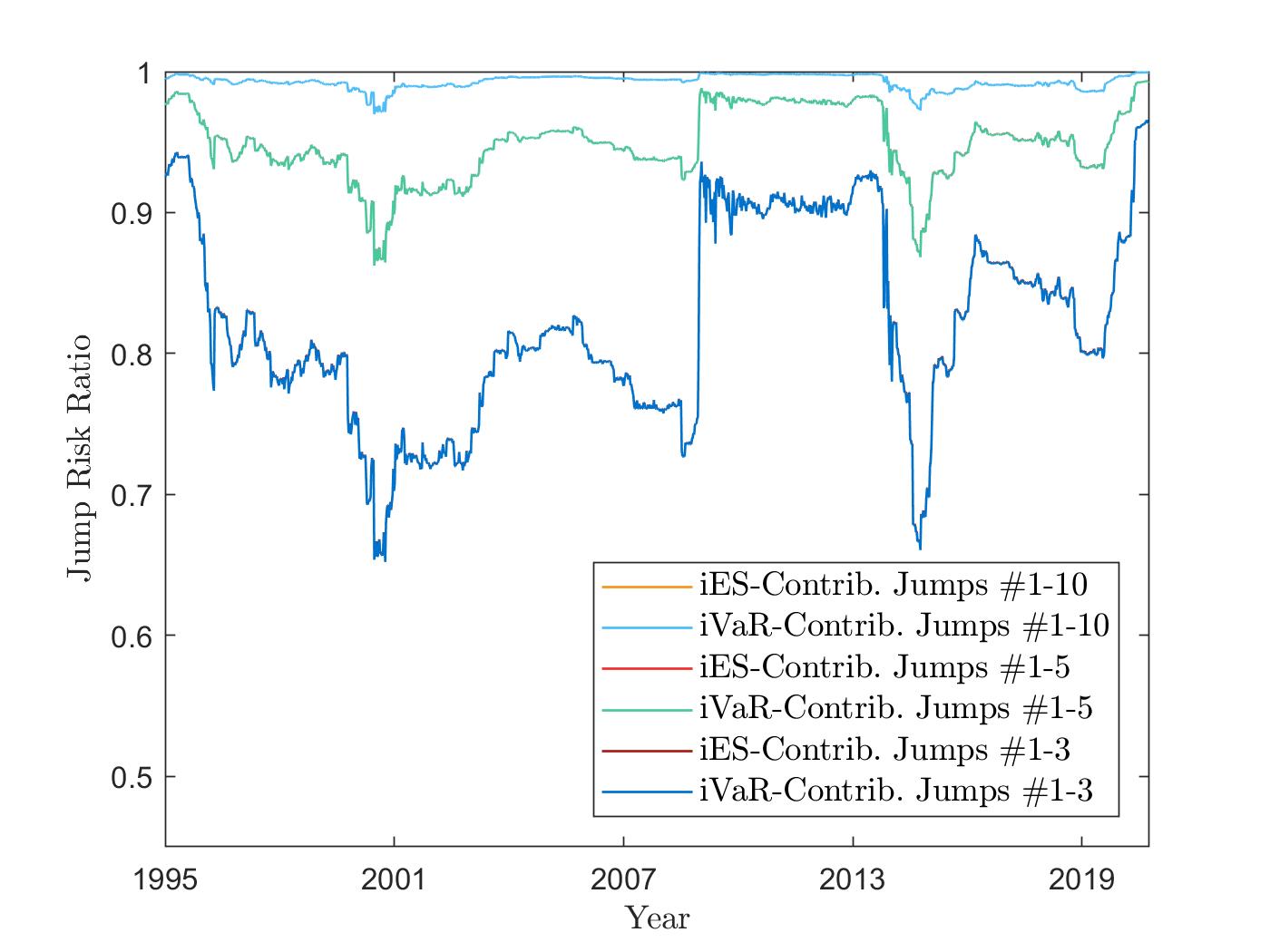}
\caption{Evolution of Jump Contributions (VG)}
\label{fig:CO1JumpPlot:sub1}
\end{subfigure}%
\begin{subfigure}{.5\linewidth}
\centering
\includegraphics[scale=.16]{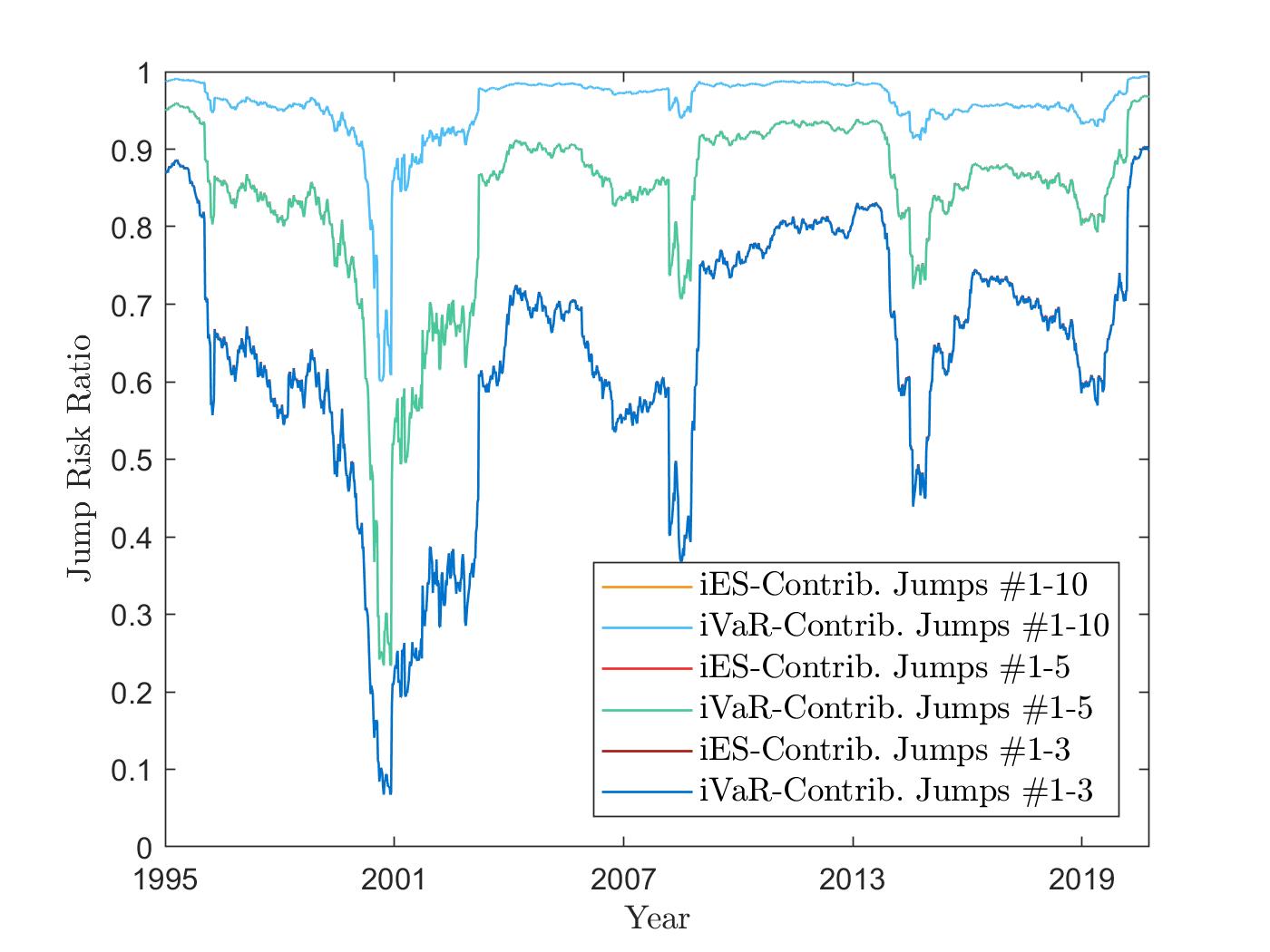}
\caption{Evolution of Jump Contributions (CGMY)}
\label{fig:CO1JumpPlot:sub2}
\end{subfigure}\\[1ex]
\begin{subfigure}{.5\linewidth}
\centering
\includegraphics[scale=.16]{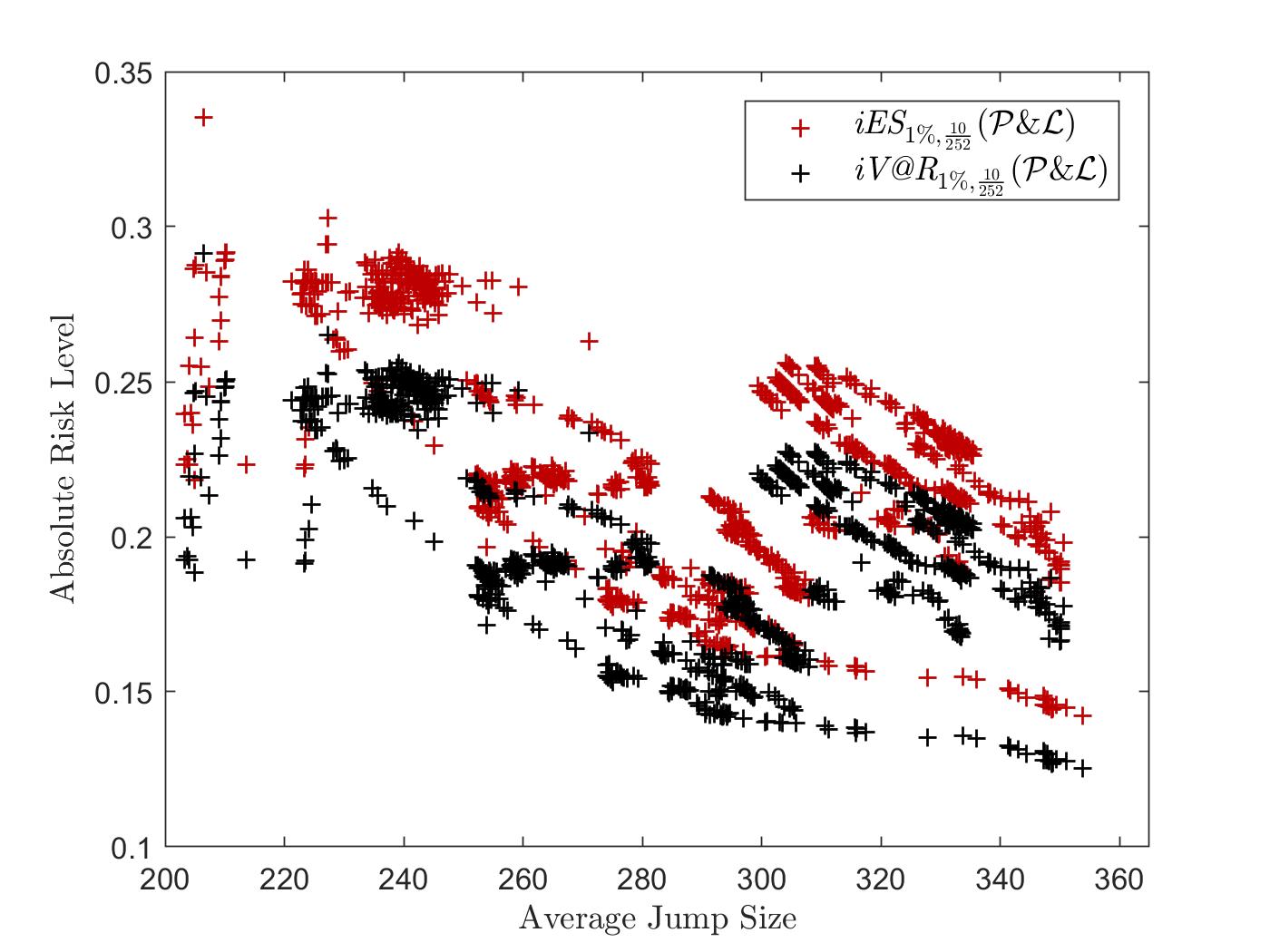}
\caption{Average Jump Size and Absolute Risk (VG)}
\label{fig:CO1JumpPlot:sub3}
\end{subfigure}
\begin{subfigure}{.5\linewidth}
\centering
\includegraphics[scale=.16]{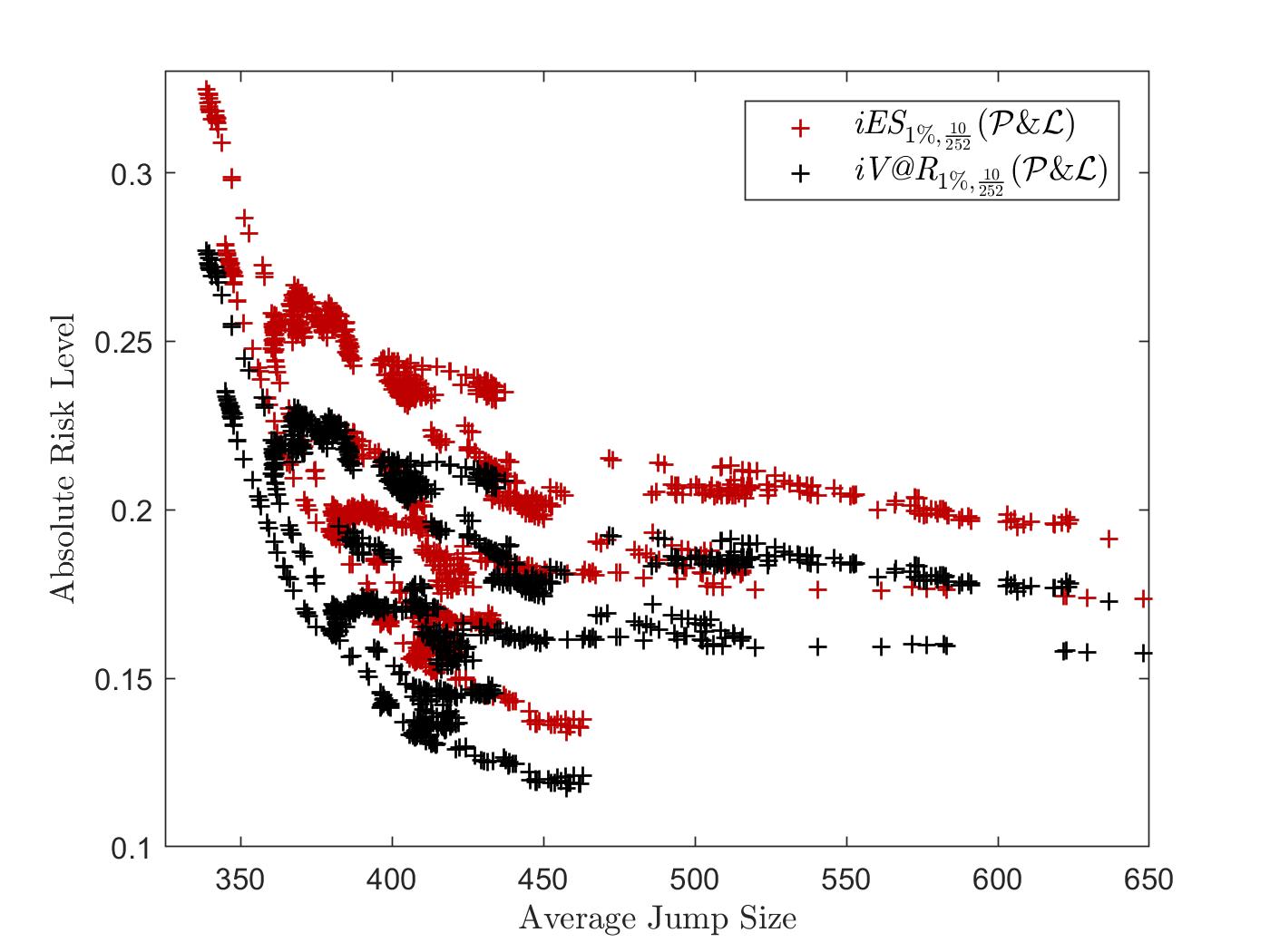}
\caption{Average Jump Size and Absolute Risk (CGMY)}
\label{fig:CO1JumpPlot:sub4}
\end{subfigure}
\caption{\textbf{Comparison of the intra-horizon risk and the contribution of certain jumps for the Brent crude oil.} Figure~\ref{fig:CO1JumpPlot:sub1} and Figure~\ref{fig:CO1JumpPlot:sub2} show the time evolution of the 10-days intra-horizon risk contributions to the 99\% loss quantile for the greatest -- in absolute size -- 3~down jumps, the greatest 5~down jumps, and the greatest 10~down jumps in the hyper-exponential jump-diffusion approximations. Additionally, Figure~\ref{fig:CO1JumpPlot:sub3} and Figure~\ref{fig:CO1JumpPlot:sub4} present the relation of the (absolute) average (down) jump size -- weighted by the probability of occurrence of each jump in the hyper-exponential jump-diffusion approximations -- to the absolute intra-horizon risk level. As earlier, the absolute risk levels correspond to negative return levels under the respective dynamics.}
\label{fig:CO1JumpPlot}
\end{figure}

\subsubsection{Intra-Horizon vs.~Point-in-Time Risk}
\noindent We now turn to the empirical risk results and start by providing a comparison of the intra-horizon and point-in-time risks inherent to a long position in the S\&P~500 index as well in the Brent crude oil from January 1995 to September 2020. To this end, we have plotted in Figure~\ref{fig:iVaRiESEvol:sub1}, Figure~\ref{fig:iVaRiESEvol:sub3}, Figure~\ref{fig:iVaRiESEvol:sub5} and Figure~\ref{fig:CO1iVaRiESEvol:sub1}, Figure~\ref{fig:CO1iVaRiESEvol:sub3}, Figure~\ref{fig:CO1iVaRiESEvol:sub5}, the time evolution of the absolute 10-days intra-horizon and point-in-time risks to the $99 \%$ quantile of the loss distribution\footnote{In our notation, this corresponds to fixing $\alpha = 1 \%$.}~calculated under the respective Lévy dynamics for the S\&P~500 index and for the Brent crude oil, respectively. These results express intra-horizon and point-in-time risks in terms of (negative) return levels, i.e.,~the graphs were obtained by computing the respective risk measures while fixing $z_{1} = z_{2} = 1$ in \textit{Scenario~2} (cf.~Section~\ref{IHRMJ}). To complement these results, we have also provided in Figure~\ref{fig:iVaRiESEvol:sub2}, Figure~\ref{fig:iVaRiESEvol:sub4}, Figure~\ref{fig:iVaRiESEvol:sub6} and Figure~\ref{fig:CO1iVaRiESEvol:sub2}, Figure~\ref{fig:CO1iVaRiESEvol:sub4}, Figure~\ref{fig:CO1iVaRiESEvol:sub6} the time evolution of the intra-horizon to point-in-time risk ratio. Finally, Figure~\ref{fig:iVaRRATIO} presents the evolution of the intra-horizon and point-in-time ratios $\omega_{1\%,\frac{10}{252}}(\mathcal{P\&L}) := \mbox{\it iV@R}_{1\%,\frac{10}{252}}(\mathcal{P\&L}) \big/ \mbox{\it iES}_{1\%,\frac{10}{252}}(\mathcal{P\&L})$ and $\omega_{1\%}\big(\mathcal{P\&L}_{\frac{10}{252}}\big) := \mbox{\it V@R}_{1\%}\big(\mathcal{P\&L}_{\frac{10}{252}}\big) \big/ \mbox{\it ES}_{1\%}\big(\mathcal{P\&L}_{\frac{10}{252}}\big)$ for all model dynamics and both underlyings. While we have chosen to follow the framework of the Basel Accords (cf.~\cite{bcbs19}) and to provide results for a 10-days horizon, we note that we do not rely on the 97.5\% quantile of the loss distribution prescribed in \cite{bcbs19}, but prefer to investigate the 99\% level. This is mainly to stay consistent with the existing literature on intra-horizon risk quantification (cf.~\cite{br04}, \cite{ro08}, \cite{bmk09}, \cite{bp10}, \cite{lv20}) and to allow for a direct comparability of our results with other articles. \vspace{1em} \\
\noindent The results in Figure~\ref{fig:iVaRiESEvol}, Figure~\ref{fig:CO1iVaRiESEvol}, and Figure~\ref{fig:iVaRRATIO} are in line with our intuition: First, the 10-days intra-horizon expected shortfall to the $99\%$ loss quantile, $\mbox{\it iES}_{1\%,\frac{10}{252}}(\mathcal{P\&L})$, exceeds at any time the intra-horizon value at risk at the same level, $\mbox{\it iV@R}_{1\%,\frac{10}{252}}(\mathcal{P\&L})$, and the same additionally holds true for the point-in-time measures. Moreover, intra-horizon risk measures always exceed their point-in-time equivalent. This becomes evident when looking at Figures~\ref{fig:iVaRiESEvol:sub1}-\ref{fig:iVaRiESEvol:sub6} and Figures~\ref{fig:CO1iVaRiESEvol:sub1}-\ref{fig:CO1iVaRiESEvol:sub6}, where the intra-horizon risk curve is always higher than its point-in-time reference and the intra-horizon to point-in-time risk ratio never falls below 1.0. In particular, Figure~\ref{fig:iVaRiESEvol:sub2}, Figure~\ref{fig:iVaRiESEvol:sub4}, Figure~\ref{fig:iVaRiESEvol:sub6}, and Figure~\ref{fig:CO1iVaRiESEvol:sub2}, Figure~\ref{fig:CO1iVaRiESEvol:sub4}, Figure~\ref{fig:CO1iVaRiESEvol:sub6} show that this ratio has a similar structure for both (intra-horizon) value at risk and (intra-horizon) expected shortfall, however, that it seems to be greater for the (intra-horizon) value at risk. Finally, we also note that intra-horizon risk is generally 5-10\% higher than point-in-time risk. Next, when investigating any of Figure~\ref{fig:iVaRiESEvol}, Figure~\ref{fig:CO1iVaRiESEvol}, and Figure~\ref{fig:iVaRRATIO}, one sees that all the intra-horizon and point-in-time measures behave similarly. In particular, all the lines inFigure~\ref{fig:iVaRiESEvol:sub1}, Figure~\ref{fig:iVaRiESEvol:sub3}, Figure~\ref{fig:iVaRiESEvol:sub5}, and Figure~\ref{fig:CO1iVaRiESEvol:sub1}, Figure~\ref{fig:CO1iVaRiESEvol:sub3}, Figure~\ref{fig:CO1iVaRiESEvol:sub5} exhibit an (almost) identical shape and seem to be obtained via a parallel shift of anyone of them. However, a closer look at these graphs reveals that the absolute difference between intra-horizon/point-in-time expected shortfall and intra-horizon/point-in-time value at risk tends to substantially increase in more severe times. That this behavior does not seem to only hold at an absolute level but also in relative terms can be seen in Figure~\ref{fig:iVaRRATIO} where the intra-horizon/point-in-time value-at-risk contribution to the intra-horizon/point-in-time expected shortfall takes its lowest values in crisis periods.

\subsubsection{Intra-Horizon Risk and Structure of Risk Across Jumps}
To finalize our discussion, we investigate the structure of intra-horizon risk across jumps.\footnote{We emphasize that similar results can be derived for point-in-time risk. However, due to the focus of the paper, we only provide intra-horizon risk results.}~To this end, we present in Figure~\ref{fig:JumpPlot} and Figure~\ref{fig:CO1JumpPlot} a comparison of intra-horizon risk and jump contributions. In particular, we have plotted in Figure~\ref{fig:JumpPlot:sub1}, Figure~\ref{fig:JumpPlot:sub2} and Figure~\ref{fig:CO1JumpPlot:sub1}, Figure~\ref{fig:CO1JumpPlot:sub2} the time evolution of the 10-days intra-horizon risk contributions to the 99\% loss quantile for the greatest -- in absolute size -- 3~down jumps, the greatest 5~down jumps, and the greatest 10~down jumps in the hyper-exponential jump-diffusion approximations of VG and CGMY. Additionally, Figure~\ref{fig:JumpPlot:sub3}, Figure~\ref{fig:JumpPlot:sub4} and Figure~\ref{fig:CO1JumpPlot:sub3}, Figure~\ref{fig:CO1JumpPlot:sub4} show the relation of the (absolute) average (down) jump size -- weighted by the probability of occurrence of each jump -- to the absolute intra-horizon risk level. The results are in line with the existing literature (cf.~e.g.,~\cite{lv20}) as well as with the observations in the previous section. First, we note that the greatest 3, 5, and 10 jumps in the hyper-exponential jump-diffusion approximations already provide a high contribution to both intra-horizon value at risk and intra-horizon expected shortfall, with a slightly higher contribution for VG than for CGMY. Additionally, looking at Figures~\ref{fig:JumpPlot:sub1}-\ref{fig:JumpPlot:sub4} and Figures~\ref{fig:CO1JumpPlot:sub1}-\ref{fig:CO1JumpPlot:sub4} reveals that the structure of risk across jumps does not differ for both of these risk measures. Indeed, while the risk contributions per jump types to both intra-horizon value at risk and intra-horizon expected shortfall are almost identical, the intra-horizon expected shortfall results in Figures~\ref{fig:JumpPlot:sub3}-\ref{fig:JumpPlot:sub4} and Figures~\ref{fig:CO1JumpPlot:sub3}-\ref{fig:CO1JumpPlot:sub4} merely replicate the shape of the intra-horizon value-at-risk results at a slightly higher risk level. This is due to the fact that the intra-horizon expected shortfall always exceeds the intra-horizon value at risk for the same time horizon and quantile. Lastly, we emphasize that Figures~\ref{fig:JumpPlot:sub3}-\ref{fig:JumpPlot:sub4} and Figures~\ref{fig:CO1JumpPlot:sub3}-\ref{fig:CO1JumpPlot:sub4} additionally present evidence of the fact that higher (absolute) average jumps generally lead to higher absolute risk levels. This is intuitively clear, since greater (absolute) average jumps immediately increase the tail of the jump distribution which likewise impacts the tail of the overall profit and loss distribution.
\section{Conclusion}
\label{Paper3CONC}
\noindent The present article extended the current literature on intra-horizon risk quantification in several directions. First, we proposed an intra-horizon analogue of the expected shortfall and discussed some of its key properties under general Lévy dynamics. The resulting (intra-horizon) risk measure is well-defined for (m)any popular class(es) of Lévy processes encountered in financial modeling and constitutes a coherent measure of risk in the sense of \cite{cd04}. Secondly, we linked our intra-horizon expected shortfall to first-passage occurrences and derived a characterization of diffusion and jump contributions to simple and maturity-randomized first-passage probabilities. These results were subsequently used to infer diffusion and jump risk contributions to the intra-horizon expected shortfall and additionally allowed us to obtain (semi-)analytical results for maturity-randomized first-passage probabilities under hyper-exponential jump-diffusion dynamics. Next, we reviewed hyper-exponential jump-diffusion approximations to Lévy processes having completely monotone jumps and proposed an adaption of the results in \cite{amp07}, \cite{jp10} that naturally preserves the diffusion vs.~jump structure of the approximated processes. We then calibrated popular Lévy processes to S\&P~500 index data as well as to Brent crude oil data and combined several of our results to analyze the intra-horizon risk inherent to a long position in these underlyings from January 1995 to September 2020. Our empirical findings revealed that even when considering large loss quantiles (i.e.,~low $\alpha$) the intra-horizon value at risk and the intra-horizon expected shortfall add conservatism to their point-in-time estimates. Additionally, they suggested that these risk measures have a very similar structure across jumps/jump clusters and that already a high contribution of their risk is due to only few, great -- in terms of the absolute jump size -- jump clusters. \vspace{1em} \\
\noindent We are convinced that the techniques and ideas presented in this paper can serve as a basis for several extensions of the theory of risk measures for stochastic processes and hope that they will further stimulate the work of the community on this topic. First, we believe that several of our ideas could be extended to introduce intra-horizon versions of spectral (point-in-time) risk measures, as introduced in \cite{ac02}. Second, the recently increasing interest in the use of expectile-based measures -- especially of the expectile value at risk -- for risk management purposes (cf.~\cite{bk14}, \cite{bd17}) may encourage the investigation of intra-horizon expectile-based risk in future research. Third, we recognize that the ideas considered in this article substantially differ from the methods of dynamic risk measures. Nevertheless, we believe that both concepts could be combined to introduce dynamic intra-horizon risk versions and we hope that this will help complementing the not yet unified theory of risk measures for stochastic processes. Finally, we think that embedding the estimation and model risk into our framework would be an interesting avenue for future reseach as well.\vspace{2em} \\
\noindent \acknow{The authors are grateful to the associate editor and two anonymous referees for their constructive comments. Additionally, the authors would like to thank Sergei Levendorskii, Tadeusz Czernik, Max Nendel, Carlo Sala, Giovanni Barone-Adesi, Matteo Burzoni, Johannes Wiesel, Felix-Benedikt Liebrich and the participants of the 9th General AMaMeF Conference, the 2019 Vienna Congress on Mathematical Finance (VCMF 2019), and the 2019 Quantitative Methods in Finance Conference (QMF 2019) for their valuable comments and suggestions.} \vspace{1em} \\
\section*{Appendices}
\renewcommand{\theequation}{A.\arabic{equation}}
\subsection*{Appendix A: Proofs - General Results}
\begin{proof}[\bf Proof of Proposition \ref{lem1}]
\noindent To start, we note that Proposition~3.2 in \cite{at02} implies that for any $T>0$ and $\alpha \in (0,1)$ the intra-horizon expected shortfall associated to the profit and loss process $(\mathcal{P\&L}_{t})_{t \in [0,T]}$, $\mbox{\textit{iES}}_{\alpha,T}(\mathcal{P\&L})$, is given by 
\begin{equation}
\label{Tasche}
\mbox{\textit{iES}}_{\alpha,T}(\mathcal{P\&L}) = -\frac{1}{\alpha} \bigg( \mathbb{E}^{
\mathbb{P}}_{z} \left[ I_{T}^{\mathcal{P\&L}} \, \mathds{1}_{\left \{ I_{T}^{\mathcal{P\&L}} \leq q_{\alpha} \left(I_{T}^{\mathcal{P\&L}}\right) \right \}} \right] - q_{\alpha}(I_{T}^{\mathcal{P\&L}}) \left[ \mathbb{P}_{z}\Big(I_{T}^{\mathcal{P\&L}} \leq q_{\alpha}\big(I_{T}^{\mathcal{P\&L}}\big)\Big) - \alpha \right]\bigg) .
\end{equation}
\noindent Therefore, we next derive an expression for the integral/expectation part in (\ref{Tasche}) and will subsequently use the result to recover (\ref{EquLem1}). Here, noting that, under $\mathbb{P}_{z}$, the inequality $I_{T}^{\mathcal{P\&L}} \leq z$ holds for any $T \geq 0$ allows us to write
\begin{align}
\mathbb{E}^{\mathbb{P}}_{z} \left[ I_{T}^{\mathcal{P\&L}} \, \mathds{1}_{\left \{ I_{T}^{\mathcal{P\&L}} \leq q_{\alpha}\left(I_{T}^{\mathcal{P\&L}}\right) \right \}} \right] & = \mathbb{E}^{\mathbb{P}}_{z} \bigg[ - \int \limits_{I_{T}^{\mathcal{P\&L}}}^{z} \mathds{1}_{\left \{ I_{T}^{\mathcal{P\&L}} \leq q_{\alpha}\left(I_{T}^{\mathcal{P\&L}}\right) \right \}} \, d\ell \bigg] + z \, \mathbb{P}_{z} \Big( I_{T}^{\mathcal{P\&L}} \leq q_{\alpha}\big(I_{T}^{\mathcal{P\&L}}\big)\Big) \nonumber \\
& = - \int \limits_{-\infty}^{z} \mathbb{P}_{z} \Big( I_{T}^{\mathcal{P\&L}} \leq \ell, \, I_{
T}^{\mathcal{P\&L}} \leq q_{\alpha}\big(I_{T}^{\mathcal{P\&L}}\big) \Big) \, d\ell + z \, \mathbb{P}_{z} \Big( I_{T}^{\mathcal{P\&L}} \leq q_{\alpha}\big(I_{T}^{\mathcal{P\&L}}\big)\Big) \nonumber \\
& = - \int \limits_{q_{\alpha}\big(I_{T}^{\mathcal{P\&L}} \big)}^{z} \mathbb{P}_{z} \Big( I_{
T}^{\mathcal{P\&L}} \leq q_{\alpha}\big(I_{T}^{\mathcal{P\&L}}\big) \Big) \, d\ell \, - \int \limits_{-\infty}^{q_{\alpha}\big(I_{T}^{\mathcal{P\&L}} \big)} \mathbb{P}_{z} \Big( I_{T}^{\mathcal{P\&L}} \leq \ell \Big) \, d\ell \nonumber \\
& \hspace{10em} + z \, \mathbb{P}_{z} \Big( I_{T}^{\mathcal{P\&L}} \leq q_{\alpha}\big(I_{T}^{\mathcal{P\&L}}\big)\Big) \nonumber \\
& = q_{\alpha}\big(I_{T}^{\mathcal{P\&L}}\big) \, \mathbb{P}_{z} \Big( I_{T}^{\mathcal{P\&L}} \leq q_{\alpha}\big(I_{T}^{\mathcal{P\&L}}\big)\Big) - \int \limits_{-\infty}^{q_{\alpha}\big(I_{T}^{\mathcal{P\&L}}\big)} \mathbb{P}_{z} \Big( I_{T}^{\mathcal{P\&L}} \leq \ell \Big) \, d\ell .
\label{Change1}
\end{align}
\noindent Hence, combining (\ref{Tasche}) and (\ref{Change1}) with the 
relation $ \mathbb{P}_{z} \Big( I_{T}^{\mathcal{P\&L}} \leq \ell \Big) = \mathbb{P}_{z} \Big( \tau^{\mathcal{P\&L},-}_{\ell} \leq T \Big)$ gives that the intra-horizon expected shortfall can be expressed in terms of first-passage probabilities, as
\begin{align}
\mbox{\textit{iES}}_{\alpha,T}(\mathcal{P\&L}) & = \frac{1}{\alpha} \int \limits_{-\infty}^{q
_{\alpha}\big(I_{T}^{\mathcal{P\&L}}\big)} \mathbb{P}_{z} \Big( \tau^{\mathcal{P\&L},-}_{\ell} \leq T \Big) \, d\ell  - q_{\alpha}\big(I_{T}^{\mathcal{P\&L}}\big),
\end{align}
\noindent which finally provides Equation (\ref{EquLem1}).
\end{proof}

\begin{proof}[\bf Proof of Lemma \ref{lem2}]
\noindent To show (\ref{lem2equa}), we rely on similar arguments to the ones used in \cite{ck11} (cf.~also~\cite{hm13}). Here, we focus on the result for the upside first-passage probabilities and show that there exists a constant $c > 1$ such that for any $x \in \mathbb{R}$ and $\mathcal{T} > 0$ we have that
\begin{equation}
\lim \limits_{\ell \uparrow \infty} \; e^{c \cdot \ell}\,\mathbb{P}_{x}^{X} \Big( M_{\mathcal{T}}^{X} \geq \ell \Big) = 0,
\label{Maxproof}
\end{equation}
\noindent where $(M_{t}^{X})_{t \geq 0}$ denotes the maximum process associated to $(X_{t})_{t \geq 0}$, i.e.,~the process defined by
\begin{equation}
M_{t}^{X} := \sup \limits_{0 \leq u \leq t} X_{u}, \hspace{1.5em} t \geq 0.
\end{equation}
\noindent Once (\ref{Maxproof}) is established, the respective convergence result for the downside first-passage probabilities is easily obtained by noting that the minimum process $(I_{t}^{X})_{t \geq 0}$ satisfies
\begin{equation}
M_{t}^{\tilde{X}} = - I_{t}^{X}, \hspace{1.5em} t \geq 0,
\end{equation}
\noindent where we have denoted by $(\tilde{X}_{t})_{t \geq 0}$ the dual process to $(X_{t})_{t \geq 0}$, i.e.,~the process that is defined by $\tilde{X}_{t} := -X_{t}$, $t \geq 0$. Therefore, we only have to prove (\ref{Maxproof}). Here, we start by recalling that the process $(Z_{t}^{\theta})_{t \geq 0}$ defined via 
\begin{equation}
Z_{t}^{\theta} := e^{\theta X_{t} - t \Phi_{X}(\theta)}, \hspace{1.5em} t \geq 0,
\end{equation}
\noindent is, for any $\theta \in \mathbb{R}$ satisfying $\mathbb{E}^{\mathbb{P}^{X}}_{0} \left[ e^{\theta X_{1}} \right] < \infty$, a well-defined martingale. Using the optional sampling theorem, this allows us to derive, in particular, that for $\theta^{\star} > 1 $ and any $x \in \mathbb{R}$, $\mathcal{T} > 0$ we have that
\begin{equation}
e^{\theta^{\star} \ell} \, \mathbb{P}_{x}^{X} \Big( M_{\mathcal{T}}^{X} \geq \ell \Big) \leq \mathbb{E}^{\mathbb{P}^{X}}_{x} \left[ \exp\big \{ \theta^{\star} X_{(\tau_{\ell}^{X,+} \wedge \mathcal{T} )} \big \} \right] \leq \left \{  \begin{array}{lc}
 e^{\Phi_{X}(\theta^{\star}) \mathcal{T} + \theta^{\star} x }, & \mbox{if} \; \Phi_{X}(\theta^{\star}) > 0 ,\\
e^{\theta^{\star} x }, & \mbox{if} \; \Phi_{X}(\theta^{\star}) \leq 0.
\end{array} \right.
\end{equation}
\noindent Therefore, we obtain in any case for $x \in \mathbb{R}$ and $\mathcal{T} >0$ that $ e^{\theta^{\star} \ell} \, \mathbb{P}_{x}^{X} \Big( M_{\mathcal{T}}^{X} \geq \ell \Big) \leq C$ for some constant $C >0$ and combining this result with the fact that $\theta^{\star} > 1$ finally gives, with $\theta_{0} > 1$ and $c > 1$ satisfying $ c \,  \theta_{0} = \theta^{\star}$, that for any $x \in \mathbb{R}$ and $\mathcal{T} > 0$
\begin{equation}
e^{c \cdot \ell}\,\mathbb{P}_{x}^{X} \Big( M_{\mathcal{T}}^{X} \geq \ell \Big) = e^{c (1-\theta_{0})\ell}\, e^{c \, \theta_{0} \cdot\ell} \, \mathbb{P}_{x}^{X} \Big( M_{\mathcal{T}}^{X} \geq \ell \Big) = e^{c (1-\theta_{0})\ell}\, e^{\theta^{\star} \ell} \, \mathbb{P}_{x}^{X} \Big( M_{\mathcal{T}}^{X} \geq \ell \Big) \rightarrow 0, \hspace{1em} \mbox{as} \; \ell \uparrow \infty,
\end{equation}
\noindent holds, hence (\ref{Maxproof}).
 \vspace{1em} \\
\noindent To prove that $\mathbb{E}^{\mathbb{P}}_{z} \left[ | I_{\mathcal{T}}^{\mathcal{P\&L}} | \right] < \infty$ holds for any $\mathcal{T} >0$, we combine the convergence results (\ref{lem2equa}) with standard techniques. First, we note that
\begin{align}
\mathbb{E}^{\mathbb{P}}_{z} \left[ | I_{\mathcal{T}}^{\mathcal{P\&L}} | \right]  & = \mathbb{E}^{\mathbb{P}}_{z} \bigg[ \int \limits_{0}^{\infty} \mathds{1}_{\{ |I_{\mathcal{T}}^{\mathcal{P\&L}}| \geq \ell \} } \, d\ell  \bigg]  \nonumber \\
& \leq  \int \limits_{0}^{\infty} \mathbb{P}_{z} \Big( I_{\mathcal{T}}^{\mathcal{P\&L}} \geq \ell \Big) \, d \ell  + \int \limits_{0}^{\infty} \mathbb{P}_{z} \Big( I_{\mathcal{T}}^{\mathcal{P\&L}} \leq -\ell \Big) \, d\ell  \label{goodEq}\\
& \leq  |z| + \int \limits_{0}^{\infty} \mathbb{P}_{z} \Big( I_{\mathcal{T}}^{\mathcal{P\&L}} \leq -\ell \Big) \, d\ell. \nonumber
\end{align}
\noindent Therefore, we only need to show the finiteness of the integral on the right hand side. Under \textit{Scenario 1}, the boundedness of $\ell \mapsto e^{c \cdot \ell} \, \mathbb{P}_{z}^{X} \Big( \tau_{-\ell}^{X,-} \leq \mathcal{T} \Big) $ on $[0,\infty)$ implies that 
\begin{align}
\int \limits_{0}^{\infty} \mathbb{P}_{z} \Big( I_{\mathcal{T}}^{\mathcal{P\&L}} \leq -\ell \Big) \, d\ell  = \int \limits_{0}^{\infty} e^{-c \cdot \ell} e^{c \cdot \ell} \, \mathbb{P}_{z}^{X} \Big( \tau_{-\ell}^{X,-} \leq \mathcal{T} \Big) \, d\ell \leq  K_{1} \int \limits_{0}^{\infty} e^{-c \cdot \ell} \, d\ell < \infty,
\label{INTarg}
\end{align}
\noindent and this already provides the required result. Therefore, we next focus on \textit{Scenario 2}. Here, we first note that for a long position the finiteness of the integral directly follows from the fact that $ I_{\mathcal{T}}^{\mathcal{P\&L}} \geq - z_{2}$. Hence, we are left with the case of a short position under \textit{Scenario 2}. In this case, similar arguments as in (\ref{INTarg}) give that
\begin{align}
\int \limits_{0}^{\infty} \mathbb{P}_{z} \Big( I_{\mathcal{T}}^{\mathcal{P\&L}} \leq -\ell \Big) \, d\ell  & = \int \limits_{0}^{\infty} e^{-c \log(z_{2} + \ell) } e^{c \log(z_{2} + \ell)} \, \mathbb{P}_{\log(z_{2}-z)}^{X} \Big( \tau_{\log(z_{2}+\ell)}^{X,+} \leq \mathcal{T} \Big) \, d\ell \nonumber \\
&  \leq  K_{2} \left( 1 + \int \limits_{1}^{\infty}  \frac{1}{(z_{2}+\ell)^{c}} \, d\ell \right) < \infty,
\end{align}
\noindent where the finiteness follows since $c > 1$. This finally gives the claim.
\end{proof}

\subsection*{Appendix B: Proofs - First-Passage Probabilities}

\begin{proof}[\bf Proof of Proposition \ref{prop2}]
\noindent We start by noting that, for any $(\mathbf{t},x,\delta) \in [0,T] \times \mathbb{R} \times \mathbb{R} $, the process $(Z_{t})_{t \in [0,\mathbf{t}]}$ defined via $Z_{t}:= (\mathbf{t}-t, x+X_{t},\delta + {\Delta X}_{t})$ is a strong Markov process with state domain given by $\mathcal{D}_{\mathbf{t}}:= [0,\mathbf{t}] \times \mathbb{R} \times \mathbb{R}$ and define, for any $\ell \in \mathbb{R}$, the following stopping domains
\begin{align}
\mathcal{S}^{+}_{\ell} := \mathcal{S}^{+}_{\ell,1} \cup \mathcal{S}^{+}_{\ell,2}, \hspace{1.5em} \mbox{with} \hspace{1.5em}  \mathcal{S}^{+}_{\ell,1} := \{0 \} \times &\, \mathbb{R} \times \mathbb{R} \hspace{1.5em} \mbox{and} \hspace{1.5em} \mathcal{S}^{+}_{\ell,2} := [0,T] \times [ \ell, \infty) \times [\delta,\infty), \nonumber \\
 \mathcal{S}_{\ell}^{\mathcal{J},+} := [0,T] \times ( \ell, \infty) \times [\delta,\infty), & \hspace{2em} \mathcal{S}_{\ell}^{0,+} := \mathcal{S}_{\ell}^{+} \setminus \mathcal{S}_{\ell}^{\mathcal{J},+} = \mathcal{S}^{0,+}_{\ell,1} \cup \mathcal{S}^{0,+}_{\ell,2}, \\
\mbox{with} \hspace{1.5em} \mathcal{S}^{0,+}_{\ell,1} = \mathcal{S}^{+}_{\ell,1} \setminus \mathcal{S}_{\ell}^{\mathcal{J},+} & \hspace{1em} \mbox{and} \hspace{1em} \mathcal{S}^{0,+}_{\ell,2} = \mathcal{S}^{+}_{\ell,2} \setminus \mathcal{S}_{\ell}^{\mathcal{J},+} ,\nonumber 
\end{align}
\begin{align}
\mathcal{S}^{-}_{\ell} := \mathcal{S}^{-}_{\ell,1} \cup \mathcal{S}^{-}_{\ell,2}, \hspace{1.5em} \mbox{with} \hspace{1.5em}  \mathcal{S}^{-}_{\ell,1} := \{0 \} \times &\, \mathbb{R} \times \mathbb{R} \hspace{1.5em} \mbox{and} \hspace{1.5em} \mathcal{S}^{-}_{\ell,2} := [0,T] \times (-\infty, \ell] \times (-\infty,\delta], \nonumber \\
 \mathcal{S}_{\ell}^{\mathcal{J},-} := [0,T] \times (-\infty, \ell) \times (-\infty,\delta], & \hspace{2em} \mathcal{S}_{\ell}^{0,-} := \mathcal{S}_{\ell}^{-} \setminus \mathcal{S}_{\ell}^{\mathcal{J},-} = \mathcal{S}^{0,-}_{\ell,1} \cup \mathcal{S}^{0,-}_{\ell,2}, \\
\mbox{with} \hspace{1.5em} \mathcal{S}^{0,-}_{\ell,1} = \mathcal{S}^{-}_{\ell,1} \setminus \mathcal{S}_{\ell}^{\mathcal{J},-} & \hspace{1em} \mbox{and} \hspace{1em} \mathcal{S}^{0,-}_{\ell,2} = \mathcal{S}^{-}_{\ell,2} \setminus \mathcal{S}_{\ell}^{\mathcal{J},-} .\nonumber 
\end{align}
\noindent Clearly, both $\mathcal{S}_{\ell}^{+}$ and $\mathcal{S}_{\ell}^{-}$ are closed in the state space $\mathcal{D}_{T}$. We therefore obtain that, for each of theses domains, the first entry times $\tau_{\mathcal{S}_{\ell}^{\pm}}$ defined via
\begin{equation}
\tau_{\mathcal{S}_{\ell}^{\pm}} := \inf \left \{ t \geq 0 : \; Z_{t} \in \mathcal{S}_{\ell}^{\pm} \right \}
\end{equation}
\noindent is a stopping time that satisfies $\tau_{\mathcal{S}_{\ell}^{\pm}} \leq \mathbf{t}$, under $\mathbb{P}_{z}^{Z}$, the measure having initial distribution $Z_{0} = z = (\mathbf{t},x,\delta)$. Using this notation, we can now re-express the first-passage probabilities $u_{X}^{\mathcal{E}_{0}^{\pm}}(\cdot)$ and $u_{X}^{\mathcal{E}_{\mathcal{J}}^{\pm}}(\cdot)$ as solutions of appropriate stopping problems. Indeed, it is easily seen that
\begin{equation}
u_{X}^{\mathcal{E}_{0}^{\pm}}(\mathcal{T},x;\ell ) = V_{0}^{\pm}\big( (\mathcal{T},x,0) \big) \hspace{2em} \mbox{and} \hspace{2em} u_{X}^{\mathcal{E}_{\mathcal{J}}^{\pm}}(\mathcal{T},x;\ell ) = V_{\mathcal{J}}^{\pm}\big( (\mathcal{T},x,0) \big),
\label{RelUV}
\end{equation}
\noindent where the value functions $V_{0}^{\pm}(\cdot)$ and $V_{\mathcal{J}}^{\pm}(\cdot)$ have the following probabilistic representations: 
\begin{align}
V_{0}^{\pm}(z) = \mathbb{E}_{z}^{\mathbb{P}^{Z}} \left[ \mathds{1}_{\mathcal{S}_{\ell,2}^{0,\pm} } \Big(Z_{\tau_{\mathcal{S}_{\ell}^{\pm}}}\Big)\right] \hspace{2em} \mbox{and} \hspace{2em} V_{\mathcal{J}}^{\pm}(z) = \mathbb{E}_{z}^{\mathbb{P}^{Z}} \left[ \mathds{1}_{ \mathcal{S}_{\ell}^{\mathcal{J},\pm}} \Big(Z_{\tau_{\mathcal{S}_{\ell}^{\pm}}}\Big)\right] .
\end{align}
\noindent Additionally, standard arguments based on the strong Markov property (cf.~\cite{pe06}, \cite{ma19a}, \cite{ma19b}) imply that, for any $\ell \in \mathbb{R}$, the functions $V_{0}^{\pm}(\cdot)$ and $V_{\mathcal{J}}^{\pm}(\cdot)$ satisfy\footnote{Note that we implicitly use the fact that the infinitesimal generator of the process $(\Delta X_{t} )_{t \in [0,\mathbf{t}]}$ vanishes. This follows since $(X_{t})_{t \in [0,\mathbf{t}]}$ is, as Feller process, quasi-left-continuous.}
\begin{align}
\partial_{\mathbf{t}} V_{0}^{\pm}\big((\mathbf{t},x,\delta)\big)  & = \mathcal{A}_{X} V_{0}^{\pm}\big((\mathbf{t},x,\delta)\big) , \hspace{1.5em} \mbox{on} \hspace{0.5em} \mathcal{D}_{T} \setminus \mathcal{S}_{\ell}^{\pm},\\
V_{0}^{\pm}\big((\mathbf{t},x,\delta)\big) & = \mathds{1}_{ \mathcal{S}_{\ell,2}^{0,\pm} } \big((\mathbf{t},x,\delta)\big), \hspace{2.7em} \mbox{on} \hspace{0.5em} \mathcal{S}_{\ell}^{\pm},
\end{align}
\noindent and
\begin{align}
\partial_{\mathbf{t}} V_{\mathcal{J}}^{\pm}\big((\mathbf{t},x,\delta)\big)  & = \mathcal{A}_{X} V_{\mathcal{J}}^{\pm}\big((\mathbf{t},x,\delta)\big) , \hspace{1.5em} \mbox{on} \hspace{0.5em} \mathcal{D}_{T} \setminus \mathcal{S}_{\ell}^{\pm} ,\\
V_{\mathcal{J}}^{\pm}\big((\mathbf{t},x,\delta)\big) & = \mathds{1}_{ \mathcal{S}_{\ell}^{\mathcal{J},\pm} } \big((\mathbf{t},x,\delta)\big), \hspace{2.7em} \mbox{on} \hspace{0.5em} \mathcal{S}_{\ell}^{\pm}.
\end{align}
\noindent Therefore, recovering $u_{X}^{\mathcal{E}_{0}^{\pm}}(\cdot)$ and $u_{X}^{\mathcal{E}_{\mathcal{J}}^{\pm}}(\cdot)$ via (\ref{RelUV}) directly gives Equations (\ref{Prop2Cauchy11})-(\ref{Prop2Cauchy13}) and (\ref{Prop2Cauchy21})-(\ref{Prop2Cauchy23}), respectively. Since Equations (\ref{Prop2Cauchy14}) and (\ref{Prop2Cauchy24}) are naturally satisfied, Proposition~\ref{prop2} follows.
\end{proof}

\begin{proof}[\bf Proof of Proposition \ref{prop3}]
\noindent We prove Proposition \ref{prop3} using a similar approach to the one adopted in the proof of Proposition \ref{prop2}. First, we recall from (\ref{PoissonAppr2}) that any of the maturity-randomized first-passage probabilities $\mathcal{LC} \big(u_{X}^{\mathcal{E}_{0}^{\pm}}\big)(\cdot)$ and $\mathcal{LC} \big(u_{X}^{\mathcal{E}_{\mathcal{J}}^{\pm}}\big)(\cdot)$ can be seen as the probability of a respective first-passage event occurring before the first jump time of an independent Poisson process $(N_{t})_{ t \geq 0}$ with intensity $\vartheta >0$. Therefore, we consider, for any $(n,x,\delta) \in \mathbb{N}_{0} \times \mathbb{R} \times \mathbb{R}$, the process $(Z_{t})_{t \geq 0}$ defined via $Z_{t} := \left(n+N_{t}, x+ X_{t}, \delta + {\Delta X}_{t} \right)$ and note that it is a strong Markov process on the state domain $\mathcal{D} := \mathbb{N}_{0} \times \mathbb{R} \times \mathbb{R}$. Additionally, we define, for $\ell \in \mathbb{R}$, the following stopping domains
\begin{align}
\mathcal{S}^{+}_{\ell} := \mathcal{S}^{+}_{\ell,1} \cup \mathcal{S}^{+}_{\ell,2}, \hspace{1.5em} \mbox{with} \hspace{1.5em}  \mathcal{S}^{+}_{\ell,1} := \mathbb{N} \times &\, \mathbb{R} \times \mathbb{R} \hspace{1.5em} \mbox{and} \hspace{1.5em} \mathcal{S}^{+}_{\ell,2} := \{0\} \times [ \ell, \infty) \times [\delta,\infty), \nonumber \\
 \mathcal{S}_{\ell}^{\mathcal{J},+} := \{0 \} \times ( \ell, \infty) \times [\delta,\infty), & \hspace{2em} \mathcal{S}_{\ell}^{0,+} := \mathcal{S}_{\ell}^{+} \setminus \mathcal{S}_{\ell}^{\mathcal{J},+} = \mathcal{S}^{0,+}_{\ell,1} \cup \mathcal{S}^{0,+}_{\ell,2}, \\
\mbox{with} \hspace{1.5em} \mathcal{S}^{0,+}_{\ell,1} = \mathcal{S}^{+}_{\ell,1} \setminus \mathcal{S}_{\ell}^{\mathcal{J},+} & \hspace{1em} \mbox{and} \hspace{1em} \mathcal{S}^{0,+}_{\ell,2} = \mathcal{S}^{+}_{\ell,2} \setminus \mathcal{S}_{\ell}^{\mathcal{J},+} ,\nonumber 
\end{align}
\begin{align}
\mathcal{S}^{-}_{\ell} := \mathcal{S}^{-}_{\ell,1} \cup \mathcal{S}^{-}_{\ell,2}, \hspace{1.5em} \mbox{with} \hspace{1.5em}  \mathcal{S}^{-}_{\ell,1} := \mathbb{N} \times & \, \mathbb{R} \times \mathbb{R} \hspace{1.5em} \mbox{and} \hspace{1.5em} \mathcal{S}^{-}_{\ell,2} := \{ 0 \} \times (-\infty, \ell] \times (-\infty,\delta], \nonumber \\
 \mathcal{S}_{\ell}^{\mathcal{J},-} := \{ 0 \} \times (-\infty, \ell) \times (-\infty,\delta], & \hspace{2em} \mathcal{S}_{\ell}^{0,-} := \mathcal{S}_{\ell}^{-} \setminus \mathcal{S}_{\ell}^{\mathcal{J},-} = \mathcal{S}^{0,-}_{\ell,1} \cup \mathcal{S}^{0,-}_{\ell,2}, \\
\mbox{with} \hspace{1.5em} \mathcal{S}^{0,-}_{\ell,1} = \mathcal{S}^{-}_{\ell,1} \setminus \mathcal{S}_{\ell}^{\mathcal{J},-} & \hspace{1em} \mbox{and} \hspace{1em} \mathcal{S}^{0,-}_{\ell,2} = \mathcal{S}^{-}_{\ell,2} \setminus \mathcal{S}_{\ell}^{\mathcal{J},-} ,\nonumber 
\end{align}
\noindent and see that both $\mathcal{S}^{+}_{\ell}$ and $\mathcal{S}^{-}_{\ell}$ form a closed set in $\mathcal{D}$.\footnote{We emphasize that several choices of a product-metric on $\mathcal{D}$ give the closedness of the set $\mathcal{S}_{\ell}^{+}$ and $\mathcal{S}_{\ell}^{-}$. In particular, one may choose on $\mathbb{N}_{0}$ the following metric
$$ d_{\mathbb{N}_{0}}(m,n) := \left \{ \begin{array}{lc}
1 + |2^{-m} - 2^{-n} |, & m\neq n, \\
0, & m =n,
\end{array} \right. $$
\noindent and consider the product-metric on $\mathcal{D}$ obtained by combining $d_{\mathbb{N}_{0}}(\cdot,\cdot)$ on $\mathbb{N}_{0}$ with the Euclidean metric on $\mathbb{R}$.}~Consequently, for each of these domains, the first entry time defined by
\begin{equation}
\tau_{\mathcal{S}_{\ell}^{\pm}} := \inf \left \{ t \geq 0 : \; Z_{t} \in \mathcal{S}_{\ell}^{\pm} \right \}
\end{equation}
\noindent is a stopping time. Furthermore, the finiteness of the first moment of the exponential distribution for any intensity parameter $\vartheta >0$ implies the $\mathbb{P}_{z}^{Z}$-almost sure finiteness of $\tau_{\mathcal{S}_{\ell}^{\pm}}$ for any $z = (n,x,\delta)$, where $\mathbb{P}_{z}^{Z}$ refers to the measure having initial distribution $Z_{0}=z$. Using this notation, we can therefore follow the line of the arguments developed in the proof of Proposition~\ref{prop2} and re-express the maturity-randomized first-passage probabilities $\mathcal{LC} \big(u_{X}^{\mathcal{E}_{0}^{\pm}}\big)(\cdot)$ and $\mathcal{LC} \big(u_{X}^{\mathcal{E}_{\mathcal{J}}^{\pm}}\big)(\cdot)$ as solutions of appropriate stopping problems:
\begin{equation}
\mathcal{LC} \big(u_{X}^{\mathcal{E}_{0}^{\pm}}\big)(\vartheta,x;\ell ) = \widehat{V}_{0}^{\pm}\big( (0,x,0) \big) \hspace{2em} \mbox{and} \hspace{2em} \mathcal{LC} \big(u_{X}^{\mathcal{E}_{\mathcal{J}}^{\pm}}\big)(\vartheta,x;\ell ) = \widehat{V}_{\mathcal{J}}^{\pm}\big( (0,x,0) \big),
\label{RelUV2}
\end{equation}
\noindent where the value functions $\widehat{V}_{0}^{\pm}(\cdot)$ and $\widehat{V}_{\mathcal{J}}^{\pm}(\cdot)$ have the following probabilistic representations: 
\begin{align}
\widehat{V}_{0}^{\pm}(z) = \mathbb{E}_{z}^{\mathbb{P}^{Z}} \left[ \mathds{1}_{\mathcal{S}_{\ell,2}^{0,\pm} } \Big(Z_{\tau_{\mathcal{S}_{\ell}^{\pm}}}\Big)\right] \hspace{2em} \mbox{and} \hspace{2em} \widehat{V}_{\mathcal{J}}^{\pm}(z) = \mathbb{E}_{z}^{\mathbb{P}^{Z}} \left[ \mathds{1}_{ \mathcal{S}_{\ell}^{\mathcal{J},\pm}} \Big(Z_{\tau_{\mathcal{S}_{\ell}^{\pm}}}\Big)\right] .
\end{align}
\noindent Additionally, standard arguments based on the strong Markov property (cf.~\cite{pe06}, \cite{ma19a}, \cite{ma19b}) imply that, for any $\ell \in \mathbb{R}$, the functions $\widehat{V}_{0}^{\pm}(\cdot)$ and $\widehat{V}_{\mathcal{J}}^{\pm}(\cdot)$ satisfy the following problems
\begin{align}
\mathcal{A}_{Z} \widehat{V}_{0}^{\pm}\big((n,x,\delta)\big)  & = 0 , \hspace{6.8em} \mbox{on} \hspace{0.5em} \mathcal{D} \setminus \mathcal{S}_{\ell}^{\pm},\\
\widehat{V}_{0}^{\pm}\big((n,x,\delta)\big) & = \mathds{1}_{ \mathcal{S}_{\ell,2}^{0,\pm} } \big((n,x,\delta)\big), \hspace{1.7em} \mbox{on} \hspace{0.5em} \mathcal{S}_{\ell}^{\pm},
\end{align}
\noindent and
\begin{align}
\mathcal{A}_{Z} \widehat{V}_{\mathcal{J}}^{\pm}\big((n,x,\delta)\big) & = 0 , \hspace{6.8em} \mbox{on} \hspace{0.5em} \mathcal{D} \setminus \mathcal{S}_{\ell}^{\pm} ,\\
\widehat{V}_{\mathcal{J}}^{\pm}\big((n,x,\delta)\big) & = \mathds{1}_{ \mathcal{S}_{\ell}^{\mathcal{J},\pm} } \big((n,x,\delta)\big), \hspace{1.7em} \mbox{on} \hspace{0.5em} \mathcal{S}_{\ell}^{\pm}.
\end{align}
\noindent where $\mathcal{A}_{Z}$ denotes the infinitesimal generator of the process $(Z_{t})_{t \geq 0}$. To complete the proof, it therefore suffices to note that (for any suitable function $V:\mathcal{D} \rightarrow \mathbb{R}$) the infinitesimal generator $\mathcal{A}_{Z}$ can be re-expressed as\footnote{Here again, we have implicitly used the quasi-left-continuity of the process $(X_{t})_{t \geq 0}$, which follows from the Feller property.}
\begin{align}
\mathcal{A}_{Z} V(z) & = \mathcal{A}_{N}^{n} V\big((n,x,\delta) \big) + \mathcal{A}_{X}^{x} V \big((n,x,\delta) \big) \\
& = \vartheta \left( V\big( (n+1,x,\delta) \big) - V\big( (n,x,\delta) \big) \right) + \mathcal{A}_{X}^{x} V\big( (n,x,\delta) \big) ,
\label{infiniFK}
\end{align}
\noindent where $\mathcal{A}_{N}$ denotes the infinitesimal generator of the Poisson process $(N_{t})_{ t \geq 0}$ and the notation $\mathcal{A}_{N}^{n}$ and $\mathcal{A}_{X}^{x}$ is used to indicate that the generators are applied to $n$ and $x$ respectively. Indeed, recovering $\mathcal{LC} \big(u_{X}^{\mathcal{E}_{0}^{\pm}}\big)(\cdot)$ and $\mathcal{LC} \big(u_{X}^{\mathcal{E}_{\mathcal{J}}^{\pm}}\big)(\cdot)$ via (\ref{RelUV2}) while noting Relation (\ref{infiniFK}) and the fact that for any $x \in \mathbb{R}$ and $\delta \in \mathbb{R}$ we have
\begin{equation}
\widehat{V}_{0}^{\pm}\big((1,x,\delta)\big) = 0 \hspace{2em} \mbox{and} \hspace{2em} \widehat{V}_{\mathcal{J}}^{\pm}\big((1,x,\delta)\big) = 0
\end{equation}
\noindent finally leads to Problems (\ref{OIDE1})-(\ref{OIDE3}) and (\ref{OIDEj1})-(\ref{OIDEj3}), respectively.
\end{proof}

\subsection*{Appendix C: Proofs - Hyper-Exponential Jump-Diffusions}

\begin{proof}[\bf Proof of Proposition \ref{prop4}]
\noindent We only provide a proof for the upside first-passage probabilities, i.e.,~for the functions $\mathcal{LC} \big(u_{X}^{\mathcal{E}_{0}^{+}}\big)(\cdot)$ and $\mathcal{LC} \big(u_{X}^{\mathcal{E}_{\mathcal{J}}^{+}}\big)(\cdot)$, and note that the downside first-passage probabilities can be derived analogously. \vspace{1em} \\
\noindent We begin by fixing $\vartheta >0$, $\ell \in \mathbb{R}$ and showing that the maturity-randomized first-passage probabilities $\mathcal{LC} \big(u_{X}^{\mathcal{E}_{0}^{+}}\big)(\cdot)$ and $\mathcal{LC} \big(u_{X}^{\mathcal{E}_{\mathcal{J}}^{+}}\big)(\cdot)$ have the structure described in (\ref{STru1}) with coefficients $(\underline{v}_{0,k})_{k \in \{ 1, \ldots , m+1 \} }$ and $(\underline{v}_{\mathcal{J},k})_{k \in \{ 1, \ldots , m+1 \} }$ satisfying the linear equations in (\ref{CoefEq1}). Here, we first note that the same arguments as in the proof of Theorem 3.2.~in \cite{ck11} imply that the general solution to OIDE (\ref{OIDE1}) takes the form
\begin{equation}
\tilde{V}(\vartheta,x;\ell) = \sum \limits_{k=1}^{m+1} \underline{w}_{k}^{\star}  e^{\beta_{k,\vartheta} \cdot x} + \sum \limits_{h=1}^{n+1} \overline{w}_{h}^{\star}  e^{\gamma_{h,\vartheta} \cdot x},
\label{FORM1}
\end{equation}
\noindent or, equivalently,
\begin{equation}
V(\vartheta,x;\ell) = \sum \limits_{k=1}^{m+1} \underline{w}_{k}  e^{\beta_{k,\vartheta} \cdot (x-\ell)} + \sum \limits_{h=1}^{n+1} \overline{w}_{h}  e^{\gamma_{h,\vartheta} \cdot (x-\ell)}.
\label{FORM2}
\end{equation}
\noindent Although both expressions (\ref{FORM1}) and (\ref{FORM2}) lead to equivalent results, we choose to follow Ansatz (\ref{FORM2}) since it will allow us to separate the dependency of the first-passage level $\ell$ from the remaining parts. This last property will prove useful in subsequent computations discussed, for instance, in Section \ref{NUMres}. \\
\noindent Next, the boundedness of the functions $\mathcal{LC} \big(u_{X}^{\mathcal{E}_{0}^{+}}\big)(\cdot)$ and $\mathcal{LC} \big(u_{X}^{\mathcal{E}_{\mathcal{J}}^{+}}\big)(\cdot)$ gives that for both of these functions one must have
\begin{equation}
\overline{w}_{1} = \ldots = \overline{w}_{n+1} = 0.
\end{equation}
\noindent Therefore, the functional form (\ref{STru1}) is obtained. We now determine the coefficients $(\underline{v}_{0,k})_{k \in \{ 1, \ldots , m+1 \} }$ and $(\underline{v}_{\mathcal{J},k})_{k \in \{ 1, \ldots , m+1 \}}$. First, the continuous-fit conditions\footnote{Note that these continuous-fit conditions are guaranteed by $\sigma_{X} >0$ and the finite jump activity characterizing compound Poisson processes, hence hyper-exponential jump-diffusions. More details can be found, for instance, in \cite{cv05}.}
\begin{equation}
\mathcal{LC} \big(u_{X}^{\mathcal{E}_{0}^{+}}\big)(\vartheta,\ell-;\ell) = \mathcal{LC} \big(u_{X}^{\mathcal{E}_{0}^{+}}\big)(\vartheta,\ell;\ell) \hspace{1.5em} \mbox{and} \hspace{1.8em} \mathcal{LC} \big(u_{X}^{\mathcal{E}_{\mathcal{J}}^{+}}\big)(\vartheta,\ell-;\ell) = \mathcal{LC} \big(u_{X}^{\mathcal{E}_{\mathcal{J}}^{+}}\big)(\vartheta,\ell;\ell)
\label{ContFit1}
\end{equation}
\noindent imply that
\begin{equation}
\sum \limits_{k=1}^{m+1} \underline{v}_{0,k}  = 1 \hspace{1.7em} \mbox{and} \hspace{2.0em} \sum \limits_{k=1}^{m+1} \underline{v}_{\mathcal{J},k}  = 0 
\label{SysEq1}
\end{equation}
\noindent must hold, respectively. Therefore, to fully determine the coefficients $(\underline{v}_{0,k})_{k \in \{ 1, \ldots , m+1 \} }$ and $(\underline{v}_{\mathcal{J},k})_{k \in \{ 1, \ldots , m+1 \}}$ at least $m$ additional equations are required in each case. We derive these equations by substituting (\ref{STru1}) back in OIDE (\ref{OIDE1}). Here, we focus on $\mathcal{LC} \big(u_{X}^{\mathcal{E}_{0}^{+}}\big)(\cdot)$ and will briefly comment on $\mathcal{LC} \big(u_{X}^{\mathcal{E}_{\mathcal{J}}^{+}}\big)(\cdot)$ afterwards. To start, we derive for $x < \ell$ that 
\begin{align}
\int \limits_{\mathbb{R}}  \mathcal{LC} & \big(u_{X}^{\mathcal{E}_{0}^{+}}\big)(\vartheta,x+y;\ell) f_{J_{1}}(y) dy \nonumber \\
& = \int \limits_{-\infty}^{0} \mathcal{LC} \big(u_{X}^{\mathcal{E}_{0}^{+}}\big)(\vartheta,x+y;\ell) \left( \sum \limits_{j=1}^{n} q_{j} \eta_{j} e^{\eta_{j}y } \right) dy + \int \limits_{0}^{\ell-x} \mathcal{LC} \big(u_{X}^{\mathcal{E}_{0}^{+}}\big)(\vartheta,x+y;\ell) \left( \sum \limits_{i=1}^{m} p_{i} \xi_{i} e^{-\xi_{i}y } \right) dy  \nonumber \\
& \hspace{10.5em} + \int \limits_{\ell-x}^{\infty} \mathcal{LC} \big(u_{X}^{\mathcal{E}_{0}^{+}}\big)(\vartheta,x+y;\ell) \left( \sum \limits_{i=1}^{m} p_{i} \xi_{i} e^{-\xi_{i}y } \right) dy \nonumber \\
& = \sum \limits_{j=1}^{n} q_{j} \eta_{j} e^{-\eta_{j} x } \int \limits_{-\infty}^{x} \mathcal{LC} \big(u_{X}^{\mathcal{E}_{0}^{+}}\big)(\vartheta,z;\ell) e^{\eta_{j} z} dz  + \sum \limits_{i=1}^{m} p_{i} \xi_{i} e^{\xi_{i} x } \int \limits_{x}^{\ell} \mathcal{LC} \big(u_{X}^{\mathcal{E}_{0}^{+}}\big)(\vartheta,z;\ell) e^{-\xi_{i} z}  dz  \nonumber \\
& \hspace{10.5em} + \sum \limits_{i=1}^{m} p_{i} \xi_{i} e^{\xi_{i} x } \int \limits_{\ell}^{\infty} \mathcal{LC} \big(u_{X}^{\mathcal{E}_{0}^{+}}\big)(\vartheta,z;\ell) e^{-\xi_{i} z} dz \nonumber \\
& = \sum \limits_{k=1}^{m+1} \sum \limits_{j=1}^{n} q_{j} \eta_{j} e^{-\eta_{j} x } \underline{v}_{0,k} \int \limits_{-\infty}^{x} e^{(\eta_{j} + \beta_{k,\vartheta}) z} \, e^{-\beta_{k,\vartheta} \cdot \ell} dz  + \sum \limits_{k=1}^{m+1} \sum \limits_{i=1}^{m} p_{i} \xi_{i} e^{\xi_{i} x } \underline{v}_{0,k} \int \limits_{x}^{\ell} e^{-(\xi_{i} - \beta_{k,\vartheta}) z} \, e^{-\beta_{k,\vartheta} \cdot \ell} dz  \\
& = \sum \limits_{k=1}^{m+1} \sum \limits_{j=1}^{n} \frac{q_{j} \eta_{j}}{\eta_{j} + \beta_{k,\vartheta}} \underline{v}_{0,k}\, e^{\beta_{k,\vartheta} \cdot (x-\ell)} + \sum \limits_{k=1}^{m+1} \sum \limits_{i=1}^{m} \frac{ p_{i} \xi_{i}}{\xi_{i} - \beta_{k,\vartheta}} \underline{v}_{0,k}\, e^{\beta_{k,\vartheta} \cdot (x-\ell) } - \sum \limits_{k=1}^{m+1} \sum \limits_{i=1}^{m} \frac{ p_{i} \xi_{i}}{\xi_{i} - \beta_{k,\vartheta}} \underline{v}_{0,k}\, e^{-\xi_{i}(\ell-x)} .
\end{align}
\noindent Therefore, using the relations
\begin{align}
\partial_{x} \mathcal{LC} \big(u_{X}^{\mathcal{E}_{0}^{+}}\big)(\vartheta,x;\ell)  = \sum \limits_{k=1}^{m+1} \underline{v}_{0,k} \,\beta_{k,\vartheta} \, e^{\beta_{k,\vartheta} \cdot (x-\ell)} \hspace{1.5em} \mbox{and} \hspace{1.5em} \partial_{x}^{2} \mathcal{LC} \big(u_{X}^{\mathcal{E}_{0}^{+}}\big)(\vartheta,x;\ell)  = \sum \limits_{k=1}^{m+1} \underline{v}_{0,k} (\beta_{k,\vartheta})^{2} e^{\beta_{k,\vartheta} \cdot (x-\ell)}
\end{align}
\noindent we arrive at the following equation for $x < \ell$:
\begin{align}
0 & = \mathcal{A}_{X} \mathcal{LC} \big(u_{X}^{\mathcal{E}_{0}^{+}}\big)(\vartheta,x;\ell ) -  \vartheta \, \mathcal{LC} \big(u_{X}^{\mathcal{E}_{0}^{+}}\big)(\vartheta,x;\ell ) \nonumber \\
& = \sum \limits_{k=1}^{m+1} \underline{v}_{0,k} e^{\beta_{k,\vartheta} \cdot (x-\ell)} \big( \Phi_{X}(\beta_{k,\vartheta}) - \vartheta \big) - \lambda \sum \limits_{i=1}^{m} p_{i} e^{-\xi_{i} (\ell- x)} \left( \sum \limits_{k=1}^{m+1} \frac{\xi_{i}}{\xi_{i} - \beta_{k,\vartheta}} \underline{v}_{0,k} \right) \nonumber \\
& = - \lambda \sum \limits_{i=1}^{m} p_{i} e^{-\xi_{i} (\ell- x)} \left( \sum \limits_{k=1}^{m+1} \frac{\xi_{i}}{\xi_{i} - \beta_{k,\vartheta}} \underline{v}_{0,k}  \right).
\end{align}
\noindent Since $\xi_{1}, \ldots, \xi_{m}$ are different from each other, we conclude that the following equation must hold:
\begin{equation}
\sum \limits_{k=1}^{m+1} \frac{\xi_{i}}{\xi_{i} - \beta_{k,\vartheta}} \underline{v}_{0,k}  = 0, \hspace{1.5em} \forall i \in \{1, \ldots, m \}
\label{SysEq2}.
\end{equation}
\noindent Hence, combining (\ref{SysEq1}) and (\ref{SysEq2}) results in $\underline{\mathbf{A}}_{\vartheta} \underline{\mathbf{v}}_{0} = \left(1,\mathbf{0}_{m} \right)^{\intercal}$ immediately. \vspace{1em} \\
\noindent To derive the corresponding system of equations for $\mathcal{LC} \big(u_{X}^{\mathcal{E}_{\mathcal{J}}^{+}}\big)(\cdot)$, we proceed as above. First, reproducing the derivation of the integral term, provides the following result for $x < \ell$:
\begin{align}
\int \limits_{\mathbb{R}}  \mathcal{LC} \big(u_{X}^{\mathcal{E}_{\mathcal{J}}^{+}}\big)(\vartheta,x+y;\ell) f_{J_{1}}(y) dy & = \sum \limits_{k=1}^{m+1} \sum \limits_{j=1}^{n} \frac{q_{j} \eta_{j}}{\eta_{j} + \beta_{k,\vartheta}} \, \underline{v}_{\mathcal{J},k}\, e^{\beta_{k,\vartheta} \cdot (x-\ell)} + \sum \limits_{k=1}^{m+1} \sum \limits_{i=1}^{m} \frac{ p_{i} \xi_{i}}{\xi_{i} - \beta_{k,\vartheta}} \, \underline{v}_{\mathcal{J},k}\, e^{\beta_{k,\vartheta} \cdot (x-\ell) }  \nonumber \\
& \hspace{4em} - \sum \limits_{k=1}^{m+1} \sum \limits_{i=1}^{m} \frac{ p_{i} \xi_{i}}{\xi_{i} - \beta_{k,\vartheta}} \,\underline{v}_{\mathcal{J},k}\, e^{-\xi_{i} (\ell-x)} + \sum \limits_{i=1}^{m} p_{i} e^{-\xi_{i} ( \ell - x) } .
\end{align}
\noindent Then, inserting the latter expression in OIDE (\ref{OIDE1}) gives that
\begin{equation}
0 = - \lambda \sum \limits_{i=1}^{m} p_{i} e^{-\xi_{i} (\ell- x)} \left( \sum \limits_{k=1}^{m+1} \frac{\xi_{i}}{\xi_{i} - \beta_{k,\vartheta}} \underline{v}_{\mathcal{J},k}  -1\right).
\end{equation}
\noindent This finally implies that
\begin{equation}
\sum \limits_{k=1}^{m+1} \frac{\xi_{i}}{\xi_{i} - \beta_{k,\vartheta}} \underline{v}_{\mathcal{J},k}  = 1, \hspace{1.5em} \forall i \in \{1, \ldots, m \},
\end{equation}
\noindent which immediately results, together with (\ref{SysEq1}), in $\underline{\mathbf{A}}_{\vartheta} \underline{\mathbf{v}}_{\mathcal{J}} = \left( 0, \mathbf{1}_{m} \right)^{\intercal}$. \vspace{1em} \\
\noindent To conclude, we note that the uniqueness of all the vectors $\underline{\mathbf{v}}_{0}$, $\underline{\mathbf{v}}_{\mathcal{J}}$, $\overline{\mathbf{v}}_{0}$ and $\overline{\mathbf{v}}_{\mathcal{J}}$ follows from the invertibility of the matrices $\underline{\mathbf{A}}_{\vartheta}$ and $\overline{\mathbf{A}}_{\vartheta}$ (cf.~\cite{ck11}).
\end{proof}
\begin{proof}[\bf Proof of Proposition \ref{prop5}]
\noindent Our proof mainly relies on arguments introduced in \cite{cc09} (cf.~also \cite{cy13}) and focuses, as in the proof of Propositions \ref{prop4}, on the upside first-passage probabilities. However, we note that the same techniques can be applied to derive the corresponding downside results. \vspace{1em} \\
\noindent We start by considering, for $\ell \in \mathbb{R}$, $x \in \mathbb{R} \setminus \overline{\mathcal{H}_{\ell}^{+}}$, and $\vartheta > 0$ the function 
\begin{equation}
b \, \mapsto \, \mathbb{E}_{x}^{\mathbb{P}^{X}} \left[ \exp \Big \{ -\vartheta \tau_{\ell}^{X,+} + b X_{\tau_{\ell}^{X,+}} \Big \} \right] 
\end{equation}
\noindent on the domain $ \mathcal{D} := \{ b \in \mathbb{C}: \, \mbox{Re}(b) \geq 0 \}$ and recall that the overshoot distribution is conditionally memoryless and independent of the first-passage time provided the overshoot is greater than zero and the exponential type of the jump distribution is specified, i.e.,~we have, for $\ell \in \mathbb{R}$, $x \in \mathbb{R} \setminus \overline{\mathcal{H}_{\ell}^{+}}$ and any $t > 0$, $y >0$,  
\begin{align}
\mathbb{P}^{X}_{x} \Big( X_{\tau_{\ell}^{X,+}} - \ell \geq  y \, \Big| \, X_{\tau_{\ell}^{X,+}} > \ell , \, J_{N_{\tau_{\ell}^{X,+}}} \sim \mbox{Exp}(\xi_{i})  \Big) = e^{-\xi_{i} y},
\end{align}
\noindent and
\begin{align}
\mathbb{P}^{X}_{x}  \Big( & \tau_{\ell}^{X,+} \leq t, \, X_{\tau_{\ell}^{X,+}} - \ell \geq  y \, \Big| \, X_{\tau_{\ell}^{X,+}} > \ell , \, J_{N_{\tau_{\ell}^{X,+}}} \sim \mbox{Exp}(\xi_{i})  \Big) \nonumber \\
& = \mathbb{P}^{X}_{x} \Big( \tau_{\ell}^{X,+} \leq t \, \Big| \, X_{\tau_{\ell}^{X,+}} > \ell, \, J_{N_{\tau_{\ell}^{X,+}}} \sim \mbox{Exp}(\xi_{i})  \Big) \cdot \mathbb{P}^{X}_{x} \Big( X_{\tau_{\ell}^{X,+}} - \ell \geq  y \, \Big| \, X_{\tau_{\ell}^{X,+}} > \ell , \, J_{N_{\tau_{\ell}^{X,+}}} \sim \mbox{Exp}(\xi_{i})  \Big).
\end{align}
\noindent This was already derived in \cite{ca09} and implies, in particular, that for any purely imaginary number $b \in \mathbb{C}$
\begin{align}
\mathbb{E}_{x}^{\mathbb{P}^{X}} \left[ \exp \Big\{ -\vartheta \tau_{\ell}^{X,+} + b X_{\tau_{\ell}^{X,+}} \Big\} \right] & = e^{b \cdot \ell} \; \mathbb{E}_{x}^{\mathbb{P}^{X}} \left[ e^{-\vartheta \tau_{\ell}^{X,+}} \mathds{1}_{ \mathcal{E}_{0}^{+}} \right]  + e^{b \cdot \ell} \, \sum \limits_{i=1}^{m} \mathbb{E}_{x}^{\mathbb{P}^{X}} \left[ e^{-\vartheta \tau_{\ell}^{X,+}} \mathds{1}_{ \mathcal{E}_{i}^{+}} \right]  \, \int \limits_{0}^{\infty} \xi_{i} e^{-(\xi_{i}-b) y} \, dy, \nonumber \\
& = e^{b \cdot \ell} \; \mathbb{E}_{x}^{\mathbb{P}^{X}} \left[ e^{-\vartheta \tau_{\ell}^{X,+}} \mathds{1}_{ \mathcal{E}_{0}^{+}} \right]  + e^{b \cdot \ell} \, \sum \limits_{i=1}^{m} \frac{\xi_{i}}{\xi_{i}- b} \; \mathbb{E}_{x}^{\mathbb{P}^{X}} \left[ e^{-\vartheta \tau_{\ell}^{X,+}} \mathds{1}_{ \mathcal{E}_{i}^{+}} \right].
\end{align}
\noindent Combining this identity with an analytic continuation argument will allow us to derive the required system of equations. Indeed, we first note that, for any $x \in \mathbb{R}$, $\vartheta > 0$ and any purely imaginary number $b \in \mathbb{C}$, the process $(M_{t})_{t \geq 0}$ defined via
\begin{align}
M_{t} := e^{-\vartheta t + b X_{t}} - e^{ b x} - \big(\Phi_{X}(b) - \vartheta \big) \int \limits_{0}^{t} e^{ -\vartheta s + b X_{s} } ds
\end{align}
\noindent is a zero-mean $\mathbb{P}_{x}^{X}$-martingale and obtain, for $x \in \mathbb{R} \setminus \overline{\mathcal{H}_{\ell}^{+}}$, via the optional sampling theorem that
\begin{align}
0 & = \mathbb{E}^{\mathbb{P}^{X}}_{x} \left[ \exp \Big\{ -\vartheta \tau_{\ell}^{X,+} + b X_{\tau_{\ell}^{X,+}} \Big \} \right]  - e^{bx} - \big(\Phi_{X}(b) - \vartheta \big) \,  \mathbb{E}^{\mathbb{P}^{X}}_{x} \Bigg[ \int \limits_{0}^{\tau_{\ell}^{X,+}} e^{- \vartheta s + b X_{s} } ds \Bigg]  \nonumber \\
& = \underbrace{e^{b \cdot \ell} \; \mathbb{E}_{x}^{\mathbb{P}^{X}} \left[ e^{-\vartheta \tau_{\ell}^{X,+}} \mathds{1}_{ \mathcal{E}_{0}^{+}} \right]  + e^{b \cdot \ell} \, \sum \limits_{i=1}^{m} \frac{\xi_{i}}{\xi_{i}- b} \; \mathbb{E}_{x}^{\mathbb{P}^{X}} \left[ e^{-\vartheta \tau_{\ell}^{X,+}} \mathds{1}_{ \mathcal{E}_{i}^{+}} \right] - e^{bx} - \big(\Phi_{X}(b) - \vartheta \big) \, \mathbb{E}^{\mathbb{P}^{X}}_{x} \Bigg[ \int \limits_{0}^{\tau_{\ell}^{X,+}} e^{- \vartheta s + b X_{s} } ds \Bigg] .}_{=: g(b) }
\label{RobEQ}                          
\end{align}
\noindent Hence, if one defines a new function by $G(b) := \prod \limits_{i=1}^{m} (\xi_{i}-b) \cdot g(b)$, one easily sees that $G(\cdot)$ is well-defined and (as a function of $b$) analytic on the full domain $\mathcal{D}$. Therefore, by the identity theorem for analytic functions (cf.~\cite{ru87}), we must have that $G(b) \equiv 0$ for all $b \in \mathcal{D}$. Accordingly, we must have that $g(b) \equiv 0$ for all $b \in \mathcal{D} \setminus \{ \xi_{1}, \ldots, \xi_{m} \}$. This finally allows us to replace $b$ in Equation (\ref{RobEQ}) by the positive roots $\beta_{1,\vartheta}, \ldots, \beta_{m+1,\vartheta}$ to obtain that
\begin{equation}
e^{\beta_{k,\vartheta} \cdot (x-\ell)} =  \mathbb{E}_{x}^{\mathbb{P}^{X}} \left[ e^{-\vartheta \tau_{\ell}^{X,+}} \mathds{1}_{ \mathcal{E}_{0}^{+}} \right]  + \sum \limits_{i=1}^{m} \frac{\xi_{i}}{\xi_{i}- \beta_{k,\vartheta}} \; \mathbb{E}_{x}^{\mathbb{P}^{X}} \left[ e^{-\vartheta \tau_{\ell}^{X,+}} \mathds{1}_{ \mathcal{E}_{i}^{+}} \right], \hspace{1.5em} \forall k \in \{1, \ldots, m+1 \},
\end{equation}
\noindent which gives, in view of (\ref{IntEqBup}), the required system of equations. 
\end{proof}

\begin{proof}[\bf Proof of Proposition \ref{prop6}]
\noindent Our derivations are based on the results obtained in Theorem~2.1 and Theorem~2.2 in~\cite{cy13}. Indeed, combining first the results of Theorem~2.2 with (\ref{IntEqBup}) and Proposition~\ref{prop5} gives that for $\ell \in \mathbb{R}$, $x \in \mathbb{R}\setminus \overline{\mathcal{H}_{\ell}^{+}}$ and $ \vartheta >0$ we have 
\begin{equation}
\mathcal{LC}(u_{X}^{\mathcal{E}_{0}^{+}})(\vartheta, x; \ell) = \mathbf{G}^{+} \, e^{\beta_{1,\vartheta}\cdot (x-\ell)} - \mathbf{G}^{+} \cdot \sum \limits_{k=2}^{m+1} \left( \sum \limits_{s=1}^{m} \frac{d^{+}_{k,s}}{\xi_{s} - \beta_{1,\vartheta}} \right) \cdot e^{\beta_{k,\vartheta}\cdot(x-\ell)},
\label{RESULT1}
\end{equation} 
\noindent and, for $i \in \{1, \ldots, m \}$,
\begin{equation}
\mathcal{LC}(u_{X}^{\mathcal{E}_{i}^{+}})(\vartheta, x; \ell) = - \frac{\mathbf{G}^{+}}{\xi_{i}} \frac{\mathbf{C}_{\vartheta}^{+}(\xi_{i})}{ \big(\mathbf{B}^{+} \big)'(\xi_{i})} \,e^{\beta_{1,\vartheta}\cdot (x-\ell)} + \frac{1}{\xi_{i}} \sum \limits_{k=2}^{m+1} \left( d_{k,i}^{+} + \mathbf{G}^{+} \,\frac{\mathbf{C}_{\vartheta}^{+}(\xi_{i})}{ \big(\mathbf{B}^{+} \big)'(\xi_{i})} \sum \limits_{s=1}^{m} \frac{d^{+}_{k,s}}{ \xi_{s} - \beta_{1,\vartheta}}  \right) \cdot e^{\beta_{k,\vartheta}\cdot(x-\ell)},
\label{RESULT2}
\end{equation}
\noindent with $\mathbf{G}^{+} := \frac{\mathbf{B}^{+}(\beta_{1,\vartheta})}{\mathbf{C}^{+}_{\vartheta}(\beta_{1,\vartheta})} $ and $\mathbf{B}^{+}(\cdot)$, $\mathbf{C}_{\vartheta}^{+}(\cdot)$ and $d_{i,j}^{+}$ defined as in (\ref{EqB&C}) and (\ref{Eqdplus}). Therefore, comparing Result (\ref{RESULT1}) with the corresponding results in Proposition~\ref{prop4} already gives that
\begin{equation}
\underline{v}_{0,1} = \mathbf{G}^{+} \hspace{1.8em} \mbox{and} \hspace{2em} \underline{v}_{0,k} = -\mathbf{G}^{+} \cdot \sum \limits_{s=1}^{m} \frac{d^{+}_{k,s}}{\xi_{s} - \beta_{1,\vartheta}} , \hspace{0.7em} k \in \{2, \ldots, m+1\} ,
\end{equation}
\noindent and substituting this identity back in (\ref{RESULT2}) finally gives the results (\ref{JuMpEq1}) and (\ref{JumpCO1}). \vspace{1em} \\
\noindent Deriving the results for the downside case can be done in the same way. First, combining Theorem~2.1 in~\cite{cy13} with (\ref{IntEqBup}) and Proposition~\ref{prop5} gives, for $\ell \in \mathbb{R}$, $x \in \mathbb{R}\setminus \overline{\mathcal{H}_{\ell}^{-}}$ and $ \vartheta >0$, that
\begin{equation}
\mathcal{LC}(u_{X}^{\mathcal{E}_{0}^{-}})(\vartheta, x; \ell) = \mathbf{G}^{-} \, e^{\gamma_{1,\vartheta}\cdot (x-\ell)} - \mathbf{G}^{-} \cdot \sum \limits_{k=2}^{n+1} \left( \sum \limits_{s=1}^{n} \frac{d^{-}_{k,s}}{\eta_{s} + \gamma_{1,\vartheta}} \right) \cdot e^{\gamma_{k,\vartheta}\cdot(x-\ell)}
\label{RESULT1down}
\end{equation} 
\noindent and, for $j \in \{1, \ldots, n \}$,
\begin{align}
\mathcal{LC}(u_{X}^{\mathcal{E}_{j}^{-}})(\vartheta, x; \ell) & = (-1)^{n} \, \frac{\mathbf{G}^{-}}{\eta_{j}} \frac{\mathbf{C}_{\vartheta}^{-}(\eta_{j})}{ \big(\mathbf{B}^{-} \big)'(-\eta_{j})} \,e^{\gamma_{1,\vartheta}\cdot (x-\ell)}  \nonumber \\
& \hspace{6.5em} + \frac{1}{\eta_{j}} \sum \limits_{k=2}^{n+1} \left( d_{k,j}^{-} + (-1)^{n-1} \, \mathbf{G}^{-} \,\frac{\mathbf{C}_{\vartheta}^{-}(\eta_{j})}{ \big(\mathbf{B}^{-} \big)'(-\eta_{j})} \sum \limits_{s=1}^{n} \frac{d^{-}_{k,s}}{ \eta_{s} + \gamma_{1,\vartheta}}  \right) \cdot e^{\gamma_{k,\vartheta}\cdot(x-\ell)}
\label{RESULT2down}
\end{align}
\noindent with $\mathbf{G}^{-} := (-1)^{n} \, \frac{\mathbf{B}^{-}(\gamma_{1,\vartheta})}{\mathbf{C}^{-}_{\vartheta}(-\gamma_{1,\vartheta})} $ and $\mathbf{B}^{-}(\cdot)$, $\mathbf{C}_{\vartheta}^{-}(\cdot)$ and $d_{i,j}^{-}$ defined as in (\ref{EqB&Cdown}) and (\ref{Eqdplusdown}). Therefore, comparing Result (\ref{RESULT1}) with the corresponding results in Proposition \ref{prop4} already gives that
\begin{equation}
\overline{v}_{0,1} = \mathbf{G}^{-} \hspace{1.8em} \mbox{and} \hspace{2em} \overline{v}_{0,k} = -\mathbf{G}^{-} \cdot \sum \limits_{s=1}^{n} \frac{d^{-}_{k,s}}{\eta_{s} + \gamma_{1,\vartheta}} , \hspace{0.7em} k \in \{2, \ldots, n+1\} 
\end{equation}
\noindent holds and substituting this identity back in (\ref{RESULT2down}) finally gives (\ref{JuMpEq2}) and (\ref{JumpCO2}).
\end{proof}

\end{document}